\documentclass[
aps,
twocolumn,
amsmath,
amssymb,
floatfix,
showpacs,
superscriptaddress,
nofootinbib,
longbibliography
]{revtex4-2}
\usepackage[T1]{fontenc}
\usepackage{mathtools}
\usepackage{braket}
\usepackage{bm}
\usepackage{MnSymbol}
\usepackage{graphicx}
\usepackage{tikz}
\usepackage{pgffor}
\usepackage{blkarray}
\usepackage{sidecap}

\usepackage[dvipsnames]{xcolor}
\usepackage{enumitem}
\usepackage[normalem]{ulem}
\usepackage{silence}

\usepackage[
colorlinks,
linkcolor=black,
citecolor=black,
urlcolor=black
]{hyperref}

\mathchardef\mhyphen="2D

\newcommand\bea{\begin{eqnarray}}
\newcommand\eea{\end{eqnarray}}
\newcommand\beq{\begin{equation}}
\newcommand\eeq{\end{equation}}

\definecolor{lime}{HTML}{A6CE39}

\DeclareRobustCommand{\orcidicon}{%
\hspace{-1.0mm}
\begin{tikzpicture}
  \draw[lime, fill=lime] (0,0)
  circle [radius=0.15]
  node[white] {{\fontfamily{qag}\selectfont \tiny \,ID}};
  \draw[white, fill=white] (-0.0525,0.095)
  circle [radius=0.007];
\end{tikzpicture}
\hspace{-3.0mm}
}

\foreach \x in {A,...,Z}{%
  \expandafter\xdef\csname orcid\x\endcsname{%
    \noexpand\href{https://orcid.org/\csname orcidauthor\x\endcsname}%
    {\noexpand\orcidicon}}%
}

\begin{document}
\title{Probing topological phase transitions via nonlinear Hall response in strained moiré dice lattice}
\author{Gourab Paul\orcidA{}}
\email[]{p.gourab@iitg.ac.in} 
\affiliation{Department of Physics, Indian Institute of Technology Guwahati, Guwahati-781039, Assam, India}
\author{Srijata Lahiri\orcidB{}}
\email[]{srijata.lahiri@iitg.ac.in} 
\affiliation{Department of Physics, Indian Institute of Technology Guwahati, Guwahati-781039, Assam, India}
\author{Bilal Tanatar\orcidC{}}
\email[]{tanatar@fen.bilkent.edu.tr}
\affiliation{Department of Physics, Bilkent University, 06800 Bilkent, Ankara, Türkiye}
\author{Saurabh Basu}
\email[]{saurabh@iitg.ac.in}
\affiliation{Department of Physics, Indian Institute of Technology Guwahati, Guwahati-781039, Assam, India}
\begin{abstract} 
Valley polarized twisted bilayer dice lattice hosts topologically nontrivial flat bands far from charge neutrality due to broken 
time reversal symmetry, whereas the ones in the vicinity of it remain topologically trivial. However, 
when both valleys are taken into consideration, the time reversal symmetry is preserved, which poses a serious hindrance to enumerate the valley specific topological phases that rely on the detection of the Berry curvature. In this work, we demonstrate that such a twisted structure with an applied uniaxial strain exhibits a nonlinear Hall effect far from charge neutrality. We ascertain that the nonlinear anomalous Hall signals can serve as a probe for topological phase transitions associated with a specific energy state that is constrained to reside at the lower edge of the middle subband and controlled via a staggered mass. Specifically, we show that the nonlinear anomalous Hall response undergoes a sign
reversal across the topological phase boundaries. By tuning the carrier density,
we compute the nonlinear Hall response obtained from the Berry curvature dipole, both in the chiral limit, and also 
when the chiral symmetry is broken. It is further seen that the nonlinear Hall effect is significantly enhanced in the broken chiral symmetry regime.
\end{abstract}

\maketitle
\noindent{T}he emergence of flat bands in long-period moiré systems with small twist angles has been under persistent focus on both theoretical and experimental fronts over the past few years~\cite{Cao2020,Cao2018,Tanaka2025,Ledwith2025,Mesple2025,Song2024,Yang2024,Devakul2021,Lei2025,Bao2025,Yang2024A,Lou2021,Sharma2021,Ma2024,Zhou2024,Gandhi2026}. The flat bands with vanishing Fermi velocity in twisted bilayers lead to strong electronic correlations, which in turn give rise to a wide range of exotic physical phenomena, such as Mott insulating states~\cite{Po2018,Chen2021,Klug2020}, unconventional superconductivity~\cite{Balents2020,Isobe2018,Yankowitz2019,Oh2021}, complex magnetic phases~\cite{Duran2022,Seo2019}, among several others. The emergence of correlated phases, anomalous quantum Hall (AQH)~\cite{Serlin2020,He2026,Tateishi2022} and fractional quantum Hall (FQH)~\cite{Xu2023,Reddy2023,Park2023} physics, which provide a direct probe of topology in electronic systems, have also been under active investigation in the context of moiré systems.

In the anomalous Hall effect, to generate a finite Hall response one needs to break the time reversal symmetry (TRS) 
which can be achieved through an external magnetic field or via magnetic dopants~\cite{Nagaosa2010}.
Moreover, the second-order Hall response, namely the nonlinear Hall effect (NLHE), has recently attracted significant interest due to its fundamental distinction from the conventional linear Hall effect, particularly in its ability to emerge in 
time-reversal symmetric systems, albeit with broken inversion symmetry ~\cite{Du2018,Low2015,Zhang2018,Tokura2018,Zhang2022}.

While TRS forces the net Berry curvature over a complete Brillouin zone (BZ) to vanish, it does not constrain the first moment of Berry curvature (also called the Berry curvature dipole) to be zero as well, thereby allowing the emergence 
of NLHE.
From an experimental standpoint, NLHE serves as an effective probe of topological phase transitions specifically in systems with TRS, owing to the complete disappearance of Berry curvature induced response.
In fact, the nonlinear Hall (NLH) response undergoes characteristic sign change across the phase boundaries, allowing efficient detection of the band inversion due to gap closing in topological systems~\cite{Zhang2022,Chakraborty2022,Hu2022}.
The NLHE has been widely reported in non-centrosymmetric materials, including corrugated graphene~\cite{Ho2021}, strained WSe$_2$~\cite{Qin2021}, TaIrTe$_4$~\cite{Kumar2021}, and MoTe$_2$~\cite{Ma2022}. In the context of moiré systems, the NLHE has been experimentally observed in twisted bilayer graphene (TBG)~\cite{Duan2022,Huang2023}, twisted bilayer WSe$_2$~\cite{Huang2023A,Cao2025}, and twisted bilayer MoS$_2$~\cite{Wu2023}. The NLHE has also been predicted on theoretical grounds in several moiré materials, including the strained TBG~\cite{Zhang2022}, twisted bilayer WTe$_2$~\cite{He2021}, twisted bilayer WSe$_2$~\cite{Hu2022}, and twisted double bilayer graphene (TDBLG)~\cite{Chakraborty2022}.

In recent years, a distinct class of moiré systems has attracted considerable attention, in which the monolayer itself hosts a flat band, exemplified by the twisted bilayer dice lattice (TBDL)~\cite{Zhou2024, Ma2024, Paul2026, Zhou2026}.
Studies on TBDL reveal that it hosts an isolated, highly degenerate flat band for all (small) values of twist angle 
$\theta$~\cite{Zhou2024}. Furthermore, on introduction of an inversion symmetry breaking Semenoff mass term, this degenerate flat band exhibits mixed topological characteristics, with multiple phase transitions being exhibited by a single non-degenerate eigenstate at the middle band edge~\cite{Paul2026}. Herein, we investigate the emergence of the NLH response in strained TBDL, which is a time reversal symmetric system with reduced crystalline symmetries and hosts highly degenerate bands near the zero energy. The bands far away from charge neutrality host direct and indirect band gaps, thereby also exhibiting topological features. To this end, the efficiency of NLHE in detecting the emergent topological phase transitions is also established.
\begin{figure}[]
    \centering
\includegraphics[width=0.48\textwidth]{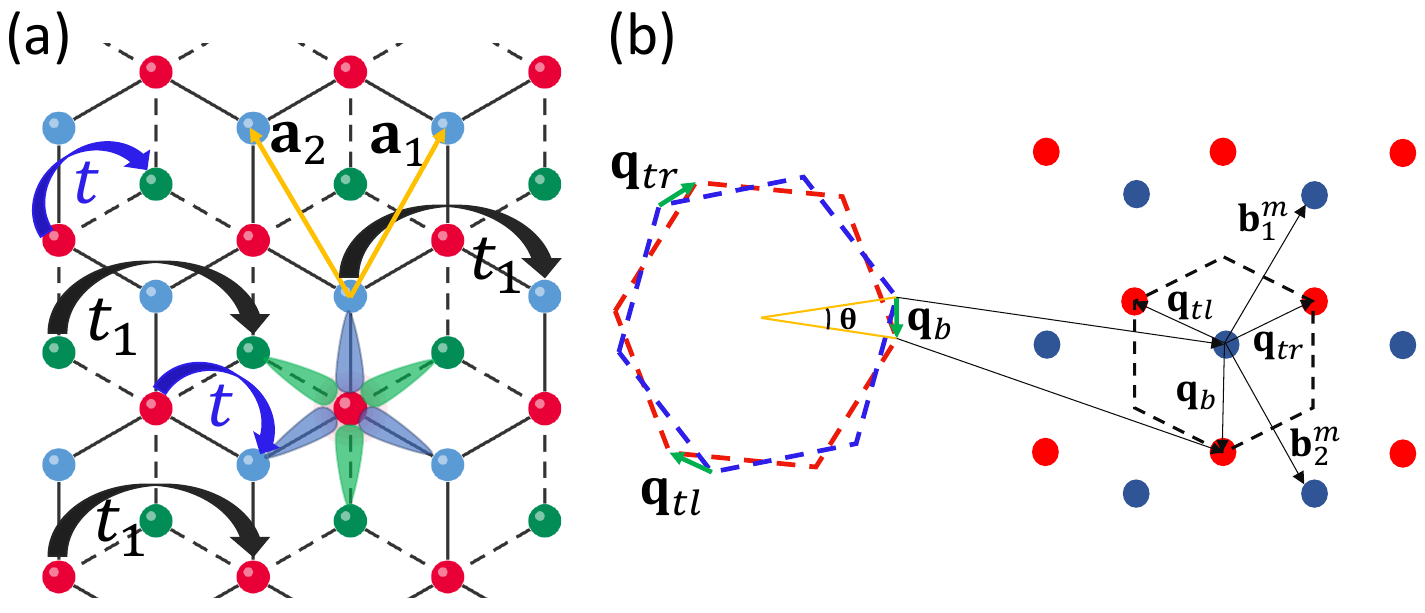}
    \caption{\textbf{Schematic diagram of the monolayer dice lattice.} (a) The blue, red, and green dots represent the $A$, $B$, and $C$ sublattice atoms, respectively. The primitive lattice vectors are denoted by $\mathbf{a}_1$ and $\mathbf{a}_2$. The NN hopping amplitude between $A$-$B$ and $B$-$C$ sublattice atoms is denoted by $t$, while the NNN hopping amplitude between $A$-$A$, $B$-$B$, and $C$-$C$ subalattice sites is represented by $t_1$. (b) Brillouin zones of the top and bottom  layers of the TBDL, shown by blue and red dashed lines, respectively, rotated by a relative twist angle $\theta$ with respect to each other. The corresponding moiré Brillouin zone (MBZ) is characterized by the moir\'e reciprocal lattice vectors $\mathbf{b}_1^m = \dfrac{8\pi \sin(\theta/2)}{\sqrt{3}a} \left( \dfrac{1}{2}, \dfrac{\sqrt{3}}{2} \right)$ and $\mathbf{b}_2^m = \dfrac{8\pi \sin(\theta/2)}{\sqrt{3}a} \left( \dfrac{1}{2}, -\dfrac{\sqrt{3}}{2} \right)$, respectively. The momentum transfer vectors connecting the nearest Dirac points of the two layers in the MBZ are denoted by $\textbf{q}_b$, $\textbf{q}_{tr}$, and $\textbf{q}_{tl}$.}\label{Model_Ham}
\end{figure}

\textcolor{black}{The remainder of this work is organized as follows. We begin by discussing the unstrained version of the TBDL followed by a systematic introduction of strain into the Hamiltonian, thereby enabling the emergence of non-zero NLH response. The broken $C_3$ symmetry is verified by analyzing the bandstructures along the high symmetry lines in the MBZ. Hence, we analyze the Chern number phase plot for the eigenstate at the middle band edge as a function of an onsite mass $\Delta$ and the twist angle $\theta$, under weakly strained condition. The strain, together with broken inversion symmetry, then leads to the emergence of non-zero NLHE which shows sign changes across the phase boundaries. The probing of different phases in the Chern phase plot, employing the NLH response is then elaborately studied. Finally, we conclude our discussion by analyzing the variation of the Berry curvature dipole (BCD) components as a function of the magnitude of strain, which further establishes the role of $C_3$ symmetry breaking in generating the second order Hall response. The main highlight of our work corresponds to the detection of topological phase boundaries of an eigenstate away from charge neutrality in strained TBDL, in the $\Delta-\theta$ plane, employing the behavior of the planar components of BCD.}

\noindent\textbf{Results}

\noindent \textcolor{blue}{\textit{Continuum model for unstrained  TBDL.-}}
We begin by considering a non-rotated monolayer (layer-1) of the dice lattice [see Fig.~\ref{Model_Ham}(a)], which can be viewed as two interpenetrating honeycomb networks (formed by the $A$-$B$ and $B$-$C$ sublattices separately) sharing a common site at sublattice $B$.
The real-space lattice vectors are defined as 
$\mathbf{a}_1 = a \left( \frac{1}{2}, \frac{\sqrt{3}}{2} \right)$ and $\mathbf{a}_2 = a \left( -\frac{1}{2}, \frac{\sqrt{3}}{2} \right)$, 
where $a = \sqrt{3}d$ is the lattice constant, and $d$ denotes the nearest-neighbor (NN) bond length.
The corresponding reciprocal lattice vectors are given by, $\mathbf{b}_1 = \frac{4\pi}{\sqrt{3}a} \left( \frac{\sqrt{3}}{2}, \frac{1}{2} \right)$ and $\mathbf{b}_2 = \frac{4\pi}{\sqrt{3}a} \left( -\frac{\sqrt{3}}{2}, \frac{1}{2} \right)$. The NN hopping amplitudes between $A\rightarrow B$ and $B \rightarrow C$ sublattices are denoted by $t$, whereas the next-nearest-neighbor (NNN) hopping amplitudes between the $A \rightarrow A$, $B\rightarrow B$, and $C \rightarrow C$ sublattice atoms are represented by $t_1$.
With the aforementioned NN and NNN hopping terms, the emergence of valley features is observed at the $K_+$ and $K_-$ points in the BZ of a monolayer dice lattice, where the band structure displays triple degeneracy, thereby facilitating a low-energy effective description in the vicinity of the valleys. 
Extensive calculations pertaining to monolayer dice lattice have been provided in the Supplementary Material (SM).

Now, we shift our attention to the composite bilayer of the dice Hamiltonian, where the top layer of the composite structure is twisted relative to the bottom.
Precisely, the top (bottom) layer is rotated by an angle $\theta/2$ (-$\theta/2$) resulting in an effective rotation of $\theta$ between the layers.
We now focus on the low-energy physics of the TBDL at small values of $\theta$. This can be approximated by the continuum model constructed employing the Bistritzer–MacDonald formalism~\cite{Bistritzer2011,Ma2024,Paul2026} as,
\begin{equation}
    H_{\zeta}(\theta) = 
\begin{pmatrix}
H_{t,\zeta}\left( \frac{\theta}{2} \right) & U_{\zeta} \\
U_{\zeta}^\dagger & H_{b,\zeta}\left( -\frac{\theta}{2} \right)
\end{pmatrix}.
\label{Twisted_Bilayer_Ham}
\end{equation}
Here, $H_{t/b,\zeta}$ ($U_{\zeta}$) denotes the intralayer (interlayer) Hamiltonian for the top/bottom layer corresponding to the valley index $\zeta$.
Thorough discussions on the construction of the intra and interlayer Hamiltonians have been provided in the SM.
It suffices to note here that for the interlayer Hamiltonian $U_{\zeta}$, $w_{1,2,3}$ represent three distinct interlayer hopping strengths. Specifically, $w_1$ corresponds to the intrasublattice hopping between $A$-$\tilde{A}$, $B$-$\tilde{B}$, and $C$-$\tilde{C}$, while $w_2$ ($w_3$) corresponds to the intersublattice hopping between $A$-$\tilde{B}$ and $B$-$\tilde{C}$ ($A$-$\tilde{C}$) sites.
The physical values of $w_{1,2,3}$ are determined by the orbital character of the electrons on each sublattice. For the TBDL, the sublattice symmetry, together with the presence of an $s$ orbital on the $B$ sublattice and $sp_2$ orbitals on the $A$ and $C$ sublattices [see Fig.~\ref{Model_Ham}(a)], lead to a hierarchy, with $w_1 \sim w_2 \gg w_3$. Under these conditions, it is assumed that the $sp_2$ orbitals, due to their $C_{3z}$ symmetry, exhibit reduced overlap of the wave functions between the $A$ and $C$ sublattice sites~\cite{Zhou2024}.

For a major part of the discussion that follows, we consider a chiral symmetric version of the model where the texture of the intralayer hopping amplitudes is exactly emulated in the interlayer hopping as well. More specifically, the suppression of the intra-sublattice hopping in the monolayer Hamiltonian, which preserves the chiral symmetry, leads to protection of the same as well in the full composite structure.
It is worth noting that an exact chiral symmetry is realized only in the presence of NN hoppings alone. Nevertheless, we include a weak NNN hopping term (with strength $t_1$), which renders the flat band of the individual monolayers become dispersive and induces a valley structure~\cite{Paul2026}, with an amplitude much smaller than the NN hopping strength $t$.
Thereby, the chiral symmetric model of the TBDL is represented by $w_1 = w_3 = 0$ and $w_2 =110.7$~meV, in which each layer of the dice lattice also possesses $\mathcal{C}_6$ symmetry, comprising of a \textit{threefold rotation} $\mathcal{C}_{3}$ and a \textit{twofold rotation} $\mathcal{C}_{2}$~\cite{Sukhachov2023}.
Since the spin–orbit coupling is effectively neglected in this system, the \textit{twofold rotation} $\mathcal{C}_2$ coincides with the $\mathcal{C}_{2z}$ symmetry. Furthermore, pristine TBDL also preserves the TRS ($\mathcal{T}$) in the absence of any magnetic (field or dopant) terms.

The presence of $\mathcal{C}_{2z}\mathcal{T}$ causes pristine TBDL to host a highly degenerate isolated flat band at energy $E = 0$ for all values of the twist angle $\theta$. This causes the Berry curvature to be ill-defined in the MBZ, unless the degeneracy is lifted~\cite{Ma2024,Paul2026,Zhou2024}.
Thereby, to induce the breaking of $\mathcal{C}_{2z}$ symmetry and hence lift the degeneracy of the flat band, TBDL can be aligned with two hBN substrates, yielding anisotropic onsite mass terms within each individual layer. Consequently, two separate nearly flat isolated bands emerge at the middle band edges, which exhibit topological phase transitions as a function of the twist angle $\theta$~\cite{Paul2026}. In this work, we break the $\mathcal{C}_{2z}$ symmetry in TBDL by introducing an onsite staggered mass term $\Delta S_3$ in the $A$ and $C$ sublattice sector analogous to the effect of hBN substrate layers, with $\Delta$ serving as a tunable parameter, while constraining the mass at the $B$ sublattice to be zero. While it is also possible to align the hBN layer such that the staggered mass term is projected onto the $A$ and $B$ sublattices, this would only shift the entire eigenspectra by a constant amount, and not lead to any change in the topological features that forms a key focus of our subsequent discussion.
\begin{figure}
    \centering
    \includegraphics[width=0.48\textwidth]{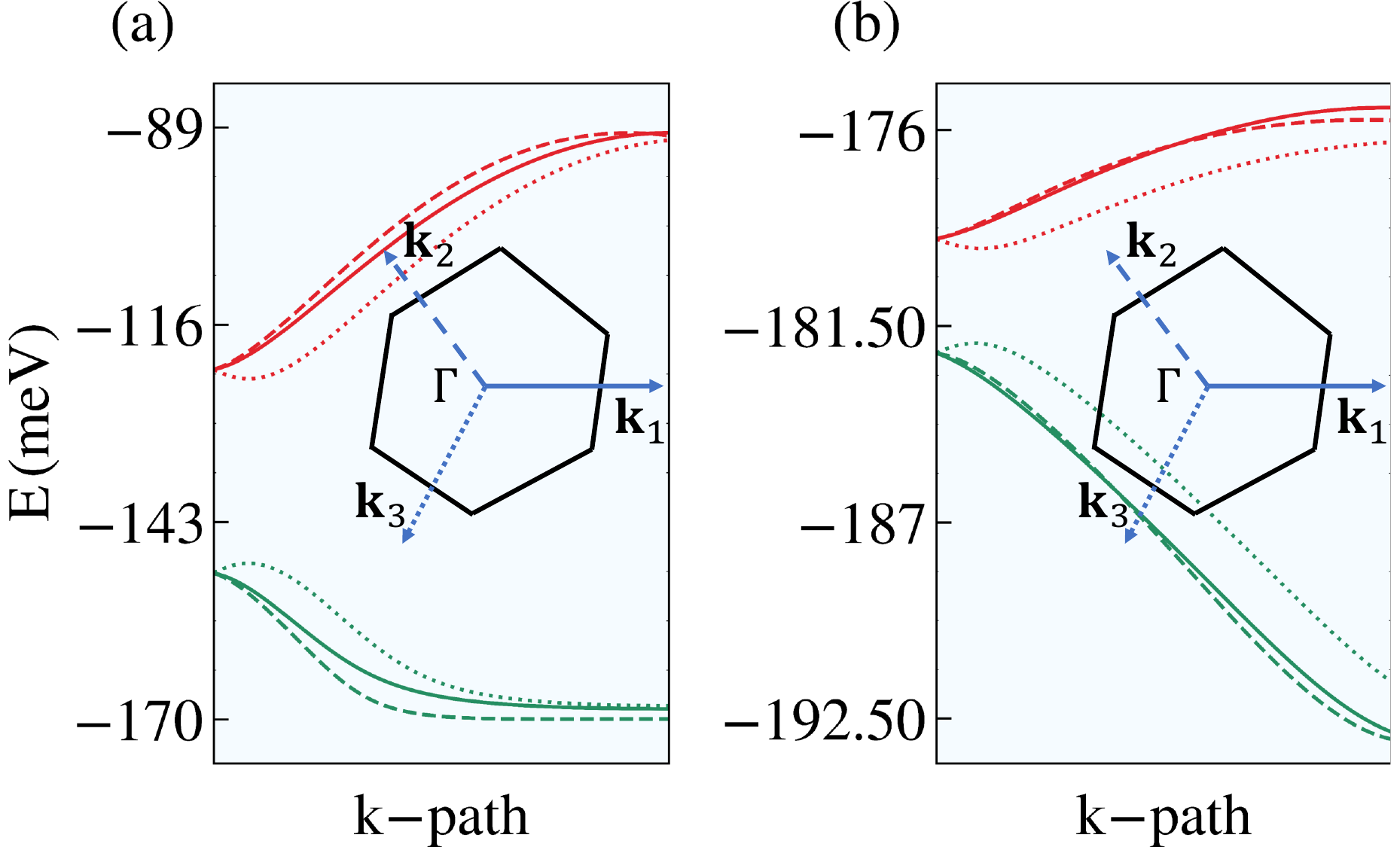}
    \caption{\textbf{Band structures of TBDL at the magic angle in presence of a uniaxial strain $\mathcal{E}_p = 0.2\%$.} Panel (a) presents the energy spectrum in the chiral limit of the system, while panel (b) shows the corresponding energy dispersions in the broken chiral symmetry regime (implemented by setting $w_1=w_2 =60$~meV), for the lowest band at the middle band edge and its nearest high energy band. The calculations are performed with $\Delta = 170$~meV for panel (a) and $\Delta = 60$~meV for panel (b). The energy dispersions along the three distinct $\mathcal{C}_3$ symmetric paths ($\Gamma$–M$_1$, $\Gamma$–M$_2$, and $\Gamma$–M$_3$), represented by solid, dashed, and dotted blue lines, deviate from one another, indicating the breaking of $\mathcal{C}_3$ symmetry in the system. Other relevant system parameters have been mentioned in the SM.}
    \label{Fig:2}
\end{figure}  

It is important to emphasize here that the continuum model of the effective moiré Hamiltonian, assumes a complete suppression of the intervalley scattering, therefore allowing a valley specific analysis of topological phases and phase transitions.
The resulting MBZ of a moiré lattice, therefore inherently exhibits broken TRS and leads to the emergence of non-trivial Berry curvature effects on inclusion of an inversion symmetry breaking onsite mass.
A detailed discussion of the valley Chern phases in the presence of a staggered onsite mass term $\Delta S_3$ is provided in the SM.
However, it is to be noted that on consideration of both the $K_+$ and $K_-$ valleys, the TRS of the full Hamiltonian is no longer broken.
Owing to this protection, the total Chern number obtained by adding the contributions of Berry curvature from both the valleys add up to zero. \textcolor{black}{As a result, the bulk band geometry as well as the topological phase transitions are difficult to detect experimentally due to the absence of circulating edge states carrying any net charge.}
However, under reduced crystalline symmetries, specifically under broken inversion and $\mathcal{C}_3$, time reversal symmetric systems exhibit emergence of NLHE owing to non vanishing BCD.
The BCD, therefore acts as a significant tool to detect the topological phase transitions in systems with protected TRS.\\
\noindent\textcolor{blue}{\textit{Continuum model for strained  TBDL.-}} As mentioned earlier, the emergence of BCD in a time reversal symmetric system requires a reduction in its crystalline symmetries, specifically inversion and $\mathcal{C}_3$.
In TBDL, we break the $\mathcal{C}_3$ symmetry, by applying a uniaxial strain which causes a shift in the location of the Dirac cones without gapping them out.
Since strains are found to be prevalent in twisted materials~\cite{Kerelsky2019, Xie2019, Choi2019, Huang2022, Pantaleon2022, Ouyang2025, Hou2025}, the $\mathcal{C}_3$ symmetry is inherenty expected to be broken in TBDL as well, thereby allowing a nonzero BCD and hence NLHE to emerge. In general, while it is possible to strain both layers separately, we restrict ourselves to the application of the same in the bottom layer only. Now, to incorporate this effect in the bottom layer Hamiltonian, we introduce a linear strain tensor $\mathbf{\mathcal{E}}$, which transforms the real space and the reciprocal space coordinates ($\mathbf{r}$, $\mathbf{q}$) in phase space as~\cite{Bi2019}, 
\begin{eqnarray}
    \mathbf{r}^\prime &=& (\mathbb{I} + \mathbf{\mathcal{E}})\,\mathbf{r} \nonumber \\
    \mathbf{q}^\prime &=& (\mathbb{I} + \mathbf{\mathcal{E}^T})^{-1}\,\mathbf{q} \approx (\mathbb{I} - \mathbf{\mathcal{E}^T})\,\mathbf{q}.
\end{eqnarray}
Without any loss of generality, we consider a uniaxial strain of magnitude $\mathcal{E}_p$ applied along an axis that makes an angle $\Phi$ with respect to the zigzag direction. The explicit form of the corresponding strain tensor can be written as~\cite{Bi2019, Pereira2009},
\begin{eqnarray}
\mathbf{\mathcal{E}} &=& R_\Phi
\begin{pmatrix}
\mathcal{E}_p & 0 \\
0 & -\sigma \mathcal{E}_p
\end{pmatrix}
R^{-1}_\Phi \nonumber \\
&=& \mathcal{E}_p
\begin{pmatrix}
\cos^2 \Phi - \sigma \sin^2 \Phi & (1+\sigma)\cos \Phi \sin \Phi \\
(1+\sigma)\cos \Phi \sin \Phi & \sin^2 \Phi - \sigma \cos^2 \Phi
\end{pmatrix},
\end{eqnarray}
where $\sigma$ is the Poisson ratio. For the dice lattice we consider $\sigma=0.165$, which is assumed to be same as that of graphene~\cite{Pereira2009}. For the bottom layer Hamiltonian, strain introduces an effective gauge field~\cite{Sun2022,Bi2019} given by,
\begin{equation}
\mathbf{A} = \frac{\beta}{d}
\left( \mathcal{E}_{xx} - \mathcal{E}_{yy}\,, -2\mathcal{E}_{xy} \right),
\end{equation}
where $\beta$ $(= 1.57)$ is the Grüneisen parameter~\cite{Bi2019}, which is also taken to be the same as graphene. The positions of the Dirac points in the bottom layer in the presence of this gauge field, are shifted to~\cite{Bi2019}
\begin{equation}
\mathbf{\mathcal{D}}_{\zeta} = \left(\mathbb{I} - \mathbf{\mathcal{E}}^{T}\right)\,\mathbf{K}_{b,\zeta} - \zeta\,\mathbf{A}. 
\end{equation}
Now, on inclusion of the staggered onsite mass term $\Delta S_3$, the bottom layer Hamiltonian can be written as,
\begin{eqnarray}
 H_{b,\zeta}^{\prime} =  v_F \,{R}_{\theta/2} \left[ (\mathbb{I} + \mathbf{\mathcal{E}}^T)\,\mathbf{q}^{\prime} + \zeta \mathbf{A} \right].(\zeta S_1, S_2) + \Delta S_3.
\end{eqnarray}
It is important to mention here that strain also affects the interlayer terms which is incorporated through the modifications in the momentum transfer associated with the interlayer hopping processes (for detailed calculations see SM).

To establish broken $\mathcal{C}_3$ symmetry in presence of strain, we now plot the band structure of TBDL at the magic angle $\theta = 1.08^\circ$ in both the chiral limit (with $w_1 = w_3 = 0$ and $w_2 = 110.7$~meV) and the broken chiral symmetry regime (by setting $w_1 = w_3 = 60$~meV and $w_2 = 110.7$~meV), in the presence of a uniaxial strain $\mathcal{E}_p = 0.2\%$, applied along the zigzag direction ($\Phi = 0$), as shown in Fig.~\ref{Fig:2}(a) and Fig.~\ref{Fig:2}(b), respectively. The energy dispersions are computed along the three distinct $\mathcal{C}_3$ symmetric paths ($\Gamma$–M$_1$, $\Gamma$–M$_2$, and $\Gamma$–M$_3$), shown by solid, dashed, and dotted blue lines in the unstrained MBZ. As evident from Fig.~\ref{Fig:2}(a) and Fig.~\ref{Fig:2}(b), the bands exhibit pronounced deviations from one another along these $\mathcal{C}_3$ symmetric paths, thereby clearly indicating at the broken $\mathcal{C}_3$ symmetry in strained TBDL. 

\begin{figure}
    \centering
    \includegraphics[width=0.45\textwidth]{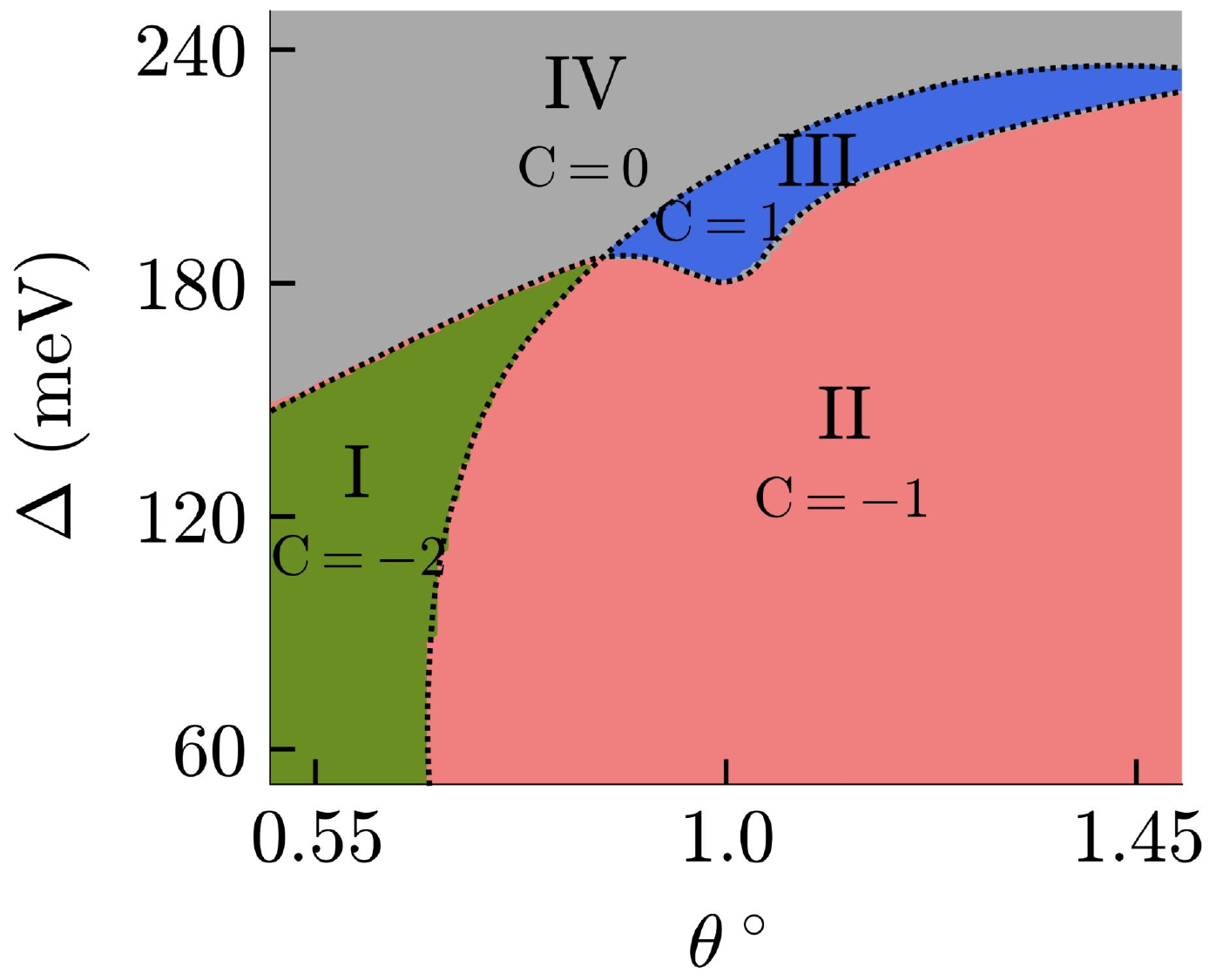}
    \caption{\textbf{Chern number phase diagram for the lowest eigenstate of the middle sub-band.} The figure shows emergent topological phases for the band index $n=162$ of strained TBDL, as a function of the onsite mass $\Delta$ and the twist angle $\theta$. The system has been strained weakly with the magnitude being $\mathcal{E}=0.2\%$.}
    \label{Fig:Phase1}
\end{figure}

We now begin our study by analyzing the topological phases that emerge in this strained TBDL as a function of the onsite potential $\Delta$ and the twist angle $\theta$. We restrict ourselves to the application of a weak strain such that the emergent strained topological phase diagram closely resembles the same in the unstrained limit, enabling the BCD to approximately map the unstrained topological phases as well.
The phase diagram depicting the Chern number of the eigenstate with eigenvalue at the edge of the middle sub-band (band index $n=162$), at a value of $\mathcal{E}_p=0.2\%$, in the $\Delta-\theta$ plane and within the chiral symmetry regime is shown in Fig. \ref{Fig:Phase1}. The emergence of multiple topological phases corresponding to Chern number $C=1, -1$ and $-2$ is observed, separated by distinct transition lines.
We observe that for high values of $\Delta$, the band transitions into a completely trivial phase while for its lower values, the $C=-1$ phase is predominant.
Now, we study the emergence of BCD in strained TBDL and analyze its variation across the phases discussed above.\\
\noindent \textcolor{blue}{\textit{Emergence of NLH response in TBDL.-}}
In NLHE, the application of an in-plane ac electric field $\mathbf{E}(t) = \mathrm{Re}\left(E\, e^{i\omega t}\right)$ with frequency $\omega$ induces a transverse Hall current $J(t) = \mathrm{Re}\left(j^{0} + j^{2\omega} e^{2 i \omega t}\right)$ with a rectified component $j^{0}_{a} = \chi_{abc}\,E_{b} E^{*}_{c}$ and a second-harmonic component $j^{2\omega}_{a} = \chi_{abc}\, E_{b}E_{c}$, with frequency $2\omega$. Here, the NLH susceptibility is given by,
\begin{eqnarray}
\chi_{abc} = - \frac{e^3 \tau}{2 (1 + i \omega \tau)}\, \epsilon_{adc}\, \mathcal{D}_{bd},
\end{eqnarray}
\begin{figure*}
    \centering
    \includegraphics[width=1\textwidth]{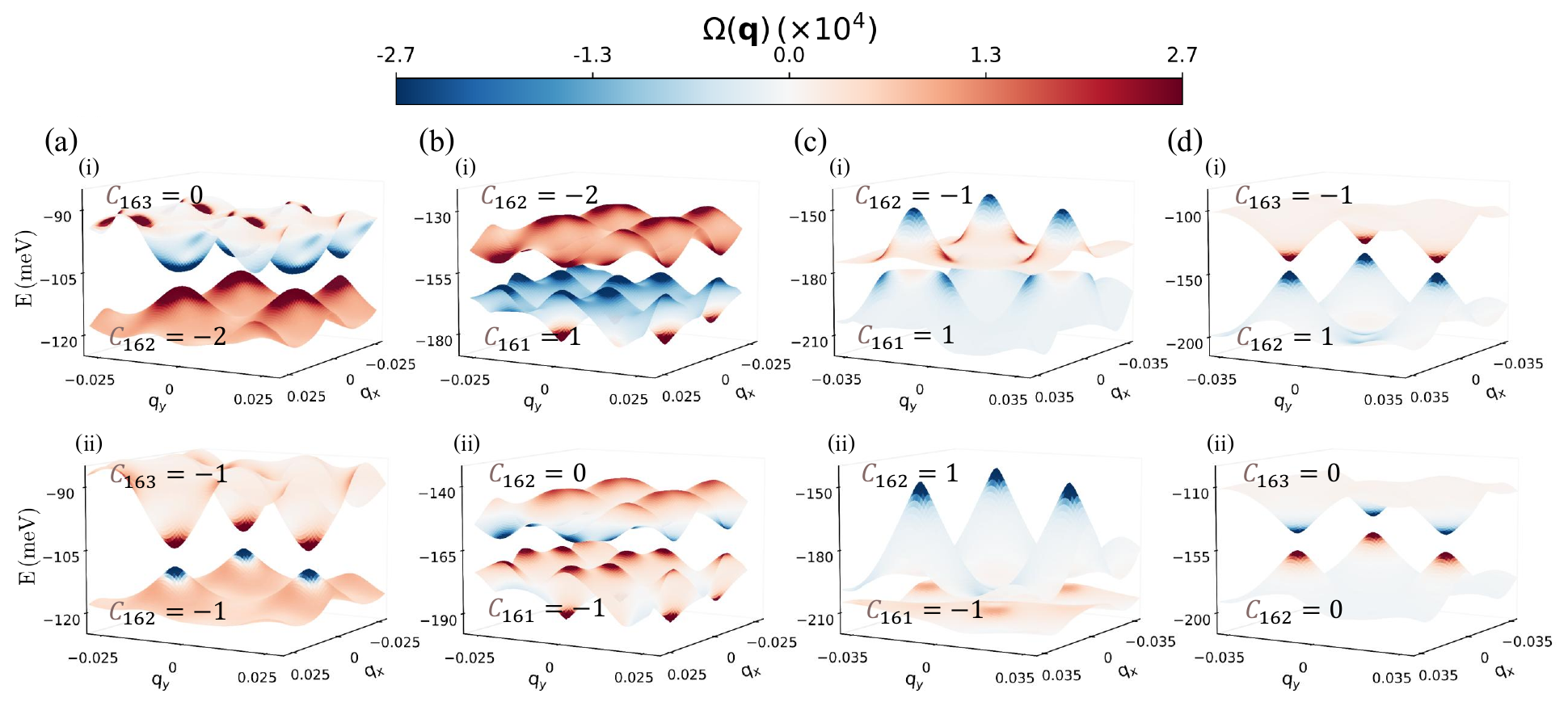}
    \caption{\textbf{Changes in Berry curvature density across phase transition lines.} The change in Berry curvature density across transitions lines from (a-i) Phase (I) $\rightarrow$ (a-ii) Phase (II), (b-i) Phase (I) $\rightarrow$ (b-ii) Phase (IV), (c-i) Phase (II) $\rightarrow$ (c-ii) Phase (III) and (d-i) Phase (III) $\rightarrow$ (d-ii) Phase (IV), corresponding to Fig. \ref{Fig:Phase1}, has been shown. The analysis has been constrained mainly to the bands with index $n=161, 162$ and $163$, which reside at the edge of the middle sub-band. The reversal in the sign of Berry curvature density (across the vertical panels) at the band anti-crossing points correspond to the occurrence of band inversion due to gap closing. The system parameters for the different phases are: (a) $\Delta = 120~\mathrm{meV}$ with (i) $\theta = 0.65^\circ$ and (ii) $\theta = 0.75^\circ$; (b) $\theta=0.6^\circ$ with (i) $\Delta=150~\mathrm{meV}$ and (ii) $\Delta=168~\mathrm{meV}$; (c) $\theta=1.08^\circ$ with (i) $\Delta=175~\mathrm{meV}$ and (ii) $\Delta=210~\mathrm{meV}$; (d) $\theta=1.08^\circ$ with (i) $\Delta=210~\mathrm{meV}$ and (ii) $\Delta=250~\mathrm{meV}$.}
    \label{Fig:BCDensity}
\end{figure*}
where $e$ denotes the electronic charge, $\tau$ is the scattering time, $\epsilon_{adc}$ is the Levi-Civita tensor, and $\mathcal{D}_{bd}$ represents the BCD. 
In the subsequent analysis, we adopt the notation $\mathcal{D}_b \equiv \mathcal{D}_{bz}$ ($b = x, y$) to represent the in-plane components of the BCD in 2D. Here, the $x$ ($y$) axis is defined as the angular bisector between the two zigzag (armchair) directions of the top and bottom layers of the dice lattice.

Having discussed the emergence of finite BCD in $C_3$ and inversion symmetry broken TBDL, we now proceed to analyze the behavior of the Berry curvature density and dipole across the phase transition boundaries shown in Fig. \ref{Fig:Phase1}. 
We track the Berry curvature of the bands occurring at the middle sub-band edge, specifically the ones with index $n = 161$, $162$ and $163$ as shown in Fig. \ref{Fig:BCDensity}.
To elaborate, Fig. \ref{Fig:BCDensity}(a), \ref{Fig:BCDensity}(b), \ref{Fig:BCDensity}(c) and \ref{Fig:BCDensity}(d) correspond to change from Phase (I) $\rightarrow$ Phase (II), Phase (I) $\rightarrow$ Phase (IV), Phase (II) $\rightarrow$ Phase (III) and Phase (III) $\rightarrow$ Phase (IV) respectively.
We observe that the density of Berry curvature undergoes sign reversal at the band anti-crossing points on either side of the phase transition, clearly depicting inversion in the bulk band due to occurrence of gap closing. 
The corresponding change in the Chern number of the bands due to sign reversal of the Berry curvature has also been mentioned in Fig. \ref{Fig:BCDensity}.

\begin{figure*}
    \centering
    \includegraphics[width=1\textwidth]{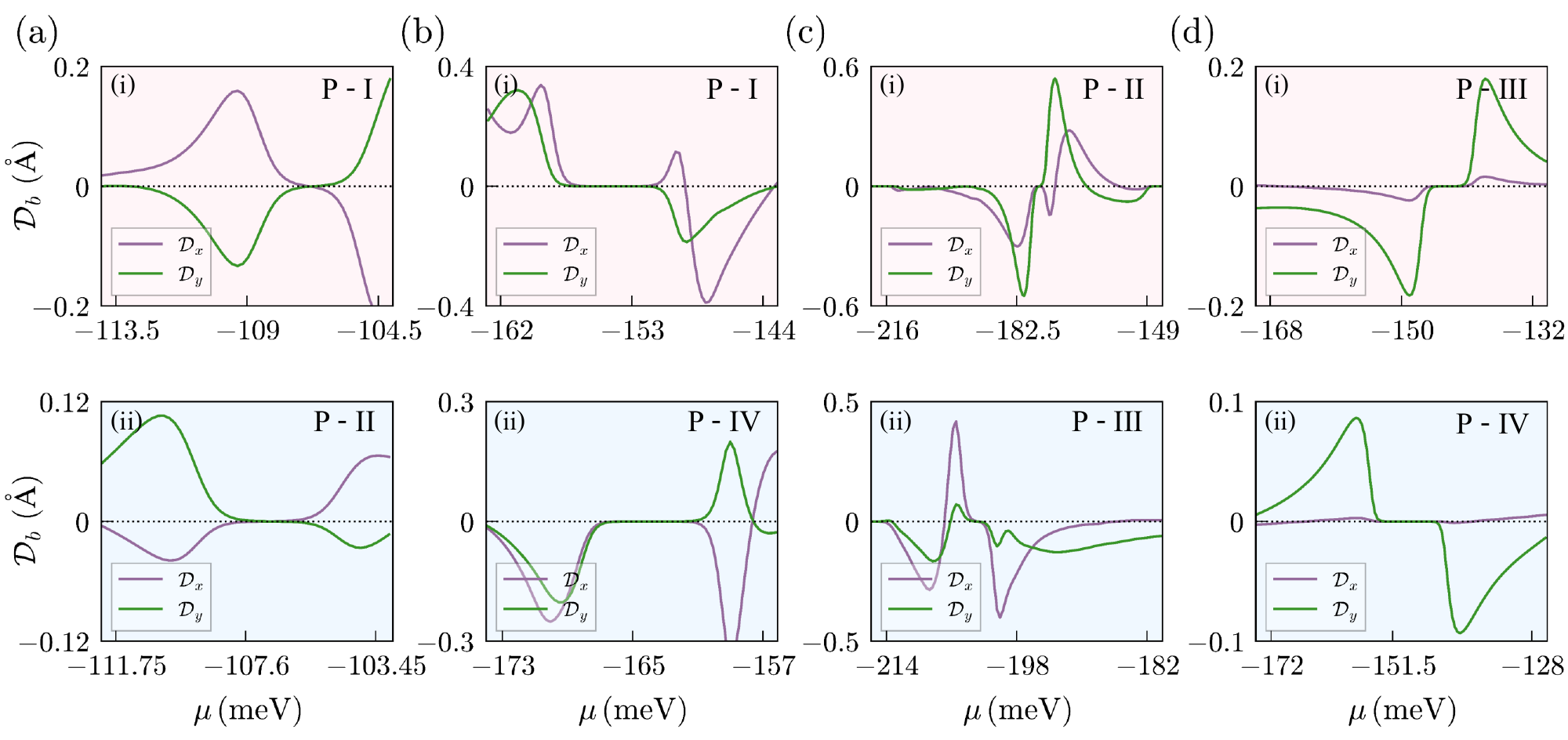}
    \caption{\textbf{Reversal in the sign of BCD across topological phases.} The rectangular components of BCD, namely $\mathcal{D}_x$ and $\mathcal{D}_y$, flip their sign on transitioning from (a) Phase (I) $\rightarrow$ Phase (II), (b) Phase (I) $\rightarrow$ Phase (IV), (c) Phase (II) $\rightarrow$ Phase (III) and (d) Phase (III) $\rightarrow$ Phase (IV), as also represented in Fig. \ref{Fig:BCDensity}, thus making it an efficient probe to detect topological phase transitions in time reversal symmetric systems.}
    \label{Fig:BCD_probe}
\end{figure*}

Having discussed the fate of Berry curvature density across gap closing points, we now look at the variation in BCD across the same phase transitions.
It is to be noted that in Fig. \ref{Fig:BCD_probe}, the analysis is restricted to the chemical potential range in which BCD components associated with the bands undergoing inversion (as shown in Fig. \ref{Fig:BCDensity}) are most prominent. Further, the rectangular components of BCD, that is $\mathcal{D}_x$ and $\mathcal{D}_y$ have been augmented with an overall factor of $4$ to account for the spin and valley degrees of freedom.
It is observed that both $\mathcal{D}_x$ and $\mathcal{D}_y$ undergo prominent sign changes corresponding to the transitions shown in Fig. \ref{Fig:BCDensity}.
This change in sign which is detectable experimentally and captures the occurrence of band inversion even in time reversal symmetric materials, aids in efficiently probing the topological phase transitions under the effect of broken inversion symmetry and $C_3$.
It is to be noted that the application of strain along the zigzag direction ($\Phi=0$) of the bottom layer (which is in turn twisted), causes the emergence of BCD in the direction perpendicular to the applied strain hence leading to the emergence of both $\mathcal{D}_x$ as well as $\mathcal{D}_y$ in the system.
\begin{figure}[]
    \centering
\includegraphics[width=0.47\textwidth]{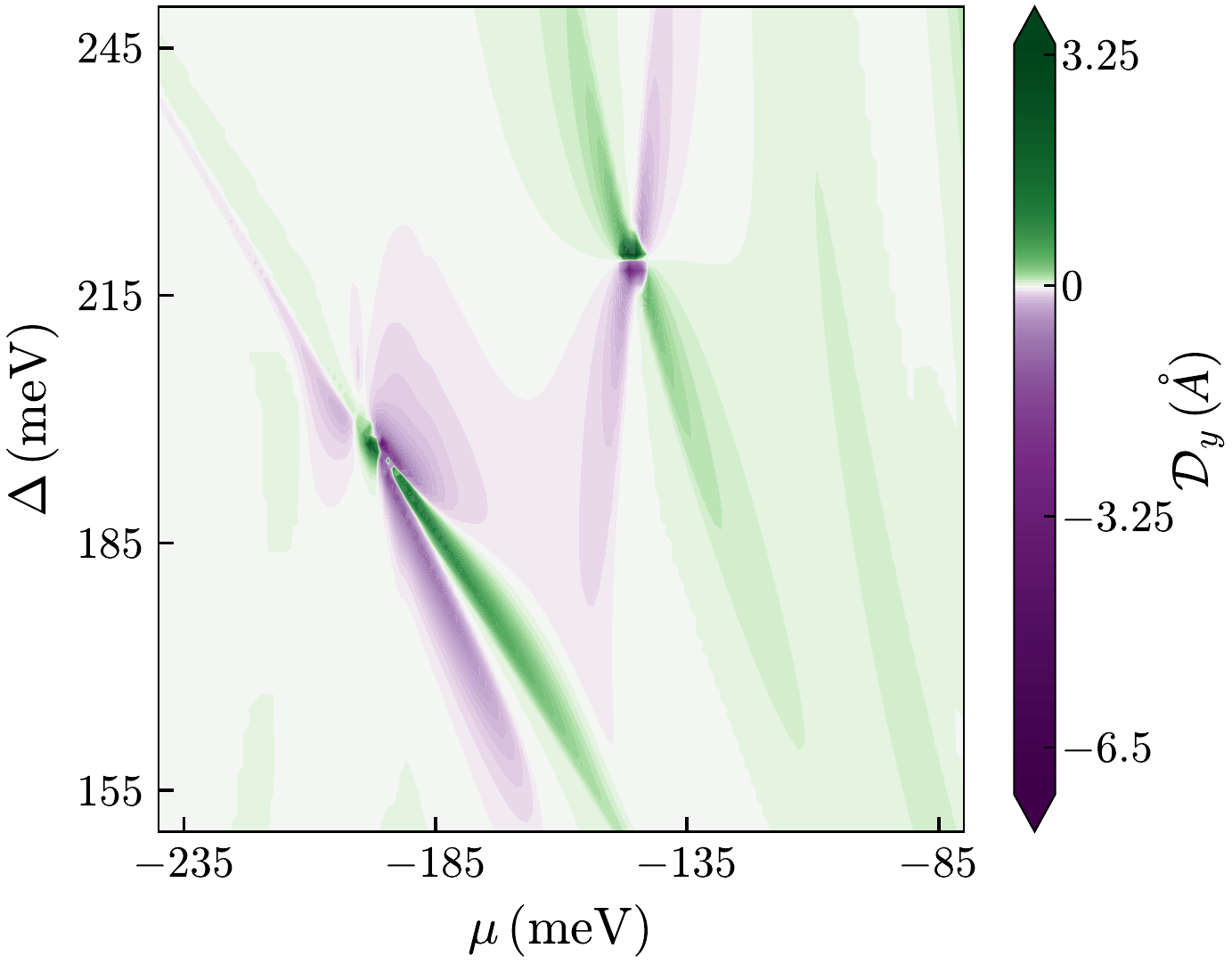}
    \caption{\textbf{Topological phase transition detected by analyzing the variation of BCD in the $\Delta-\mu$ plane.} The butterfly like features in the phase plot of BCD, in the $\Delta-\theta$ plane capture points of topological phase transition which directly correspond to Fig. \ref{Fig:Phase1}. Along a vertical line (fixed $\mu$ but changing $\Delta$), the reversal in the sign of BCD indicates at a topological phase transition. The first (second) butterfly like feature corresponds to the phase transition occurring at $\Delta=195\mathrm{~meV}$ ($219.2\mathrm{~meV}$), at $\theta=1.08^\circ$ in Fig. \ref{Fig:Phase1}}
    \label{Fig: Phase2}
\end{figure}
To observe the reversal in sign of the BCD components across a phase transition point more clearly, we look at a phase plot of the BCD component $\mathcal{D}_y$ as a function of the onsite mass $\Delta$ and the chemical potential $\mu$ at the magic angle $\theta=1.08^\circ$, as shown in Fig. \ref{Fig: Phase2}. The butterfly like structures in the colormap of $\mathcal{D}_y$ indicate points of phase transition which can be correlated directly with the Chern phase plot in Fig. \ref{Fig:Phase1}. The lower two lobes of each butterfly along a horizontal line (keeping $\Delta$ fixed) correspond to the value of $\mathcal{D}_y$ for the band at that particular value of $\mu$. The change in sign of $\mathcal{D}_y$ along a vertical line (keeping $\mu$ fixed) corresponds to the reversal in sign of BCD for the same band as a function of $\Delta$. In Fig. \ref{Fig: Phase2} the transition from Phase (II) $\rightarrow$ Phase (III) and Phase (III) $\rightarrow$ Phase (IV) have been explicitly shown. As clearly seen from Fig. \ref{Fig:Phase1} and Fig. \ref{Fig: Phase2}, the transitions occur around $\Delta=195 \mathrm{~meV}$ and $219.2 \mathrm{~meV}$. This establishes the experimental relevance behind analyzing BCD as a probe for detecting topological phase transitions. Similar phase plots corresponding to the other transitions are provided in the SM.\\
\noindent\textcolor{blue}{\textit{Strain dependence of BCD.-}} The gate dependency of the BCD components $\mathcal{D}_x$ and $\mathcal{D}_y$ for the different values of strain applied along the zigzag direction ($\Phi = 0$) is shown in Fig.~\ref{fig:5}.
Again, we plot $\mathcal{D}_x$ and $\mathcal{D}_y$ for the two distinct cases of the system, namely the chiral limit and the broken chiral regime, as shown in Fig. \ref{fig:5}(a), (c) and Fig. \ref{fig:5}(b), (d) respectively.
It is clear from Fig.~\ref{fig:5} that the components of BCD, $\mathcal{D}_x$ and $\mathcal{D}_y$, exhibit oscillatory behavior in the vicinity of the Fermi energy, and both undergo sign reversals within a very narrow region of the chemical potential $\mu$. The maximum value of the BCD emerges near the band anti-crossing points. This can be verified from the strained bandstructures shown in the SM. Evidently, we obtain a maximal value of the BCD of the order $1$~\AA\ in the chiral limit of the system [see Fig.~\ref{fig:5}(c)] and approximately $6$~\AA\ in the broken chiral symmetry regime [see Fig.~\ref{fig:5}(d)]. The order of magnitude of the BCD obtained in TBDL is comparable to that observed in previous experiments~\cite{Ma2019, Kang2019}, which is of the order $1$~\AA.
\begin{figure}[]
    \centering
\includegraphics[width=0.485\textwidth]{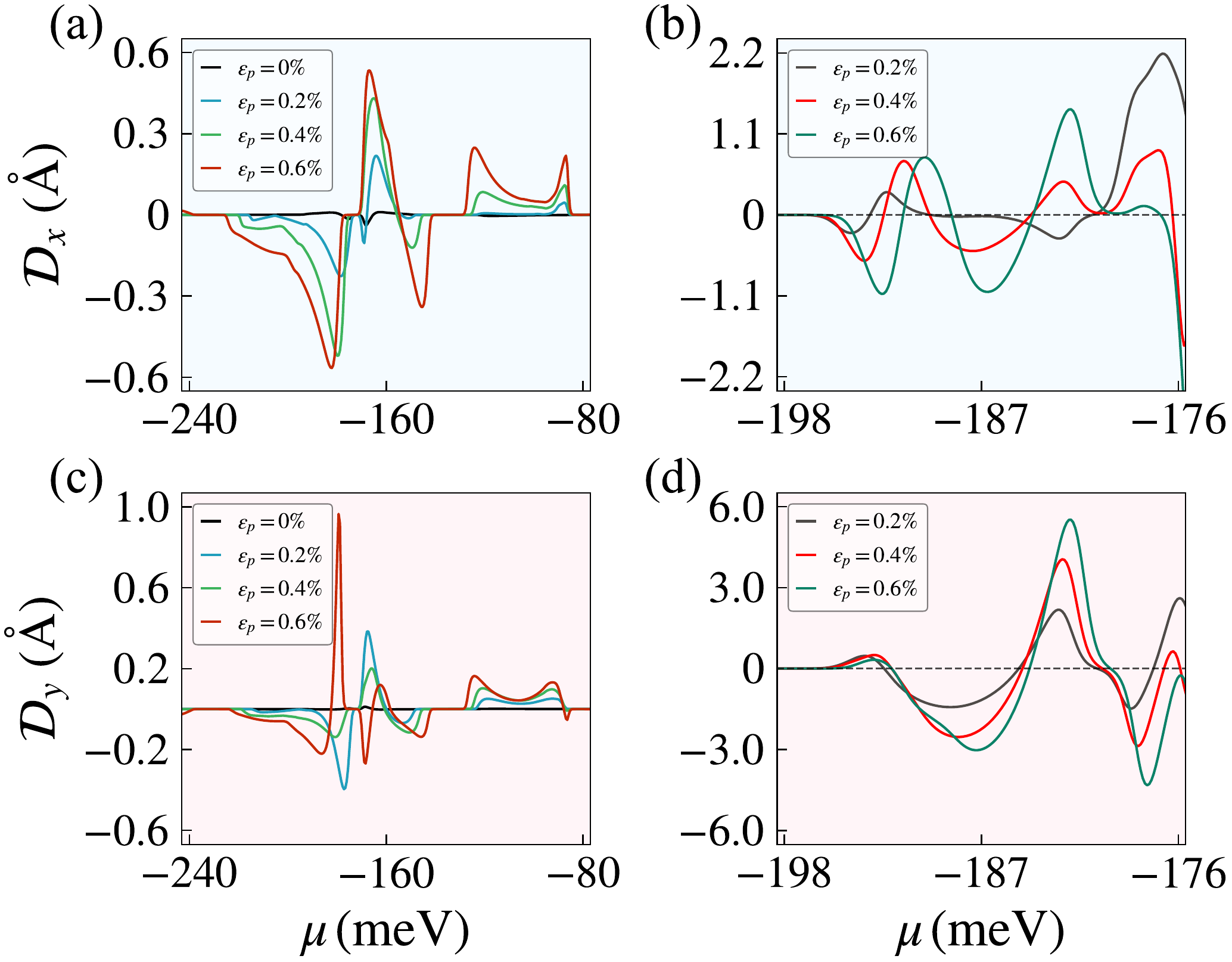}
    \caption{\textbf{Variation of BCD with respect to strain.} BCD components $\mathcal{D}_x$ and $\mathcal{D}_y$ for a fixed twist angle $\theta = 1.08^\circ$, with different strain values applied along the zigzag direction ($\Phi = 0^\circ$), with the temperature being set at $T=4$~K. Panels (a) and (c) represent the dipole components corresponding to the chiral limit of the system ($w_1=w_3=0$ and $w_2=110.7$~meV) with $\Delta=170$~meV. On the other hand, panels (b) and (d) denote the dipole components corresponding to the broken chiral symmetry regime of the system ($w_1 = w_3 = 60$~meV and $w_2 = 110.7$~meV) with $\Delta = 60$~meV.}
    \label{fig:5}
\end{figure}
The variation of BCD with respect to the magnitude of strain clearly shows enhancement in the strength of BCD as the strain is increased, with the dipole strength being nearly zero at low values of strain. This emphasizes further on the role of $C_3$ symmetry breaking in the emergence of non-zero dipole signature of the Berry curvature. Furthermore, the dip-to-peak and peak-to-dip behavior of $\mathcal{D}_x$ and $\mathcal{D}_y$ as a function of the chemical potential $\mu$ results from the reversal in the sign of the Berry curvature density near the band anticrossing points. This causes the BCD components to change sign as $\mu$ scans through the different bands. We re-emphasize here that we have constrained our analysis to the lower edge of the middle sub-band of TBDL since the huge degeneracy of the bands near charge neutrality forbids any relevant discussion on the emergence of BCD. It should also be pointed out that the magnitude of $\mathcal{D}_y$, as shown in Fig. \ref{Fig:BCD_probe}, is larger than $\mathcal{D}_x$ as a consequence of the strain being applied along the zigzag direction of the bottom layer. A discussion on the temperature dependence of $\mathcal{D}_x$ as well as $\mathcal{D}_y$ has been provided in the SM.\\
\noindent \textcolor{blue}{\textit{Material realization.-}} It is well known that the dice lattice can be experimentally realized by growing trilayers of cubic lattices, such as SrTiO$_3$/SrIrO$_3$/SrTiO$_3$, along the (111) direction~\cite{Okamoto2018}. Other potential realizations of the dice lattice include Co molecules on Cu(111) surfaces~\cite{Tassi2024,Khajetoorians2019}, LaAlO$_3$/SrTiO$_3$ (111) quantum wells~\cite{Doenning2013,Soni2020}, and YCl~\cite{Geng2026}. Additional candidate platforms may arise in the MXene and MSene material families~\cite{Qiao2025}, such as Zr$_2$CS$_2$~\cite{Papadopoulou2022}. Beyond electronic systems, synthetic realizations of the dice lattice have also been proposed in optical lattices~\cite{Cheng2020,Andrijauskas2015,Moller2012} and cold-atom systems~\cite{Rizzi2006,Bercioux2009}, where such intricate hopping processes can be engineered with high precision.

\noindent {\textbf{Discussions}}\\
We theoretically investigate the emergence of multiple topological phase transitions at the edge of the middle band in TBDL in the presence of an inversion symmetry breaking onsite mass term ($\Delta$) and a $C_3$ symmetry breaking strain. The broken inversion and $C_3$ symmetries, in turn, cause the emergence of BCD which acts as an experimental probe to detect topological phase transitions in systems with TRS where band geometry is otherwise difficult to diagnose. We show that the components of BCD correctly capture the phase transitions occurring in the relevant bands, by complete reversal in their sign before and after the transition line is crossed. This flip in the sign of Berry curvature as well as dipole is indicative of the occurrence of band inversion in the system due to gap closure. We also show the variation of the strength of dipole as a function of the applied strain which shows negligible values of dipole strength at low values of strain thereby establishing the importance of $C_3$ symmetry breaking in generating non-zero BCD. The calculated BCD in TBDL is about 1~\AA\ in the chiral limit and increases to approximately $6$~\AA\ in the broken chiral symmetry regime of the system. In other strained systems, such as bilayer graphene~\cite{Battilomo2019} and MoS$_2$~\cite{Son2019}, the strain induced BCD is typically of the order $\sim 10^{-2}$~\AA. On the other hand, in bilayer and multilayer WTe$_2$ systems~\cite{Du2018, Wang2019}, the BCD is of the order $\sim 0.2$~\AA\ and $\sim 1$~\AA, respectively. Thus, the system considered in this work records approximately an order of magnitude larger value of BCD in the broken chiral limit. However, other strained twisted bilayer systems, such as twisted bilayer WSe$_2$~\cite{Hu2022}, TBG~\cite{Zhang2022}, and twisted bilayer WTe$_2$~\cite{He2021}, exhibit significantly large BCD values of approximately $3$~\AA, $200$~\AA, and $1500$~\AA, respectively.

\noindent\textcolor{black}{\textbf{Methods}}\\ 
\noindent \textbf{Berry curvature and dipole}.-
In two dimensions (2D), the BCD is expressed as~\cite{Sodemann2015},
\begin{eqnarray}
\mathcal{D}_{bd} = \sum_{n, \zeta} \int_{\mathbf{q}} \frac{d^2q}{(2\pi)^2}f_0\, \partial_b \Omega_d,
\end{eqnarray}
where the summation is performed over the band index $n$ and the valley index $\zeta$. The operator $\partial_b \equiv \frac{\partial}{\partial q_b}$ denotes partial differentiation with respect to the momentum component $q_b$. Additionally, $f_0$ represents the equilibrium Fermi–Dirac distribution function and $\Omega_d$ corresponds to the $z$-component of the Berry curvature ($d=z$) whose expression is given by~\cite{Sodemann2015,Hu2022},
\begin{eqnarray}
\Omega_{n}(\mathbf{q}) = i \left\langle \partial_{\mathbf{q}} u_{n, \mathbf{q}} \,\middle| \,\times  \,\middle|\, \partial_{\mathbf{q}} u_{n, \mathbf{q}} \right\rangle,
\end{eqnarray}
Here $|u_{n, \mathbf{q}}\rangle$ denotes the periodic part of the Bloch wave function at momentum $\mathbf{q}$ for the band index $n$. Also the Chern number for an isolated
band can be evaluated from Berry curvature as~\cite{Thouless1997, Debnath2025, Lahiri2024},
\begin{equation}
    C_n = \frac{1}{2\pi} \int_{\text{MBZ}} d \mathbf{q} \,\Omega_{n}(\mathbf{q})
\end{equation}

\noindent\textcolor{black}{\textbf{Data availability}}\\ The data that support the findings of this work are not publicly available. The data are available from the authors upon reasonable request.

\noindent\textcolor{black}{\textbf{Acknowledgments}}\\
GP and SL sincerely thank Dr. Kuntal Bhattacharyya and Ms. Shreya Debnath for their valuable comments and suggestions. SL acknowledges financial support from the MoE, Govt. of India, through the Prime Minister’s Research Fellowship (PMRF) scheme in May 2022. GP and SL further acknowledge the National Supercomputing Mission (NSM) for providing computing resources of `PARAM Kamrupa' at IIT Guwahati, which is implemented by C-DAC and supported by the Ministry of Electronics and Information Technology (MeitY) and Department of Science and Technology (DST), Government of India. BT is supported by The Scientific and Technological Research Council of T{\"u}rkiye (T{\"U}B{\.I}TAK)
under Grant No. 125F435 and Turkish Academy of Sciences (TUBA) under Grant No. AD-2026. SB acknowledges 
support from T{\"U}B{\.I}TAK-BIDEB.

\noindent\textcolor{black}{\textbf{Author contributions}}\\
GP and SL carried out the theoretical analysis, numerical simulations, and data analysis. They jointly wrote the manuscript, with input from all authors during the editing process. GP and SL contributed equally to all aspects of this work. SB and BT supervised the project. All authors discussed the results and provided feedback on the manuscript.
\bibliography{mainNotes}

\begin{thebibliography}{92}%
\makeatletter
\providecommand \@ifxundefined [1]{%
 \@ifx{#1\undefined}
}%
\providecommand \@ifnum [1]{%
 \ifnum #1\expandafter \@firstoftwo
 \else \expandafter \@secondoftwo
 \fi
}%
\providecommand \@ifx [1]{%
 \ifx #1\expandafter \@firstoftwo
 \else \expandafter \@secondoftwo
 \fi
}%
\providecommand \natexlab [1]{#1}%
\providecommand \enquote  [1]{``#1''}%
\providecommand \bibnamefont  [1]{#1}%
\providecommand \bibfnamefont [1]{#1}%
\providecommand \citenamefont [1]{#1}%
\providecommand \href@noop [0]{\@secondoftwo}%
\providecommand \href [0]{\begingroup \@sanitize@url \@href}%
\providecommand \@href[1]{\@@startlink{#1}\@@href}%
\providecommand \@@href[1]{\endgroup#1\@@endlink}%
\providecommand \@sanitize@url [0]{\catcode `\\12\catcode `\$12\catcode
  `\&12\catcode `\#12\catcode `\^12\catcode `\_12\catcode `\%12\relax}%
\providecommand \@@startlink[1]{}%
\providecommand \@@endlink[0]{}%
\providecommand \url  [0]{\begingroup\@sanitize@url \@url }%
\providecommand \@url [1]{\endgroup\@href {#1}{\urlprefix }}%
\providecommand \urlprefix  [0]{URL }%
\providecommand \Eprint [0]{\href }%
\providecommand \doibase [0]{https://doi.org/}%
\providecommand \selectlanguage [0]{\@gobble}%
\providecommand \bibinfo  [0]{\@secondoftwo}%
\providecommand \bibfield  [0]{\@secondoftwo}%
\providecommand \translation [1]{[#1]}%
\providecommand \BibitemOpen [0]{}%
\providecommand \bibitemStop [0]{}%
\providecommand \bibitemNoStop [0]{.\EOS\space}%
\providecommand \EOS [0]{\spacefactor3000\relax}%
\providecommand \BibitemShut  [1]{\csname bibitem#1\endcsname}%
\let\auto@bib@innerbib\@empty
\bibitem [{\citenamefont {Cao}\ \emph {et~al.}(2020)\citenamefont {Cao},
  \citenamefont {Chowdhury}, \citenamefont {Rodan-Legrain}, \citenamefont
  {Rubies-Bigorda}, \citenamefont {Watanabe}, \citenamefont {Taniguchi},
  \citenamefont {Senthil},\ and\ \citenamefont {Jarillo-Herrero}}]{Cao2020}%
  \BibitemOpen
  \bibfield  {author} {\bibinfo {author} {\bibfnamefont {Y.}~\bibnamefont
  {Cao}}, \bibinfo {author} {\bibfnamefont {D.}~\bibnamefont {Chowdhury}},
  \bibinfo {author} {\bibfnamefont {D.}~\bibnamefont {Rodan-Legrain}}, \bibinfo
  {author} {\bibfnamefont {O.}~\bibnamefont {Rubies-Bigorda}}, \bibinfo
  {author} {\bibfnamefont {K.}~\bibnamefont {Watanabe}}, \bibinfo {author}
  {\bibfnamefont {T.}~\bibnamefont {Taniguchi}}, \bibinfo {author}
  {\bibfnamefont {T.}~\bibnamefont {Senthil}},\ and\ \bibinfo {author}
  {\bibfnamefont {P.}~\bibnamefont {Jarillo-Herrero}},\ }\bibfield  {title}
  {\bibinfo {title} {Strange metal in magic-angle graphene with near planckian
  dissipation},\ }\href {https://doi.org/10.1103/PhysRevLett.124.076801}
  {\bibfield  {journal} {\bibinfo  {journal} {Phys. Rev. Lett.}\ }\textbf
  {\bibinfo {volume} {124}},\ \bibinfo {pages} {076801} (\bibinfo {year}
  {2020})}\BibitemShut {NoStop}%
\bibitem [{\citenamefont {Cao}\ \emph {et~al.}(2018)\citenamefont {Cao},
  \citenamefont {Fatemi}, \citenamefont {Demir}, \citenamefont {Fang},
  \citenamefont {Tomarken}, \citenamefont {Luo}, \citenamefont
  {Sanchez-Yamagishi}, \citenamefont {Watanabe}, \citenamefont {Taniguchi},
  \citenamefont {Kaxiras}, \citenamefont {Ashoori},\ and\ \citenamefont
  {Jarillo-Herrero}}]{Cao2018}%
  \BibitemOpen
  \bibfield  {author} {\bibinfo {author} {\bibfnamefont {Y.}~\bibnamefont
  {Cao}}, \bibinfo {author} {\bibfnamefont {V.}~\bibnamefont {Fatemi}},
  \bibinfo {author} {\bibfnamefont {A.}~\bibnamefont {Demir}}, \bibinfo
  {author} {\bibfnamefont {S.}~\bibnamefont {Fang}}, \bibinfo {author}
  {\bibfnamefont {S.~L.}\ \bibnamefont {Tomarken}}, \bibinfo {author}
  {\bibfnamefont {J.~Y.}\ \bibnamefont {Luo}}, \bibinfo {author} {\bibfnamefont
  {J.~D.}\ \bibnamefont {Sanchez-Yamagishi}}, \bibinfo {author} {\bibfnamefont
  {K.}~\bibnamefont {Watanabe}}, \bibinfo {author} {\bibfnamefont
  {T.}~\bibnamefont {Taniguchi}}, \bibinfo {author} {\bibfnamefont
  {E.}~\bibnamefont {Kaxiras}}, \bibinfo {author} {\bibfnamefont {R.~C.}\
  \bibnamefont {Ashoori}},\ and\ \bibinfo {author} {\bibfnamefont
  {P.}~\bibnamefont {Jarillo-Herrero}},\ }\bibfield  {title} {\bibinfo {title}
  {Correlated insulator behaviour at half-filling in magic-angle graphene
  superlattices},\ }\href {https://doi.org/10.1038/nature26154} {\bibfield
  {journal} {\bibinfo  {journal} {Nature}\ }\textbf {\bibinfo {volume} {556}},\
  \bibinfo {pages} {80} (\bibinfo {year} {2018})}\BibitemShut {NoStop}%
\bibitem [{\citenamefont {Tanaka}\ \emph {et~al.}(2025)\citenamefont {Tanaka},
  \citenamefont {Wang}, \citenamefont {Dinh}, \citenamefont {Rodan-Legrain},
  \citenamefont {Zaman}, \citenamefont {Hays}, \citenamefont {Almanakly},
  \citenamefont {Kannan}, \citenamefont {Kim}, \citenamefont {Niedzielski},
  \citenamefont {Serniak}, \citenamefont {Schwartz}, \citenamefont {Watanabe},
  \citenamefont {Taniguchi}, \citenamefont {Orlando}, \citenamefont
  {Gustavsson}, \citenamefont {Grover}, \citenamefont {Jarillo-Herrero},\ and\
  \citenamefont {Oliver}}]{Tanaka2025}%
  \BibitemOpen
  \bibfield  {author} {\bibinfo {author} {\bibfnamefont {M.}~\bibnamefont
  {Tanaka}}, \bibinfo {author} {\bibfnamefont {J.~I.-J.}\ \bibnamefont {Wang}},
  \bibinfo {author} {\bibfnamefont {T.~H.}\ \bibnamefont {Dinh}}, \bibinfo
  {author} {\bibfnamefont {D.}~\bibnamefont {Rodan-Legrain}}, \bibinfo {author}
  {\bibfnamefont {S.}~\bibnamefont {Zaman}}, \bibinfo {author} {\bibfnamefont
  {M.}~\bibnamefont {Hays}}, \bibinfo {author} {\bibfnamefont {A.}~\bibnamefont
  {Almanakly}}, \bibinfo {author} {\bibfnamefont {B.}~\bibnamefont {Kannan}},
  \bibinfo {author} {\bibfnamefont {D.~K.}\ \bibnamefont {Kim}}, \bibinfo
  {author} {\bibfnamefont {B.~M.}\ \bibnamefont {Niedzielski}}, \bibinfo
  {author} {\bibfnamefont {K.}~\bibnamefont {Serniak}}, \bibinfo {author}
  {\bibfnamefont {M.~E.}\ \bibnamefont {Schwartz}}, \bibinfo {author}
  {\bibfnamefont {K.}~\bibnamefont {Watanabe}}, \bibinfo {author}
  {\bibfnamefont {T.}~\bibnamefont {Taniguchi}}, \bibinfo {author}
  {\bibfnamefont {T.~P.}\ \bibnamefont {Orlando}}, \bibinfo {author}
  {\bibfnamefont {S.}~\bibnamefont {Gustavsson}}, \bibinfo {author}
  {\bibfnamefont {J.~A.}\ \bibnamefont {Grover}}, \bibinfo {author}
  {\bibfnamefont {P.}~\bibnamefont {Jarillo-Herrero}},\ and\ \bibinfo {author}
  {\bibfnamefont {W.~D.}\ \bibnamefont {Oliver}},\ }\bibfield  {title}
  {\bibinfo {title} {Superfluid stiffness of magic-angle twisted bilayer
  graphene},\ }\href {https://doi.org/10.1038/s41586-024-08494-7} {\bibfield
  {journal} {\bibinfo  {journal} {Nature}\ }\textbf {\bibinfo {volume} {638}},\
  \bibinfo {pages} {99} (\bibinfo {year} {2025})}\BibitemShut {NoStop}%
\bibitem [{\citenamefont {Ledwith}\ \emph {et~al.}(2025)\citenamefont
  {Ledwith}, \citenamefont {Dong}, \citenamefont {Vishwanath},\ and\
  \citenamefont {Khalaf}}]{Ledwith2025}%
  \BibitemOpen
  \bibfield  {author} {\bibinfo {author} {\bibfnamefont {P.~J.}\ \bibnamefont
  {Ledwith}}, \bibinfo {author} {\bibfnamefont {J.}~\bibnamefont {Dong}},
  \bibinfo {author} {\bibfnamefont {A.}~\bibnamefont {Vishwanath}},\ and\
  \bibinfo {author} {\bibfnamefont {E.}~\bibnamefont {Khalaf}},\ }\bibfield
  {title} {\bibinfo {title} {Nonlocal moments and mott semimetal in the chern
  bands of twisted bilayer graphene},\ }\href
  {https://doi.org/10.1103/PhysRevX.15.021087} {\bibfield  {journal} {\bibinfo
  {journal} {Phys. Rev. X}\ }\textbf {\bibinfo {volume} {15}},\ \bibinfo
  {pages} {021087} (\bibinfo {year} {2025})}\BibitemShut {NoStop}%
\bibitem [{\citenamefont {Mesple}\ \emph {et~al.}(2025)\citenamefont {Mesple},
  \citenamefont {Mallet}, \citenamefont {Trambly~de Laissardi\`ere},\ and\
  \citenamefont {et~al.}}]{Mesple2025}%
  \BibitemOpen
  \bibfield  {author} {\bibinfo {author} {\bibfnamefont {F.}~\bibnamefont
  {Mesple}}, \bibinfo {author} {\bibfnamefont {P.}~\bibnamefont {Mallet}},
  \bibinfo {author} {\bibfnamefont {G.}~\bibnamefont {Trambly~de
  Laissardi\`ere}},\ and\ \bibinfo {author} {\bibnamefont {et~al.}},\
  }\bibfield  {title} {\bibinfo {title} {Experimental evidence of the
  topological obstruction in twisted bilayer graphene},\ }\href
  {https://doi.org/10.1038/s41467-025-66257-y} {\bibfield  {journal} {\bibinfo
  {journal} {Nature Communications}\ }\textbf {\bibinfo {volume} {16}},\
  \bibinfo {pages} {11478} (\bibinfo {year} {2025})}\BibitemShut {NoStop}%
\bibitem [{\citenamefont {Song}\ \emph {et~al.}(2024)\citenamefont {Song},
  \citenamefont {Qi}, \citenamefont {Liebman},\ and\ \citenamefont
  {Narang}}]{Song2024}%
  \BibitemOpen
  \bibfield  {author} {\bibinfo {author} {\bibfnamefont {Z.}~\bibnamefont
  {Song}}, \bibinfo {author} {\bibfnamefont {J.}~\bibnamefont {Qi}}, \bibinfo
  {author} {\bibfnamefont {O.}~\bibnamefont {Liebman}},\ and\ \bibinfo {author}
  {\bibfnamefont {P.}~\bibnamefont {Narang}},\ }\bibfield  {title} {\bibinfo
  {title} {Collective spin in twisted bilayer materials},\ }\href
  {https://doi.org/10.1103/PhysRevB.110.024401} {\bibfield  {journal} {\bibinfo
   {journal} {Phys. Rev. B}\ }\textbf {\bibinfo {volume} {110}},\ \bibinfo
  {pages} {024401} (\bibinfo {year} {2024})}\BibitemShut {NoStop}%
\bibitem [{\citenamefont {Yang}\ \emph
  {et~al.}(2024{\natexlab{a}})\citenamefont {Yang}, \citenamefont {Hu},
  \citenamefont {Wu}, \citenamefont {He}, \citenamefont {Zhou}, \citenamefont
  {Xue}, \citenamefont {He}, \citenamefont {Hu}, \citenamefont {Chen},
  \citenamefont {Gong}, \citenamefont {Zhang}, \citenamefont {Tan},
  \citenamefont {Hern{\'a}ndez},\ and\ \citenamefont {Xie}}]{Yang2024}%
  \BibitemOpen
  \bibfield  {author} {\bibinfo {author} {\bibfnamefont {H.}~\bibnamefont
  {Yang}}, \bibinfo {author} {\bibfnamefont {R.}~\bibnamefont {Hu}}, \bibinfo
  {author} {\bibfnamefont {H.}~\bibnamefont {Wu}}, \bibinfo {author}
  {\bibfnamefont {X.}~\bibnamefont {He}}, \bibinfo {author} {\bibfnamefont
  {Y.}~\bibnamefont {Zhou}}, \bibinfo {author} {\bibfnamefont {Y.}~\bibnamefont
  {Xue}}, \bibinfo {author} {\bibfnamefont {K.}~\bibnamefont {He}}, \bibinfo
  {author} {\bibfnamefont {W.}~\bibnamefont {Hu}}, \bibinfo {author}
  {\bibfnamefont {H.}~\bibnamefont {Chen}}, \bibinfo {author} {\bibfnamefont
  {M.}~\bibnamefont {Gong}}, \bibinfo {author} {\bibfnamefont {X.}~\bibnamefont
  {Zhang}}, \bibinfo {author} {\bibfnamefont {P.-H.}\ \bibnamefont {Tan}},
  \bibinfo {author} {\bibfnamefont {E.~R.}\ \bibnamefont {Hern{\'a}ndez}},\
  and\ \bibinfo {author} {\bibfnamefont {Y.}~\bibnamefont {Xie}},\ }\bibfield
  {title} {\bibinfo {title} {Identification and structural characterization of
  twisted atomically thin bilayer materials by deep learning},\ }\href
  {https://doi.org/10.1021/acs.nanolett.3c04815} {\bibfield  {journal}
  {\bibinfo  {journal} {Nano Letters}\ }\textbf {\bibinfo {volume} {24}},\
  \bibinfo {pages} {2789} (\bibinfo {year} {2024}{\natexlab{a}})}\BibitemShut
  {NoStop}%
\bibitem [{\citenamefont {Devakul}\ \emph {et~al.}(2021)\citenamefont
  {Devakul}, \citenamefont {Cr{\'e}pel}, \citenamefont {Zhang},\ and\
  \citenamefont {et~al.}}]{Devakul2021}%
  \BibitemOpen
  \bibfield  {author} {\bibinfo {author} {\bibfnamefont {T.}~\bibnamefont
  {Devakul}}, \bibinfo {author} {\bibfnamefont {V.}~\bibnamefont {Cr{\'e}pel}},
  \bibinfo {author} {\bibfnamefont {Y.}~\bibnamefont {Zhang}},\ and\ \bibinfo
  {author} {\bibnamefont {et~al.}},\ }\bibfield  {title} {\bibinfo {title}
  {Magic in twisted transition metal dichalcogenide bilayers},\ }\href
  {https://doi.org/10.1038/s41467-021-27042-9} {\bibfield  {journal} {\bibinfo
  {journal} {Nature Communications}\ }\textbf {\bibinfo {volume} {12}},\
  \bibinfo {pages} {6730} (\bibinfo {year} {2021})}\BibitemShut {NoStop}%
\bibitem [{\citenamefont {Lei}\ \emph {et~al.}(2025)\citenamefont {Lei},
  \citenamefont {Mahon},\ and\ \citenamefont {MacDonald}}]{Lei2025}%
  \BibitemOpen
  \bibfield  {author} {\bibinfo {author} {\bibfnamefont {C.}~\bibnamefont
  {Lei}}, \bibinfo {author} {\bibfnamefont {P.~T.}\ \bibnamefont {Mahon}},\
  and\ \bibinfo {author} {\bibfnamefont {A.~H.}\ \bibnamefont {MacDonald}},\
  }\bibfield  {title} {\bibinfo {title} {Moir\'e band theory for m-valley
  twisted transition metal dichalcogenides},\ }\href
  {https://doi.org/10.1103/5zt2-scbg} {\bibfield  {journal} {\bibinfo
  {journal} {Phys. Rev. Lett.}\ }\textbf {\bibinfo {volume} {135}},\ \bibinfo
  {pages} {196402} (\bibinfo {year} {2025})}\BibitemShut {NoStop}%
\bibitem [{\citenamefont {Bao}\ \emph {et~al.}(2025)\citenamefont {Bao},
  \citenamefont {Wang}, \citenamefont {Liu},\ and\ \citenamefont
  {Wang}}]{Bao2025}%
  \BibitemOpen
  \bibfield  {author} {\bibinfo {author} {\bibfnamefont {K.}~\bibnamefont
  {Bao}}, \bibinfo {author} {\bibfnamefont {H.}~\bibnamefont {Wang}}, \bibinfo
  {author} {\bibfnamefont {Z.}~\bibnamefont {Liu}},\ and\ \bibinfo {author}
  {\bibfnamefont {J.}~\bibnamefont {Wang}},\ }\bibfield  {title} {\bibinfo
  {title} {Anisotropic moir\'e band flattening in twisted bilayers of m-valley
  mxenes},\ }\href {https://doi.org/10.1103/l88b-67d2} {\bibfield  {journal}
  {\bibinfo  {journal} {Phys. Rev. B}\ }\textbf {\bibinfo {volume} {112}},\
  \bibinfo {pages} {L041406} (\bibinfo {year} {2025})}\BibitemShut {NoStop}%
\bibitem [{\citenamefont {Yang}\ \emph
  {et~al.}(2024{\natexlab{b}})\citenamefont {Yang}, \citenamefont {Duan},
  \citenamefont {Li},\ and\ \citenamefont {Liu}}]{Yang2024A}%
  \BibitemOpen
  \bibfield  {author} {\bibinfo {author} {\bibfnamefont {Y.}~\bibnamefont
  {Yang}}, \bibinfo {author} {\bibfnamefont {Z.}~\bibnamefont {Duan}}, \bibinfo
  {author} {\bibfnamefont {H.}~\bibnamefont {Li}},\ and\ \bibinfo {author}
  {\bibfnamefont {S.}~\bibnamefont {Liu}},\ }\bibfield  {title} {\bibinfo
  {title} {Advances in twisted transition metal dichalcogenides: synthesis,
  characterization, and properties},\ }\href
  {https://doi.org/10.1088/2515-7639/ad2b7b} {\bibfield  {journal} {\bibinfo
  {journal} {Journal of Physics: Materials}\ }\textbf {\bibinfo {volume} {7}},\
  \bibinfo {pages} {022002} (\bibinfo {year} {2024}{\natexlab{b}})}\BibitemShut
  {NoStop}%
\bibitem [{\citenamefont {Lou}\ \emph {et~al.}(2021)\citenamefont {Lou},
  \citenamefont {Zhao}, \citenamefont {Minkov}, \citenamefont {Guo},
  \citenamefont {Orenstein},\ and\ \citenamefont {Fan}}]{Lou2021}%
  \BibitemOpen
  \bibfield  {author} {\bibinfo {author} {\bibfnamefont {B.}~\bibnamefont
  {Lou}}, \bibinfo {author} {\bibfnamefont {N.}~\bibnamefont {Zhao}}, \bibinfo
  {author} {\bibfnamefont {M.}~\bibnamefont {Minkov}}, \bibinfo {author}
  {\bibfnamefont {C.}~\bibnamefont {Guo}}, \bibinfo {author} {\bibfnamefont
  {M.}~\bibnamefont {Orenstein}},\ and\ \bibinfo {author} {\bibfnamefont
  {S.}~\bibnamefont {Fan}},\ }\bibfield  {title} {\bibinfo {title} {Theory for
  twisted bilayer photonic crystal slabs},\ }\href
  {https://doi.org/10.1103/PhysRevLett.126.136101} {\bibfield  {journal}
  {\bibinfo  {journal} {Phys. Rev. Lett.}\ }\textbf {\bibinfo {volume} {126}},\
  \bibinfo {pages} {136101} (\bibinfo {year} {2021})}\BibitemShut {NoStop}%
\bibitem [{\citenamefont {Sharma}\ \emph {et~al.}(2021)\citenamefont {Sharma},
  \citenamefont {Yudhistira}, \citenamefont {Chakraborty},\ and\ \citenamefont
  {et~al.}}]{Sharma2021}%
  \BibitemOpen
  \bibfield  {author} {\bibinfo {author} {\bibfnamefont {G.}~\bibnamefont
  {Sharma}}, \bibinfo {author} {\bibfnamefont {I.}~\bibnamefont {Yudhistira}},
  \bibinfo {author} {\bibfnamefont {N.}~\bibnamefont {Chakraborty}},\ and\
  \bibinfo {author} {\bibnamefont {et~al.}},\ }\bibfield  {title} {\bibinfo
  {title} {Carrier transport theory for twisted bilayer graphene in the
  metallic regime},\ }\href {https://doi.org/10.1038/s41467-021-25864-1}
  {\bibfield  {journal} {\bibinfo  {journal} {Nature Communications}\ }\textbf
  {\bibinfo {volume} {12}},\ \bibinfo {pages} {5737} (\bibinfo {year}
  {2021})}\BibitemShut {NoStop}%
\bibitem [{\citenamefont {Ma}\ \emph {et~al.}(2024)\citenamefont {Ma},
  \citenamefont {Chen}, \citenamefont {Yu},\ and\ \citenamefont
  {Luo}}]{Ma2024}%
  \BibitemOpen
  \bibfield  {author} {\bibinfo {author} {\bibfnamefont {D.}~\bibnamefont
  {Ma}}, \bibinfo {author} {\bibfnamefont {Y.~G.}\ \bibnamefont {Chen}},
  \bibinfo {author} {\bibfnamefont {Y.}~\bibnamefont {Yu}},\ and\ \bibinfo
  {author} {\bibfnamefont {X.}~\bibnamefont {Luo}},\ }\bibfield  {title}
  {\bibinfo {title} {Moir\'e semiconductors on the twisted bilayer dice
  lattice},\ }\href {https://doi.org/10.1103/PhysRevB.109.155159} {\bibfield
  {journal} {\bibinfo  {journal} {Phys. Rev. B}\ }\textbf {\bibinfo {volume}
  {109}},\ \bibinfo {pages} {155159} (\bibinfo {year} {2024})}\BibitemShut
  {NoStop}%
\bibitem [{\citenamefont {Zhou}\ \emph {et~al.}(2024)\citenamefont {Zhou},
  \citenamefont {Hung}, \citenamefont {Wang},\ and\ \citenamefont
  {Bansil}}]{Zhou2024}%
  \BibitemOpen
  \bibfield  {author} {\bibinfo {author} {\bibfnamefont {X.}~\bibnamefont
  {Zhou}}, \bibinfo {author} {\bibfnamefont {Y.~C.}\ \bibnamefont {Hung}},
  \bibinfo {author} {\bibfnamefont {B.}~\bibnamefont {Wang}},\ and\ \bibinfo
  {author} {\bibfnamefont {A.}~\bibnamefont {Bansil}},\ }\bibfield  {title}
  {\bibinfo {title} {Generation of isolated flat bands with tunable numbers
  through moir\'e engineering},\ }\href
  {https://doi.org/10.1103/PhysRevLett.133.236401} {\bibfield  {journal}
  {\bibinfo  {journal} {Phys. Rev. Lett.}\ }\textbf {\bibinfo {volume} {133}},\
  \bibinfo {pages} {236401} (\bibinfo {year} {2024})}\BibitemShut {NoStop}%
\bibitem [{\citenamefont {Gandhi}\ \emph {et~al.}(2026)\citenamefont {Gandhi},
  \citenamefont {Paul}, \citenamefont {Lahiri},\ and\ \citenamefont
  {Basu}}]{Gandhi2026}%
  \BibitemOpen
  \bibfield  {author} {\bibinfo {author} {\bibfnamefont {S.}~\bibnamefont
  {Gandhi}}, \bibinfo {author} {\bibfnamefont {G.}~\bibnamefont {Paul}},
  \bibinfo {author} {\bibfnamefont {S.}~\bibnamefont {Lahiri}},\ and\ \bibinfo
  {author} {\bibfnamefont {S.}~\bibnamefont {Basu}},\ }\href@noop {} {\bibinfo
  {title} {Emergence of non-hermitian magic angles and topological phase
  transitions in twisted bilayer lattices}} (\bibinfo {year} {2026}),\ \Eprint
  {https://arxiv.org/abs/2603.29779} {arXiv:2603.29779 [cond-mat.mes-hall]}
  \BibitemShut {NoStop}%
\bibitem [{\citenamefont {Po}\ \emph {et~al.}(2018)\citenamefont {Po},
  \citenamefont {Zou}, \citenamefont {Vishwanath},\ and\ \citenamefont
  {Senthil}}]{Po2018}%
  \BibitemOpen
  \bibfield  {author} {\bibinfo {author} {\bibfnamefont {H.~C.}\ \bibnamefont
  {Po}}, \bibinfo {author} {\bibfnamefont {L.}~\bibnamefont {Zou}}, \bibinfo
  {author} {\bibfnamefont {A.}~\bibnamefont {Vishwanath}},\ and\ \bibinfo
  {author} {\bibfnamefont {T.}~\bibnamefont {Senthil}},\ }\bibfield  {title}
  {\bibinfo {title} {Origin of mott insulating behavior and superconductivity
  in twisted bilayer graphene},\ }\href
  {https://doi.org/10.1103/PhysRevX.8.031089} {\bibfield  {journal} {\bibinfo
  {journal} {Phys. Rev. X}\ }\textbf {\bibinfo {volume} {8}},\ \bibinfo {pages}
  {031089} (\bibinfo {year} {2018})}\BibitemShut {NoStop}%
\bibitem [{\citenamefont {Chen}\ \emph {et~al.}(2021)\citenamefont {Chen},
  \citenamefont {Liao}, \citenamefont {Chen} \emph {et~al.}}]{Chen2021}%
  \BibitemOpen
  \bibfield  {author} {\bibinfo {author} {\bibfnamefont {B.~B.}\ \bibnamefont
  {Chen}}, \bibinfo {author} {\bibfnamefont {Y.~D.}\ \bibnamefont {Liao}},
  \bibinfo {author} {\bibfnamefont {Z.}~\bibnamefont {Chen}}, \emph {et~al.},\
  }\bibfield  {title} {\bibinfo {title} {Realization of topological mott
  insulator in a twisted bilayer graphene lattice model},\ }\href
  {https://doi.org/10.1038/s41467-021-25438-1} {\bibfield  {journal} {\bibinfo
  {journal} {Nature Communications}\ }\textbf {\bibinfo {volume} {12}},\
  \bibinfo {pages} {5480} (\bibinfo {year} {2021})}\BibitemShut {NoStop}%
\bibitem [{\citenamefont {Klug}(2020)}]{Klug2020}%
  \BibitemOpen
  \bibfield  {author} {\bibinfo {author} {\bibfnamefont {M.~J.}\ \bibnamefont
  {Klug}},\ }\bibfield  {title} {\bibinfo {title} {Charge order and mott
  insulating ground states in small-angle twisted bilayer graphene},\ }\href
  {https://doi.org/10.1088/1367-2630/ab950c} {\bibfield  {journal} {\bibinfo
  {journal} {New Journal of Physics}\ }\textbf {\bibinfo {volume} {22}},\
  \bibinfo {pages} {073016} (\bibinfo {year} {2020})}\BibitemShut {NoStop}%
\bibitem [{\citenamefont {Balents}\ \emph {et~al.}(2020)\citenamefont
  {Balents}, \citenamefont {Dean}, \citenamefont {Efetov},\ and\ \citenamefont
  {Young}}]{Balents2020}%
  \BibitemOpen
  \bibfield  {author} {\bibinfo {author} {\bibfnamefont {L.}~\bibnamefont
  {Balents}}, \bibinfo {author} {\bibfnamefont {C.~R.}\ \bibnamefont {Dean}},
  \bibinfo {author} {\bibfnamefont {D.~K.}\ \bibnamefont {Efetov}},\ and\
  \bibinfo {author} {\bibfnamefont {A.~F.}\ \bibnamefont {Young}},\ }\bibfield
  {title} {\bibinfo {title} {Superconductivity and strong correlations in
  moiré flat bands},\ }\href {https://doi.org/10.1038/s41567-020-0906-9}
  {\bibfield  {journal} {\bibinfo  {journal} {Nature Physics}\ }\textbf
  {\bibinfo {volume} {16}},\ \bibinfo {pages} {725} (\bibinfo {year}
  {2020})}\BibitemShut {NoStop}%
\bibitem [{\citenamefont {Isobe}\ \emph {et~al.}(2018)\citenamefont {Isobe},
  \citenamefont {Yuan},\ and\ \citenamefont {Fu}}]{Isobe2018}%
  \BibitemOpen
  \bibfield  {author} {\bibinfo {author} {\bibfnamefont {H.}~\bibnamefont
  {Isobe}}, \bibinfo {author} {\bibfnamefont {N.~F.~Q.}\ \bibnamefont {Yuan}},\
  and\ \bibinfo {author} {\bibfnamefont {L.}~\bibnamefont {Fu}},\ }\bibfield
  {title} {\bibinfo {title} {Unconventional superconductivity and density waves
  in twisted bilayer graphene},\ }\href
  {https://doi.org/10.1103/PhysRevX.8.041041} {\bibfield  {journal} {\bibinfo
  {journal} {Phys. Rev. X}\ }\textbf {\bibinfo {volume} {8}},\ \bibinfo {pages}
  {041041} (\bibinfo {year} {2018})}\BibitemShut {NoStop}%
\bibitem [{\citenamefont {Yankowitz}\ \emph {et~al.}(2019)\citenamefont
  {Yankowitz}, \citenamefont {Chen}, \citenamefont {Polshyn}, \citenamefont
  {Zhang}, \citenamefont {Watanabe}, \citenamefont {Taniguchi}, \citenamefont
  {Graf}, \citenamefont {Young},\ and\ \citenamefont {Dean}}]{Yankowitz2019}%
  \BibitemOpen
  \bibfield  {author} {\bibinfo {author} {\bibfnamefont {M.}~\bibnamefont
  {Yankowitz}}, \bibinfo {author} {\bibfnamefont {S.}~\bibnamefont {Chen}},
  \bibinfo {author} {\bibfnamefont {H.}~\bibnamefont {Polshyn}}, \bibinfo
  {author} {\bibfnamefont {Y.}~\bibnamefont {Zhang}}, \bibinfo {author}
  {\bibfnamefont {K.}~\bibnamefont {Watanabe}}, \bibinfo {author}
  {\bibfnamefont {T.}~\bibnamefont {Taniguchi}}, \bibinfo {author}
  {\bibfnamefont {D.}~\bibnamefont {Graf}}, \bibinfo {author} {\bibfnamefont
  {A.~F.}\ \bibnamefont {Young}},\ and\ \bibinfo {author} {\bibfnamefont
  {C.~R.}\ \bibnamefont {Dean}},\ }\bibfield  {title} {\bibinfo {title} {Tuning
  superconductivity in twisted bilayer graphene},\ }\href
  {https://doi.org/10.1126/science.aav1910} {\bibfield  {journal} {\bibinfo
  {journal} {Science}\ }\textbf {\bibinfo {volume} {363}},\ \bibinfo {pages}
  {1059} (\bibinfo {year} {2019})}\BibitemShut {NoStop}%
\bibitem [{\citenamefont {Oh}\ \emph {et~al.}(2021)\citenamefont {Oh},
  \citenamefont {Nuckolls}, \citenamefont {Wong},\ and\ \citenamefont
  {et~al.}}]{Oh2021}%
  \BibitemOpen
  \bibfield  {author} {\bibinfo {author} {\bibfnamefont {M.~J.}\ \bibnamefont
  {Oh}}, \bibinfo {author} {\bibfnamefont {K.~P.}\ \bibnamefont {Nuckolls}},
  \bibinfo {author} {\bibfnamefont {D.}~\bibnamefont {Wong}},\ and\ \bibinfo
  {author} {\bibnamefont {et~al.}},\ }\bibfield  {title} {\bibinfo {title}
  {Evidence for unconventional superconductivity in twisted bilayer graphene},\
  }\href {https://doi.org/10.1038/s41586-021-04121-x} {\bibfield  {journal}
  {\bibinfo  {journal} {Nature}\ }\textbf {\bibinfo {volume} {600}},\ \bibinfo
  {pages} {240} (\bibinfo {year} {2021})}\BibitemShut {NoStop}%
\bibitem [{\citenamefont {Morales-Dur\'an}\ \emph {et~al.}(2022)\citenamefont
  {Morales-Dur\'an}, \citenamefont {Hu}, \citenamefont {Potasz},\ and\
  \citenamefont {MacDonald}}]{Duran2022}%
  \BibitemOpen
  \bibfield  {author} {\bibinfo {author} {\bibfnamefont {N.}~\bibnamefont
  {Morales-Dur\'an}}, \bibinfo {author} {\bibfnamefont {N.~C.}\ \bibnamefont
  {Hu}}, \bibinfo {author} {\bibfnamefont {P.}~\bibnamefont {Potasz}},\ and\
  \bibinfo {author} {\bibfnamefont {A.~H.}\ \bibnamefont {MacDonald}},\
  }\bibfield  {title} {\bibinfo {title} {Nonlocal interactions in moir\'e
  hubbard systems},\ }\href {https://doi.org/10.1103/PhysRevLett.128.217202}
  {\bibfield  {journal} {\bibinfo  {journal} {Phys. Rev. Lett.}\ }\textbf
  {\bibinfo {volume} {128}},\ \bibinfo {pages} {217202} (\bibinfo {year}
  {2022})}\BibitemShut {NoStop}%
\bibitem [{\citenamefont {Seo}\ \emph {et~al.}(2019)\citenamefont {Seo},
  \citenamefont {Kotov},\ and\ \citenamefont {Uchoa}}]{Seo2019}%
  \BibitemOpen
  \bibfield  {author} {\bibinfo {author} {\bibfnamefont {K.}~\bibnamefont
  {Seo}}, \bibinfo {author} {\bibfnamefont {V.~N.}\ \bibnamefont {Kotov}},\
  and\ \bibinfo {author} {\bibfnamefont {B.}~\bibnamefont {Uchoa}},\ }\bibfield
   {title} {\bibinfo {title} {Ferromagnetic mott state in twisted graphene
  bilayers at the magic angle},\ }\href
  {https://doi.org/10.1103/PhysRevLett.122.246402} {\bibfield  {journal}
  {\bibinfo  {journal} {Phys. Rev. Lett.}\ }\textbf {\bibinfo {volume} {122}},\
  \bibinfo {pages} {246402} (\bibinfo {year} {2019})}\BibitemShut {NoStop}%
\bibitem [{\citenamefont {Serlin}\ \emph {et~al.}(2020)\citenamefont {Serlin},
  \citenamefont {Tschirhart}, \citenamefont {Polshyn}, \citenamefont {Zhang},
  \citenamefont {Zhu}, \citenamefont {Watanabe}, \citenamefont {Taniguchi},
  \citenamefont {Balents},\ and\ \citenamefont {Young}}]{Serlin2020}%
  \BibitemOpen
  \bibfield  {author} {\bibinfo {author} {\bibfnamefont {M.}~\bibnamefont
  {Serlin}}, \bibinfo {author} {\bibfnamefont {C.~L.}\ \bibnamefont
  {Tschirhart}}, \bibinfo {author} {\bibfnamefont {H.}~\bibnamefont {Polshyn}},
  \bibinfo {author} {\bibfnamefont {Y.}~\bibnamefont {Zhang}}, \bibinfo
  {author} {\bibfnamefont {J.}~\bibnamefont {Zhu}}, \bibinfo {author}
  {\bibfnamefont {K.}~\bibnamefont {Watanabe}}, \bibinfo {author}
  {\bibfnamefont {T.}~\bibnamefont {Taniguchi}}, \bibinfo {author}
  {\bibfnamefont {L.}~\bibnamefont {Balents}},\ and\ \bibinfo {author}
  {\bibfnamefont {A.~F.}\ \bibnamefont {Young}},\ }\bibfield  {title} {\bibinfo
  {title} {Intrinsic quantized anomalous hall effect in a moiré
  heterostructure},\ }\href {https://doi.org/10.1126/science.aay5533}
  {\bibfield  {journal} {\bibinfo  {journal} {Science}\ }\textbf {\bibinfo
  {volume} {367}},\ \bibinfo {pages} {900} (\bibinfo {year}
  {2020})}\BibitemShut {NoStop}%
\bibitem [{\citenamefont {He}\ \emph {et~al.}(2026)\citenamefont {He},
  \citenamefont {Gong}, \citenamefont {Li}, \citenamefont {Liu}, \citenamefont
  {Mu},\ and\ \citenamefont {An}}]{He2026}%
  \BibitemOpen
  \bibfield  {author} {\bibinfo {author} {\bibfnamefont {H.}~\bibnamefont
  {He}}, \bibinfo {author} {\bibfnamefont {Z.}~\bibnamefont {Gong}}, \bibinfo
  {author} {\bibfnamefont {S.}~\bibnamefont {Li}}, \bibinfo {author}
  {\bibfnamefont {J.-J.}\ \bibnamefont {Liu}}, \bibinfo {author} {\bibfnamefont
  {H.-Y.}\ \bibnamefont {Mu}},\ and\ \bibinfo {author} {\bibfnamefont {X.-T.}\
  \bibnamefont {An}},\ }\bibfield  {title} {\bibinfo {title} {Multiple quantum
  spin hall states and topological current divider in twisted bilayer
  ${\mathrm{wse}}_{2}$},\ }\href {https://doi.org/10.1103/z9w9-9vkw} {\bibfield
   {journal} {\bibinfo  {journal} {Phys. Rev. B}\ }\textbf {\bibinfo {volume}
  {113}},\ \bibinfo {pages} {045423} (\bibinfo {year} {2026})}\BibitemShut
  {NoStop}%
\bibitem [{\citenamefont {Tateishi}\ and\ \citenamefont
  {Hirayama}(2022)}]{Tateishi2022}%
  \BibitemOpen
  \bibfield  {author} {\bibinfo {author} {\bibfnamefont {I.}~\bibnamefont
  {Tateishi}}\ and\ \bibinfo {author} {\bibfnamefont {M.}~\bibnamefont
  {Hirayama}},\ }\bibfield  {title} {\bibinfo {title} {Quantum spin hall effect
  from multiscale band inversion in twisted bilayer
  ${\mathrm{bi}}_{2}{({\mathrm{Te}}_{1\ensuremath{-}x}{\mathrm{Se}}_{x})}_{3}$},\
  }\href {https://doi.org/10.1103/PhysRevResearch.4.043045} {\bibfield
  {journal} {\bibinfo  {journal} {Phys. Rev. Res.}\ }\textbf {\bibinfo {volume}
  {4}},\ \bibinfo {pages} {043045} (\bibinfo {year} {2022})}\BibitemShut
  {NoStop}%
\bibitem [{\citenamefont {Xu}\ \emph {et~al.}(2023)\citenamefont {Xu},
  \citenamefont {Sun}, \citenamefont {Jia}, \citenamefont {Liu}, \citenamefont
  {Xu}, \citenamefont {Li}, \citenamefont {Gu}, \citenamefont {Watanabe},
  \citenamefont {Taniguchi}, \citenamefont {Tong}, \citenamefont {Jia},
  \citenamefont {Shi}, \citenamefont {Jiang}, \citenamefont {Zhang},
  \citenamefont {Liu},\ and\ \citenamefont {Li}}]{Xu2023}%
  \BibitemOpen
  \bibfield  {author} {\bibinfo {author} {\bibfnamefont {F.}~\bibnamefont
  {Xu}}, \bibinfo {author} {\bibfnamefont {Z.}~\bibnamefont {Sun}}, \bibinfo
  {author} {\bibfnamefont {T.}~\bibnamefont {Jia}}, \bibinfo {author}
  {\bibfnamefont {C.}~\bibnamefont {Liu}}, \bibinfo {author} {\bibfnamefont
  {C.}~\bibnamefont {Xu}}, \bibinfo {author} {\bibfnamefont {C.}~\bibnamefont
  {Li}}, \bibinfo {author} {\bibfnamefont {Y.}~\bibnamefont {Gu}}, \bibinfo
  {author} {\bibfnamefont {K.}~\bibnamefont {Watanabe}}, \bibinfo {author}
  {\bibfnamefont {T.}~\bibnamefont {Taniguchi}}, \bibinfo {author}
  {\bibfnamefont {B.}~\bibnamefont {Tong}}, \bibinfo {author} {\bibfnamefont
  {J.}~\bibnamefont {Jia}}, \bibinfo {author} {\bibfnamefont {Z.}~\bibnamefont
  {Shi}}, \bibinfo {author} {\bibfnamefont {S.}~\bibnamefont {Jiang}}, \bibinfo
  {author} {\bibfnamefont {Y.}~\bibnamefont {Zhang}}, \bibinfo {author}
  {\bibfnamefont {X.}~\bibnamefont {Liu}},\ and\ \bibinfo {author}
  {\bibfnamefont {T.}~\bibnamefont {Li}},\ }\bibfield  {title} {\bibinfo
  {title} {Observation of integer and fractional quantum anomalous hall effects
  in twisted bilayer ${\mathrm{mote}}_{2}$},\ }\href
  {https://doi.org/10.1103/PhysRevX.13.031037} {\bibfield  {journal} {\bibinfo
  {journal} {Phys. Rev. X}\ }\textbf {\bibinfo {volume} {13}},\ \bibinfo
  {pages} {031037} (\bibinfo {year} {2023})}\BibitemShut {NoStop}%
\bibitem [{\citenamefont {Reddy}\ \emph {et~al.}(2023)\citenamefont {Reddy},
  \citenamefont {Alsallom}, \citenamefont {Zhang}, \citenamefont {Devakul},\
  and\ \citenamefont {Fu}}]{Reddy2023}%
  \BibitemOpen
  \bibfield  {author} {\bibinfo {author} {\bibfnamefont {A.~P.}\ \bibnamefont
  {Reddy}}, \bibinfo {author} {\bibfnamefont {F.}~\bibnamefont {Alsallom}},
  \bibinfo {author} {\bibfnamefont {Y.}~\bibnamefont {Zhang}}, \bibinfo
  {author} {\bibfnamefont {T.}~\bibnamefont {Devakul}},\ and\ \bibinfo {author}
  {\bibfnamefont {L.}~\bibnamefont {Fu}},\ }\bibfield  {title} {\bibinfo
  {title} {Fractional quantum anomalous hall states in twisted bilayer
  ${\mathrm{mote}}_{2}$ and ${\mathrm{wse}}_{2}$},\ }\href
  {https://doi.org/10.1103/PhysRevB.108.085117} {\bibfield  {journal} {\bibinfo
   {journal} {Phys. Rev. B}\ }\textbf {\bibinfo {volume} {108}},\ \bibinfo
  {pages} {085117} (\bibinfo {year} {2023})}\BibitemShut {NoStop}%
\bibitem [{\citenamefont {Park}\ \emph {et~al.}(2023)\citenamefont {Park},
  \citenamefont {Cai}, \citenamefont {Anderson},\ and\ \citenamefont
  {et~al.}}]{Park2023}%
  \BibitemOpen
  \bibfield  {author} {\bibinfo {author} {\bibfnamefont {H.}~\bibnamefont
  {Park}}, \bibinfo {author} {\bibfnamefont {J.}~\bibnamefont {Cai}}, \bibinfo
  {author} {\bibfnamefont {E.}~\bibnamefont {Anderson}},\ and\ \bibinfo
  {author} {\bibnamefont {et~al.}},\ }\bibfield  {title} {\bibinfo {title}
  {Observation of fractionally quantized anomalous hall effect},\ }\href
  {https://doi.org/10.1038/s41586-023-06536-0} {\bibfield  {journal} {\bibinfo
  {journal} {Nature}\ }\textbf {\bibinfo {volume} {622}},\ \bibinfo {pages}
  {74} (\bibinfo {year} {2023})}\BibitemShut {NoStop}%
\bibitem [{\citenamefont {Nagaosa}\ \emph {et~al.}(2010)\citenamefont
  {Nagaosa}, \citenamefont {Sinova}, \citenamefont {Onoda}, \citenamefont
  {MacDonald},\ and\ \citenamefont {Ong}}]{Nagaosa2010}%
  \BibitemOpen
  \bibfield  {author} {\bibinfo {author} {\bibfnamefont {N.}~\bibnamefont
  {Nagaosa}}, \bibinfo {author} {\bibfnamefont {J.}~\bibnamefont {Sinova}},
  \bibinfo {author} {\bibfnamefont {S.}~\bibnamefont {Onoda}}, \bibinfo
  {author} {\bibfnamefont {A.~H.}\ \bibnamefont {MacDonald}},\ and\ \bibinfo
  {author} {\bibfnamefont {N.~P.}\ \bibnamefont {Ong}},\ }\bibfield  {title}
  {\bibinfo {title} {Anomalous hall effect},\ }\href
  {https://doi.org/10.1103/RevModPhys.82.1539} {\bibfield  {journal} {\bibinfo
  {journal} {Rev. Mod. Phys.}\ }\textbf {\bibinfo {volume} {82}},\ \bibinfo
  {pages} {1539} (\bibinfo {year} {2010})}\BibitemShut {NoStop}%
\bibitem [{\citenamefont {Du}\ \emph {et~al.}(2018)\citenamefont {Du},
  \citenamefont {Wang}, \citenamefont {Lu},\ and\ \citenamefont
  {Xie}}]{Du2018}%
  \BibitemOpen
  \bibfield  {author} {\bibinfo {author} {\bibfnamefont {Z.~Z.}\ \bibnamefont
  {Du}}, \bibinfo {author} {\bibfnamefont {C.~M.}\ \bibnamefont {Wang}},
  \bibinfo {author} {\bibfnamefont {H.-Z.}\ \bibnamefont {Lu}},\ and\ \bibinfo
  {author} {\bibfnamefont {X.~C.}\ \bibnamefont {Xie}},\ }\bibfield  {title}
  {\bibinfo {title} {Band signatures for strong nonlinear hall effect in
  bilayer ${\mathrm{wte}}_{2}$},\ }\href
  {https://doi.org/10.1103/PhysRevLett.121.266601} {\bibfield  {journal}
  {\bibinfo  {journal} {Phys. Rev. Lett.}\ }\textbf {\bibinfo {volume} {121}},\
  \bibinfo {pages} {266601} (\bibinfo {year} {2018})}\BibitemShut {NoStop}%
\bibitem [{\citenamefont {Low}\ \emph {et~al.}(2015)\citenamefont {Low},
  \citenamefont {Jiang},\ and\ \citenamefont {Guinea}}]{Low2015}%
  \BibitemOpen
  \bibfield  {author} {\bibinfo {author} {\bibfnamefont {T.}~\bibnamefont
  {Low}}, \bibinfo {author} {\bibfnamefont {Y.}~\bibnamefont {Jiang}},\ and\
  \bibinfo {author} {\bibfnamefont {F.}~\bibnamefont {Guinea}},\ }\bibfield
  {title} {\bibinfo {title} {Topological currents in black phosphorus with
  broken inversion symmetry},\ }\href
  {https://doi.org/10.1103/PhysRevB.92.235447} {\bibfield  {journal} {\bibinfo
  {journal} {Phys. Rev. B}\ }\textbf {\bibinfo {volume} {92}},\ \bibinfo
  {pages} {235447} (\bibinfo {year} {2015})}\BibitemShut {NoStop}%
\bibitem [{\citenamefont {Zhang}\ \emph {et~al.}(2018)\citenamefont {Zhang},
  \citenamefont {van~den Brink}, \citenamefont {Felser},\ and\ \citenamefont
  {Yan}}]{Zhang2018}%
  \BibitemOpen
  \bibfield  {author} {\bibinfo {author} {\bibfnamefont {Y.}~\bibnamefont
  {Zhang}}, \bibinfo {author} {\bibfnamefont {J.}~\bibnamefont {van~den
  Brink}}, \bibinfo {author} {\bibfnamefont {C.}~\bibnamefont {Felser}},\ and\
  \bibinfo {author} {\bibfnamefont {B.}~\bibnamefont {Yan}},\ }\bibfield
  {title} {\bibinfo {title} {Electrically tuneable nonlinear anomalous hall
  effect in two-dimensional transition-metal dichalcogenides wte$_2$ and
  mote$_2$},\ }\href {https://doi.org/10.1088/2053-1583/aad1ae} {\bibfield
  {journal} {\bibinfo  {journal} {2D Materials}\ }\textbf {\bibinfo {volume}
  {5}},\ \bibinfo {pages} {044001} (\bibinfo {year} {2018})}\BibitemShut
  {NoStop}%
\bibitem [{\citenamefont {Tokura}\ and\ \citenamefont
  {Nagaosa}(2018)}]{Tokura2018}%
  \BibitemOpen
  \bibfield  {author} {\bibinfo {author} {\bibfnamefont {Y.}~\bibnamefont
  {Tokura}}\ and\ \bibinfo {author} {\bibfnamefont {N.}~\bibnamefont
  {Nagaosa}},\ }\bibfield  {title} {\bibinfo {title} {Nonreciprocal responses
  from non-centrosymmetric quantum materials},\ }\href
  {https://doi.org/10.1038/s41467-018-05759-4} {\bibfield  {journal} {\bibinfo
  {journal} {Nature Communications}\ }\textbf {\bibinfo {volume} {9}},\
  \bibinfo {pages} {3740} (\bibinfo {year} {2018})}\BibitemShut {NoStop}%
\bibitem [{\citenamefont {Zhang}\ \emph {et~al.}(2022)\citenamefont {Zhang},
  \citenamefont {Xiao}, \citenamefont {Zhou}, \citenamefont {Hu}, \citenamefont
  {Xie}, \citenamefont {Yan},\ and\ \citenamefont {Law}}]{Zhang2022}%
  \BibitemOpen
  \bibfield  {author} {\bibinfo {author} {\bibfnamefont {C.-P.}\ \bibnamefont
  {Zhang}}, \bibinfo {author} {\bibfnamefont {J.}~\bibnamefont {Xiao}},
  \bibinfo {author} {\bibfnamefont {B.~T.}\ \bibnamefont {Zhou}}, \bibinfo
  {author} {\bibfnamefont {J.-X.}\ \bibnamefont {Hu}}, \bibinfo {author}
  {\bibfnamefont {Y.-M.}\ \bibnamefont {Xie}}, \bibinfo {author} {\bibfnamefont
  {B.}~\bibnamefont {Yan}},\ and\ \bibinfo {author} {\bibfnamefont {K.~T.}\
  \bibnamefont {Law}},\ }\bibfield  {title} {\bibinfo {title} {Giant nonlinear
  hall effect in strained twisted bilayer graphene},\ }\href
  {https://doi.org/10.1103/PhysRevB.106.L041111} {\bibfield  {journal}
  {\bibinfo  {journal} {Phys. Rev. B}\ }\textbf {\bibinfo {volume} {106}},\
  \bibinfo {pages} {L041111} (\bibinfo {year} {2022})}\BibitemShut {NoStop}%
\bibitem [{\citenamefont {Chakraborty}\ \emph {et~al.}(2022)\citenamefont
  {Chakraborty}, \citenamefont {Das}, \citenamefont {Sinha}, \citenamefont
  {Adak}, \citenamefont {Deshmukh},\ and\ \citenamefont
  {Agarwal}}]{Chakraborty2022}%
  \BibitemOpen
  \bibfield  {author} {\bibinfo {author} {\bibfnamefont {A.}~\bibnamefont
  {Chakraborty}}, \bibinfo {author} {\bibfnamefont {K.}~\bibnamefont {Das}},
  \bibinfo {author} {\bibfnamefont {S.}~\bibnamefont {Sinha}}, \bibinfo
  {author} {\bibfnamefont {P.~C.}\ \bibnamefont {Adak}}, \bibinfo {author}
  {\bibfnamefont {M.~M.}\ \bibnamefont {Deshmukh}},\ and\ \bibinfo {author}
  {\bibfnamefont {A.}~\bibnamefont {Agarwal}},\ }\bibfield  {title} {\bibinfo
  {title} {Nonlinear anomalous hall effects probe topological phase transitions
  in twisted double bilayer graphene},\ }\href
  {https://doi.org/10.1088/2053-1583/ac8f2c} {\bibfield  {journal} {\bibinfo
  {journal} {2D Materials}\ }\textbf {\bibinfo {volume} {9}},\ \bibinfo {pages}
  {045020} (\bibinfo {year} {2022})}\BibitemShut {NoStop}%
\bibitem [{\citenamefont {Hu}\ \emph {et~al.}(2022)\citenamefont {Hu},
  \citenamefont {Zhang}, \citenamefont {Xie},\ and\ \citenamefont
  {et~al.}}]{Hu2022}%
  \BibitemOpen
  \bibfield  {author} {\bibinfo {author} {\bibfnamefont {J.~X.}\ \bibnamefont
  {Hu}}, \bibinfo {author} {\bibfnamefont {C.~P.}\ \bibnamefont {Zhang}},
  \bibinfo {author} {\bibfnamefont {Y.~M.}\ \bibnamefont {Xie}},\ and\ \bibinfo
  {author} {\bibnamefont {et~al.}},\ }\bibfield  {title} {\bibinfo {title}
  {Nonlinear hall effects in strained twisted bilayer wse$_2$},\ }\href
  {https://doi.org/10.1038/s42005-022-01034-7} {\bibfield  {journal} {\bibinfo
  {journal} {Communications Physics}\ }\textbf {\bibinfo {volume} {5}},\
  \bibinfo {pages} {255} (\bibinfo {year} {2022})}\BibitemShut {NoStop}%
\bibitem [{\citenamefont {Ho}\ \emph {et~al.}(2021)\citenamefont {Ho},
  \citenamefont {Chang}, \citenamefont {Hsieh},\ and\ \citenamefont
  {et~al.}}]{Ho2021}%
  \BibitemOpen
  \bibfield  {author} {\bibinfo {author} {\bibfnamefont {S.-C.}\ \bibnamefont
  {Ho}}, \bibinfo {author} {\bibfnamefont {C.-H.}\ \bibnamefont {Chang}},
  \bibinfo {author} {\bibfnamefont {Y.-C.}\ \bibnamefont {Hsieh}},\ and\
  \bibinfo {author} {\bibnamefont {et~al.}},\ }\bibfield  {title} {\bibinfo
  {title} {Hall effects in artificially corrugated bilayer graphene without
  breaking time-reversal symmetry},\ }\href
  {https://doi.org/10.1038/s41928-021-00537-5} {\bibfield  {journal} {\bibinfo
  {journal} {Nature Electronics}\ }\textbf {\bibinfo {volume} {4}},\ \bibinfo
  {pages} {116} (\bibinfo {year} {2021})}\BibitemShut {NoStop}%
\bibitem [{\citenamefont {Qin}\ \emph {et~al.}(2021)\citenamefont {Qin},
  \citenamefont {Zhu}, \citenamefont {Ye}, \citenamefont {Xu}, \citenamefont
  {Song}, \citenamefont {Liang}, \citenamefont {Liu},\ and\ \citenamefont
  {Liao}}]{Qin2021}%
  \BibitemOpen
  \bibfield  {author} {\bibinfo {author} {\bibfnamefont {M.-S.}\ \bibnamefont
  {Qin}}, \bibinfo {author} {\bibfnamefont {P.-F.}\ \bibnamefont {Zhu}},
  \bibinfo {author} {\bibfnamefont {X.-G.}\ \bibnamefont {Ye}}, \bibinfo
  {author} {\bibfnamefont {W.-Z.}\ \bibnamefont {Xu}}, \bibinfo {author}
  {\bibfnamefont {Z.-H.}\ \bibnamefont {Song}}, \bibinfo {author}
  {\bibfnamefont {J.}~\bibnamefont {Liang}}, \bibinfo {author} {\bibfnamefont
  {K.}~\bibnamefont {Liu}},\ and\ \bibinfo {author} {\bibfnamefont {Z.-M.}\
  \bibnamefont {Liao}},\ }\bibfield  {title} {\bibinfo {title} {Strain tunable
  berry curvature dipole, orbital magnetization and nonlinear hall effect in
  wse$_2$ monolayer},\ }\href {https://doi.org/10.1088/0256-307X/38/1/017301}
  {\bibfield  {journal} {\bibinfo  {journal} {Chinese Physics Letters}\
  }\textbf {\bibinfo {volume} {38}},\ \bibinfo {pages} {017301} (\bibinfo
  {year} {2021})}\BibitemShut {NoStop}%
\bibitem [{\citenamefont {Kumar}\ \emph {et~al.}(2021)\citenamefont {Kumar},
  \citenamefont {Hsu}, \citenamefont {Sharma},\ and\ \citenamefont
  {et~al.}}]{Kumar2021}%
  \BibitemOpen
  \bibfield  {author} {\bibinfo {author} {\bibfnamefont {D.}~\bibnamefont
  {Kumar}}, \bibinfo {author} {\bibfnamefont {C.-H.}\ \bibnamefont {Hsu}},
  \bibinfo {author} {\bibfnamefont {R.}~\bibnamefont {Sharma}},\ and\ \bibinfo
  {author} {\bibnamefont {et~al.}},\ }\bibfield  {title} {\bibinfo {title}
  {Room-temperature nonlinear hall effect and wireless radiofrequency
  rectification in weyl semimetal tairte$_4$},\ }\href
  {https://doi.org/10.1038/s41565-020-00839-3} {\bibfield  {journal} {\bibinfo
  {journal} {Nature Nanotechnology}\ }\textbf {\bibinfo {volume} {16}},\
  \bibinfo {pages} {421} (\bibinfo {year} {2021})}\BibitemShut {NoStop}%
\bibitem [{\citenamefont {Ma}\ \emph {et~al.}(2022)\citenamefont {Ma},
  \citenamefont {Chen}, \citenamefont {Yananose}, \citenamefont {Zhou},
  \citenamefont {Wang}, \citenamefont {Li}, \citenamefont {Zhu}, \citenamefont
  {Wu}, \citenamefont {Xu}, \citenamefont {Yu}, \citenamefont {Qiu},
  \citenamefont {Stroppa},\ and\ \citenamefont {Loh}}]{Ma2022}%
  \BibitemOpen
  \bibfield  {author} {\bibinfo {author} {\bibfnamefont {T.}~\bibnamefont
  {Ma}}, \bibinfo {author} {\bibfnamefont {H.}~\bibnamefont {Chen}}, \bibinfo
  {author} {\bibfnamefont {K.}~\bibnamefont {Yananose}}, \bibinfo {author}
  {\bibfnamefont {X.}~\bibnamefont {Zhou}}, \bibinfo {author} {\bibfnamefont
  {L.}~\bibnamefont {Wang}}, \bibinfo {author} {\bibfnamefont {R.}~\bibnamefont
  {Li}}, \bibinfo {author} {\bibfnamefont {Z.}~\bibnamefont {Zhu}}, \bibinfo
  {author} {\bibfnamefont {Z.}~\bibnamefont {Wu}}, \bibinfo {author}
  {\bibfnamefont {Q.-H.}\ \bibnamefont {Xu}}, \bibinfo {author} {\bibfnamefont
  {J.}~\bibnamefont {Yu}}, \bibinfo {author} {\bibfnamefont {C.-W.}\
  \bibnamefont {Qiu}}, \bibinfo {author} {\bibfnamefont {A.}~\bibnamefont
  {Stroppa}},\ and\ \bibinfo {author} {\bibfnamefont {K.~P.}\ \bibnamefont
  {Loh}},\ }\bibfield  {title} {\bibinfo {title} {Growth of bilayer mote$_2$
  single crystals with strong nonlinear hall effect},\ }\href
  {https://doi.org/10.1038/s41467-022-33201-3} {\bibfield  {journal} {\bibinfo
  {journal} {Nature Communications}\ }\textbf {\bibinfo {volume} {13}},\
  \bibinfo {pages} {5465} (\bibinfo {year} {2022})}\BibitemShut {NoStop}%
\bibitem [{\citenamefont {Duan}\ \emph {et~al.}(2022)\citenamefont {Duan},
  \citenamefont {Jian}, \citenamefont {Gao}, \citenamefont {Peng},
  \citenamefont {Zhong}, \citenamefont {Feng}, \citenamefont {Mao},\ and\
  \citenamefont {Yao}}]{Duan2022}%
  \BibitemOpen
  \bibfield  {author} {\bibinfo {author} {\bibfnamefont {J.}~\bibnamefont
  {Duan}}, \bibinfo {author} {\bibfnamefont {Y.}~\bibnamefont {Jian}}, \bibinfo
  {author} {\bibfnamefont {Y.}~\bibnamefont {Gao}}, \bibinfo {author}
  {\bibfnamefont {H.}~\bibnamefont {Peng}}, \bibinfo {author} {\bibfnamefont
  {J.}~\bibnamefont {Zhong}}, \bibinfo {author} {\bibfnamefont
  {Q.}~\bibnamefont {Feng}}, \bibinfo {author} {\bibfnamefont {J.}~\bibnamefont
  {Mao}},\ and\ \bibinfo {author} {\bibfnamefont {Y.}~\bibnamefont {Yao}},\
  }\bibfield  {title} {\bibinfo {title} {Giant second-order nonlinear hall
  effect in twisted bilayer graphene},\ }\href
  {https://doi.org/10.1103/PhysRevLett.129.186801} {\bibfield  {journal}
  {\bibinfo  {journal} {Phys. Rev. Lett.}\ }\textbf {\bibinfo {volume} {129}},\
  \bibinfo {pages} {186801} (\bibinfo {year} {2022})}\BibitemShut {NoStop}%
\bibitem [{\citenamefont {Huang}\ \emph
  {et~al.}(2023{\natexlab{a}})\citenamefont {Huang}, \citenamefont {Wu},
  \citenamefont {Zhang}, \citenamefont {Feng}, \citenamefont {Zhou},
  \citenamefont {Wang}, \citenamefont {Chen}, \citenamefont {Cheng},
  \citenamefont {Sun}, \citenamefont {Meng},\ and\ \citenamefont
  {Wang}}]{Huang2023}%
  \BibitemOpen
  \bibfield  {author} {\bibinfo {author} {\bibfnamefont {M.}~\bibnamefont
  {Huang}}, \bibinfo {author} {\bibfnamefont {Z.}~\bibnamefont {Wu}}, \bibinfo
  {author} {\bibfnamefont {X.}~\bibnamefont {Zhang}}, \bibinfo {author}
  {\bibfnamefont {X.}~\bibnamefont {Feng}}, \bibinfo {author} {\bibfnamefont
  {Z.}~\bibnamefont {Zhou}}, \bibinfo {author} {\bibfnamefont {S.}~\bibnamefont
  {Wang}}, \bibinfo {author} {\bibfnamefont {Y.}~\bibnamefont {Chen}}, \bibinfo
  {author} {\bibfnamefont {C.}~\bibnamefont {Cheng}}, \bibinfo {author}
  {\bibfnamefont {K.}~\bibnamefont {Sun}}, \bibinfo {author} {\bibfnamefont
  {Z.~Y.}\ \bibnamefont {Meng}},\ and\ \bibinfo {author} {\bibfnamefont
  {N.}~\bibnamefont {Wang}},\ }\bibfield  {title} {\bibinfo {title} {Intrinsic
  nonlinear hall effect and gate-switchable berry curvature sliding in twisted
  bilayer graphene},\ }\href {https://doi.org/10.1103/PhysRevLett.131.066301}
  {\bibfield  {journal} {\bibinfo  {journal} {Phys. Rev. Lett.}\ }\textbf
  {\bibinfo {volume} {131}},\ \bibinfo {pages} {066301} (\bibinfo {year}
  {2023}{\natexlab{a}})}\BibitemShut {NoStop}%
\bibitem [{\citenamefont {Huang}\ \emph
  {et~al.}(2023{\natexlab{b}})\citenamefont {Huang}, \citenamefont {Wu},
  \citenamefont {Hu}, \citenamefont {Cai}, \citenamefont {Li}, \citenamefont
  {An}, \citenamefont {Feng}, \citenamefont {Ye}, \citenamefont {Lin},
  \citenamefont {Law},\ and\ \citenamefont {Wang}}]{Huang2023A}%
  \BibitemOpen
  \bibfield  {author} {\bibinfo {author} {\bibfnamefont {M.}~\bibnamefont
  {Huang}}, \bibinfo {author} {\bibfnamefont {Z.}~\bibnamefont {Wu}}, \bibinfo
  {author} {\bibfnamefont {J.}~\bibnamefont {Hu}}, \bibinfo {author}
  {\bibfnamefont {X.}~\bibnamefont {Cai}}, \bibinfo {author} {\bibfnamefont
  {E.}~\bibnamefont {Li}}, \bibinfo {author} {\bibfnamefont {L.}~\bibnamefont
  {An}}, \bibinfo {author} {\bibfnamefont {X.}~\bibnamefont {Feng}}, \bibinfo
  {author} {\bibfnamefont {Z.}~\bibnamefont {Ye}}, \bibinfo {author}
  {\bibfnamefont {N.}~\bibnamefont {Lin}}, \bibinfo {author} {\bibfnamefont
  {K.~T.}\ \bibnamefont {Law}},\ and\ \bibinfo {author} {\bibfnamefont
  {N.}~\bibnamefont {Wang}},\ }\bibfield  {title} {\bibinfo {title} {Giant
  nonlinear hall effect in twisted bilayer wse$_2$},\ }\href
  {https://doi.org/10.1093/nsr/nwac232} {\bibfield  {journal} {\bibinfo
  {journal} {National Science Review}\ }\textbf {\bibinfo {volume} {10}},\
  \bibinfo {pages} {nwac232} (\bibinfo {year}
  {2023}{\natexlab{b}})}\BibitemShut {NoStop}%
\bibitem [{\citenamefont {Cao}\ \emph {et~al.}(2025)\citenamefont {Cao},
  \citenamefont {Wang}, \citenamefont {Liu}, \citenamefont {Zhao},
  \citenamefont {Yu},\ and\ \citenamefont {Liao}}]{Cao2025}%
  \BibitemOpen
  \bibfield  {author} {\bibinfo {author} {\bibfnamefont {Z.-Y.}\ \bibnamefont
  {Cao}}, \bibinfo {author} {\bibfnamefont {A.-Q.}\ \bibnamefont {Wang}},
  \bibinfo {author} {\bibfnamefont {X.-Y.}\ \bibnamefont {Liu}}, \bibinfo
  {author} {\bibfnamefont {T.-Y.}\ \bibnamefont {Zhao}}, \bibinfo {author}
  {\bibfnamefont {D.}~\bibnamefont {Yu}},\ and\ \bibinfo {author}
  {\bibfnamefont {Z.-M.}\ \bibnamefont {Liao}},\ }\bibfield  {title} {\bibinfo
  {title} {Nonlinear hall effect and scaling law analysis in twisted bilayer
  $\mathrm{WS}{\mathrm{e}}_{2}$},\ }\href
  {https://doi.org/10.1103/PhysRevB.111.125407} {\bibfield  {journal} {\bibinfo
   {journal} {Phys. Rev. B}\ }\textbf {\bibinfo {volume} {111}},\ \bibinfo
  {pages} {125407} (\bibinfo {year} {2025})}\BibitemShut {NoStop}%
\bibitem [{\citenamefont {Wu}\ \emph {et~al.}(2023)\citenamefont {Wu},
  \citenamefont {Xu}, \citenamefont {Wang}, \citenamefont {Chu}, \citenamefont
  {Li}, \citenamefont {Tang}, \citenamefont {Liu}, \citenamefont {Tian},
  \citenamefont {Ji}, \citenamefont {Liu}, \citenamefont {Yuan}, \citenamefont
  {Huang}, \citenamefont {Zhao}, \citenamefont {Zan}, \citenamefont {Watanabe},
  \citenamefont {Taniguchi}, \citenamefont {Shi}, \citenamefont {Gu},
  \citenamefont {Xu}, \citenamefont {Xian}, \citenamefont {Yang}, \citenamefont
  {Du},\ and\ \citenamefont {Zhang}}]{Wu2023}%
  \BibitemOpen
  \bibfield  {author} {\bibinfo {author} {\bibfnamefont {F.}~\bibnamefont
  {Wu}}, \bibinfo {author} {\bibfnamefont {Q.}~\bibnamefont {Xu}}, \bibinfo
  {author} {\bibfnamefont {Q.}~\bibnamefont {Wang}}, \bibinfo {author}
  {\bibfnamefont {Y.}~\bibnamefont {Chu}}, \bibinfo {author} {\bibfnamefont
  {L.}~\bibnamefont {Li}}, \bibinfo {author} {\bibfnamefont {J.}~\bibnamefont
  {Tang}}, \bibinfo {author} {\bibfnamefont {J.}~\bibnamefont {Liu}}, \bibinfo
  {author} {\bibfnamefont {J.}~\bibnamefont {Tian}}, \bibinfo {author}
  {\bibfnamefont {Y.}~\bibnamefont {Ji}}, \bibinfo {author} {\bibfnamefont
  {L.}~\bibnamefont {Liu}}, \bibinfo {author} {\bibfnamefont {Y.}~\bibnamefont
  {Yuan}}, \bibinfo {author} {\bibfnamefont {Z.}~\bibnamefont {Huang}},
  \bibinfo {author} {\bibfnamefont {J.}~\bibnamefont {Zhao}}, \bibinfo {author}
  {\bibfnamefont {X.}~\bibnamefont {Zan}}, \bibinfo {author} {\bibfnamefont
  {K.}~\bibnamefont {Watanabe}}, \bibinfo {author} {\bibfnamefont
  {T.}~\bibnamefont {Taniguchi}}, \bibinfo {author} {\bibfnamefont
  {D.}~\bibnamefont {Shi}}, \bibinfo {author} {\bibfnamefont {G.}~\bibnamefont
  {Gu}}, \bibinfo {author} {\bibfnamefont {Y.}~\bibnamefont {Xu}}, \bibinfo
  {author} {\bibfnamefont {L.}~\bibnamefont {Xian}}, \bibinfo {author}
  {\bibfnamefont {W.}~\bibnamefont {Yang}}, \bibinfo {author} {\bibfnamefont
  {L.}~\bibnamefont {Du}},\ and\ \bibinfo {author} {\bibfnamefont
  {G.}~\bibnamefont {Zhang}},\ }\href@noop {} {\bibinfo {title}
  {Room-temperature correlated states in twisted bilayer \text{MoS}$_2$}}
  (\bibinfo {year} {2023}),\ \Eprint {https://arxiv.org/abs/2311.16655}
  {arXiv:2311.16655 [cond-mat.mtrl-sci]} \BibitemShut {NoStop}%
\bibitem [{\citenamefont {He}\ and\ \citenamefont {Weng}(2021)}]{He2021}%
  \BibitemOpen
  \bibfield  {author} {\bibinfo {author} {\bibfnamefont {Z.}~\bibnamefont
  {He}}\ and\ \bibinfo {author} {\bibfnamefont {H.}~\bibnamefont {Weng}},\
  }\bibfield  {title} {\bibinfo {title} {Giant nonlinear hall effect in twisted
  bilayer wte$_2$},\ }\href {https://doi.org/10.1038/s41535-021-00403-9}
  {\bibfield  {journal} {\bibinfo  {journal} {npj Quantum Materials}\ }\textbf
  {\bibinfo {volume} {6}},\ \bibinfo {pages} {101} (\bibinfo {year}
  {2021})}\BibitemShut {NoStop}%
\bibitem [{\citenamefont {Paul}\ \emph {et~al.}(2026)\citenamefont {Paul},
  \citenamefont {Lahiri}, \citenamefont {Bhattacharyya},\ and\ \citenamefont
  {Basu}}]{Paul2026}%
  \BibitemOpen
  \bibfield  {author} {\bibinfo {author} {\bibfnamefont {G.}~\bibnamefont
  {Paul}}, \bibinfo {author} {\bibfnamefont {S.}~\bibnamefont {Lahiri}},
  \bibinfo {author} {\bibfnamefont {K.}~\bibnamefont {Bhattacharyya}},\ and\
  \bibinfo {author} {\bibfnamefont {S.}~\bibnamefont {Basu}},\ }\bibfield
  {title} {\bibinfo {title} {Emergent topology of flat bands in a twisted
  bilayer $\ensuremath{\alpha}\text{\ensuremath{-}}{T}_{3}$ lattice},\ }\href
  {https://doi.org/10.1103/pyjd-jyrc} {\bibfield  {journal} {\bibinfo
  {journal} {Phys. Rev. B}\ }\textbf {\bibinfo {volume} {113}},\ \bibinfo
  {pages} {035145} (\bibinfo {year} {2026})}\BibitemShut {NoStop}%
\bibitem [{\citenamefont {Zhou}\ \emph {et~al.}(2026)\citenamefont {Zhou},
  \citenamefont {Hung},\ and\ \citenamefont {Bansil}}]{Zhou2026}%
  \BibitemOpen
  \bibfield  {author} {\bibinfo {author} {\bibfnamefont {X.}~\bibnamefont
  {Zhou}}, \bibinfo {author} {\bibfnamefont {Y.-C.}\ \bibnamefont {Hung}},\
  and\ \bibinfo {author} {\bibfnamefont {A.}~\bibnamefont {Bansil}},\
  }\bibfield  {title} {\bibinfo {title} {Quantum geometry of moir\'e flat bands
  beyond the valley paradigm},\ }\href@noop {} {\bibfield  {journal} {\bibinfo
  {journal} {arXiv preprint arXiv:2603.20852}\ } (\bibinfo {year} {2026})},\
  \Eprint {https://arxiv.org/abs/2603.20852} {arXiv:2603.20852
  [cond-mat.mes-hall]} \BibitemShut {NoStop}%
\bibitem [{\citenamefont {Bistritzer}\ and\ \citenamefont
  {MacDonald}(2011)}]{Bistritzer2011}%
  \BibitemOpen
  \bibfield  {author} {\bibinfo {author} {\bibfnamefont {R.}~\bibnamefont
  {Bistritzer}}\ and\ \bibinfo {author} {\bibfnamefont {A.~H.}\ \bibnamefont
  {MacDonald}},\ }\bibfield  {title} {\bibinfo {title} {Moir{\'e} bands in
  twisted double-layer graphene},\ }\href
  {https://doi.org/10.1073/pnas.1108174108} {\bibfield  {journal} {\bibinfo
  {journal} {Proceedings of the National Academy of Sciences}\ }\textbf
  {\bibinfo {volume} {108}},\ \bibinfo {pages} {12233} (\bibinfo {year}
  {2011})}\BibitemShut {NoStop}%
\bibitem [{\citenamefont {Sukhachov}\ \emph {et~al.}(2023)\citenamefont
  {Sukhachov}, \citenamefont {Oriekhov},\ and\ \citenamefont
  {Gorbar}}]{Sukhachov2023}%
  \BibitemOpen
  \bibfield  {author} {\bibinfo {author} {\bibfnamefont {P.~O.}\ \bibnamefont
  {Sukhachov}}, \bibinfo {author} {\bibfnamefont {D.~O.}\ \bibnamefont
  {Oriekhov}},\ and\ \bibinfo {author} {\bibfnamefont {E.~V.}\ \bibnamefont
  {Gorbar}},\ }\bibfield  {title} {\bibinfo {title} {Stackings and effective
  models of bilayer dice lattices},\ }\href
  {https://doi.org/10.1103/PhysRevB.108.075166} {\bibfield  {journal} {\bibinfo
   {journal} {Phys. Rev. B}\ }\textbf {\bibinfo {volume} {108}},\ \bibinfo
  {pages} {075166} (\bibinfo {year} {2023})}\BibitemShut {NoStop}%
\bibitem [{\citenamefont {Kerelsky}\ \emph {et~al.}(2019)\citenamefont
  {Kerelsky}, \citenamefont {McGilly}, \citenamefont {Kennes}, \citenamefont
  {Xian}, \citenamefont {Yankowitz}, \citenamefont {Chen}, \citenamefont
  {Watanabe}, \citenamefont {Taniguchi}, \citenamefont {Hone}, \citenamefont
  {Dean}, \citenamefont {Rubio},\ and\ \citenamefont
  {Pasupathy}}]{Kerelsky2019}%
  \BibitemOpen
  \bibfield  {author} {\bibinfo {author} {\bibfnamefont {A.}~\bibnamefont
  {Kerelsky}}, \bibinfo {author} {\bibfnamefont {L.~J.}\ \bibnamefont
  {McGilly}}, \bibinfo {author} {\bibfnamefont {D.~M.}\ \bibnamefont {Kennes}},
  \bibinfo {author} {\bibfnamefont {L.}~\bibnamefont {Xian}}, \bibinfo {author}
  {\bibfnamefont {M.}~\bibnamefont {Yankowitz}}, \bibinfo {author}
  {\bibfnamefont {S.}~\bibnamefont {Chen}}, \bibinfo {author} {\bibfnamefont
  {K.}~\bibnamefont {Watanabe}}, \bibinfo {author} {\bibfnamefont
  {T.}~\bibnamefont {Taniguchi}}, \bibinfo {author} {\bibfnamefont
  {J.}~\bibnamefont {Hone}}, \bibinfo {author} {\bibfnamefont {C.}~\bibnamefont
  {Dean}}, \bibinfo {author} {\bibfnamefont {A.}~\bibnamefont {Rubio}},\ and\
  \bibinfo {author} {\bibfnamefont {A.~N.}\ \bibnamefont {Pasupathy}},\
  }\bibfield  {title} {\bibinfo {title} {Maximized electron interactions at the
  magic angle in twisted bilayer graphene},\ }\href
  {https://doi.org/10.1038/s41586-019-1431-9} {\bibfield  {journal} {\bibinfo
  {journal} {Nature}\ }\textbf {\bibinfo {volume} {572}},\ \bibinfo {pages}
  {95} (\bibinfo {year} {2019})}\BibitemShut {NoStop}%
\bibitem [{\citenamefont {Xie}\ \emph {et~al.}(2019)\citenamefont {Xie},
  \citenamefont {Lian}, \citenamefont {J{\"a}ck}, \citenamefont {Liu},
  \citenamefont {Chiu}, \citenamefont {Watanabe}, \citenamefont {Taniguchi},
  \citenamefont {Bernevig},\ and\ \citenamefont {Yazdani}}]{Xie2019}%
  \BibitemOpen
  \bibfield  {author} {\bibinfo {author} {\bibfnamefont {Y.}~\bibnamefont
  {Xie}}, \bibinfo {author} {\bibfnamefont {B.}~\bibnamefont {Lian}}, \bibinfo
  {author} {\bibfnamefont {B.}~\bibnamefont {J{\"a}ck}}, \bibinfo {author}
  {\bibfnamefont {X.}~\bibnamefont {Liu}}, \bibinfo {author} {\bibfnamefont
  {C.-L.}\ \bibnamefont {Chiu}}, \bibinfo {author} {\bibfnamefont
  {K.}~\bibnamefont {Watanabe}}, \bibinfo {author} {\bibfnamefont
  {T.}~\bibnamefont {Taniguchi}}, \bibinfo {author} {\bibfnamefont {B.~A.}\
  \bibnamefont {Bernevig}},\ and\ \bibinfo {author} {\bibfnamefont
  {A.}~\bibnamefont {Yazdani}},\ }\bibfield  {title} {\bibinfo {title}
  {Spectroscopic signatures of many-body correlations in magic-angle twisted
  bilayer graphene},\ }\href {https://doi.org/10.1038/s41586-019-1422-x}
  {\bibfield  {journal} {\bibinfo  {journal} {Nature}\ }\textbf {\bibinfo
  {volume} {572}},\ \bibinfo {pages} {101} (\bibinfo {year}
  {2019})}\BibitemShut {NoStop}%
\bibitem [{\citenamefont {Choi}\ \emph {et~al.}(2019)\citenamefont {Choi},
  \citenamefont {Kemmer}, \citenamefont {Peng}, \citenamefont {Thomson},
  \citenamefont {Arora}, \citenamefont {Polski}, \citenamefont {Zhang},
  \citenamefont {Ren}, \citenamefont {Alicea}, \citenamefont {Refael},
  \citenamefont {von Oppen}, \citenamefont {Watanabe}, \citenamefont
  {Taniguchi},\ and\ \citenamefont {Nadj-Perge}}]{Choi2019}%
  \BibitemOpen
  \bibfield  {author} {\bibinfo {author} {\bibfnamefont {Y.}~\bibnamefont
  {Choi}}, \bibinfo {author} {\bibfnamefont {J.}~\bibnamefont {Kemmer}},
  \bibinfo {author} {\bibfnamefont {Y.}~\bibnamefont {Peng}}, \bibinfo {author}
  {\bibfnamefont {A.}~\bibnamefont {Thomson}}, \bibinfo {author} {\bibfnamefont
  {H.}~\bibnamefont {Arora}}, \bibinfo {author} {\bibfnamefont
  {R.}~\bibnamefont {Polski}}, \bibinfo {author} {\bibfnamefont
  {Y.}~\bibnamefont {Zhang}}, \bibinfo {author} {\bibfnamefont
  {H.}~\bibnamefont {Ren}}, \bibinfo {author} {\bibfnamefont {J.}~\bibnamefont
  {Alicea}}, \bibinfo {author} {\bibfnamefont {G.}~\bibnamefont {Refael}},
  \bibinfo {author} {\bibfnamefont {F.}~\bibnamefont {von Oppen}}, \bibinfo
  {author} {\bibfnamefont {K.}~\bibnamefont {Watanabe}}, \bibinfo {author}
  {\bibfnamefont {T.}~\bibnamefont {Taniguchi}},\ and\ \bibinfo {author}
  {\bibfnamefont {S.}~\bibnamefont {Nadj-Perge}},\ }\bibfield  {title}
  {\bibinfo {title} {Electronic correlations in twisted bilayer graphene near
  the magic angle},\ }\href {https://doi.org/10.1038/s41567-019-0606-5}
  {\bibfield  {journal} {\bibinfo  {journal} {Nature Physics}\ }\textbf
  {\bibinfo {volume} {15}},\ \bibinfo {pages} {1174} (\bibinfo {year}
  {2019})}\BibitemShut {NoStop}%
\bibitem [{\citenamefont {Huang}\ \emph {et~al.}(2022)\citenamefont {Huang},
  \citenamefont {Wu}, \citenamefont {Hu}, \citenamefont {Cai}, \citenamefont
  {Li}, \citenamefont {An}, \citenamefont {Feng}, \citenamefont {Ye},
  \citenamefont {Lin}, \citenamefont {Law},\ and\ \citenamefont
  {Wang}}]{Huang2022}%
  \BibitemOpen
  \bibfield  {author} {\bibinfo {author} {\bibfnamefont {M.}~\bibnamefont
  {Huang}}, \bibinfo {author} {\bibfnamefont {Z.}~\bibnamefont {Wu}}, \bibinfo
  {author} {\bibfnamefont {J.}~\bibnamefont {Hu}}, \bibinfo {author}
  {\bibfnamefont {X.}~\bibnamefont {Cai}}, \bibinfo {author} {\bibfnamefont
  {E.}~\bibnamefont {Li}}, \bibinfo {author} {\bibfnamefont {L.}~\bibnamefont
  {An}}, \bibinfo {author} {\bibfnamefont {X.}~\bibnamefont {Feng}}, \bibinfo
  {author} {\bibfnamefont {Z.}~\bibnamefont {Ye}}, \bibinfo {author}
  {\bibfnamefont {N.}~\bibnamefont {Lin}}, \bibinfo {author} {\bibfnamefont
  {K.~T.}\ \bibnamefont {Law}},\ and\ \bibinfo {author} {\bibfnamefont
  {N.}~\bibnamefont {Wang}},\ }\bibfield  {title} {\bibinfo {title} {Giant
  nonlinear hall effect in twisted bilayer wse$_2$},\ }\href
  {https://doi.org/10.1093/nsr/nwac232} {\bibfield  {journal} {\bibinfo
  {journal} {Natl. Sci. Rev.}\ }\textbf {\bibinfo {volume} {10}},\ \bibinfo
  {pages} {nwac232} (\bibinfo {year} {2022})}\BibitemShut {NoStop}%
\bibitem [{\citenamefont {Pantale\'on}\ \emph {et~al.}(2022)\citenamefont
  {Pantale\'on}, \citenamefont {Phong}, \citenamefont {Naumis},\ and\
  \citenamefont {Guinea}}]{Pantaleon2022}%
  \BibitemOpen
  \bibfield  {author} {\bibinfo {author} {\bibfnamefont {P.~A.}\ \bibnamefont
  {Pantale\'on}}, \bibinfo {author} {\bibfnamefont {V.~o.~T.}\ \bibnamefont
  {Phong}}, \bibinfo {author} {\bibfnamefont {G.~G.}\ \bibnamefont {Naumis}},\
  and\ \bibinfo {author} {\bibfnamefont {F.}~\bibnamefont {Guinea}},\
  }\bibfield  {title} {\bibinfo {title} {Interaction-enhanced topological hall
  effects in strained twisted bilayer graphene},\ }\href
  {https://doi.org/10.1103/PhysRevB.106.L161101} {\bibfield  {journal}
  {\bibinfo  {journal} {Phys. Rev. B}\ }\textbf {\bibinfo {volume} {106}},\
  \bibinfo {pages} {L161101} (\bibinfo {year} {2022})}\BibitemShut {NoStop}%
\bibitem [{\citenamefont {Ouyang}\ \emph {et~al.}(2025)\citenamefont {Ouyang},
  \citenamefont {Yu}, \citenamefont {Li}, \citenamefont {Jia}, \citenamefont
  {Wang}, \citenamefont {Xiao}, \citenamefont {Zhang}, \citenamefont {Hu},
  \citenamefont {Pantale{\'o}n}, \citenamefont {Zhan}, \citenamefont {Zhou},
  \citenamefont {Guinea}, \citenamefont {Xue},\ and\ \citenamefont
  {Li}}]{Ouyang2025}%
  \BibitemOpen
  \bibfield  {author} {\bibinfo {author} {\bibfnamefont {P.}~\bibnamefont
  {Ouyang}}, \bibinfo {author} {\bibfnamefont {J.}~\bibnamefont {Yu}}, \bibinfo
  {author} {\bibfnamefont {Q.}~\bibnamefont {Li}}, \bibinfo {author}
  {\bibfnamefont {G.}~\bibnamefont {Jia}}, \bibinfo {author} {\bibfnamefont
  {Y.}~\bibnamefont {Wang}}, \bibinfo {author} {\bibfnamefont {K.}~\bibnamefont
  {Xiao}}, \bibinfo {author} {\bibfnamefont {H.}~\bibnamefont {Zhang}},
  \bibinfo {author} {\bibfnamefont {Z.}~\bibnamefont {Hu}}, \bibinfo {author}
  {\bibfnamefont {P.~A.}\ \bibnamefont {Pantale{\'o}n}}, \bibinfo {author}
  {\bibfnamefont {Z.}~\bibnamefont {Zhan}}, \bibinfo {author} {\bibfnamefont
  {S.}~\bibnamefont {Zhou}}, \bibinfo {author} {\bibfnamefont {F.}~\bibnamefont
  {Guinea}}, \bibinfo {author} {\bibfnamefont {Q.~K.}\ \bibnamefont {Xue}},\
  and\ \bibinfo {author} {\bibfnamefont {W.}~\bibnamefont {Li}},\ }\bibfield
  {title} {\bibinfo {title} {Structural and electronic signatures of
  strain-tunable marginally twisted bilayer graphene},\ }\href
  {https://doi.org/10.1093/nsr/nwaf568} {\bibfield  {journal} {\bibinfo
  {journal} {Natl. Sci. Rev.}\ }\textbf {\bibinfo {volume} {13}},\ \bibinfo
  {pages} {nwaf568} (\bibinfo {year} {2025})}\BibitemShut {NoStop}%
\bibitem [{\citenamefont {Hou}\ \emph {et~al.}(2025)\citenamefont {Hou},
  \citenamefont {Zhou}, \citenamefont {Xue}, \citenamefont {Yu}, \citenamefont
  {Han}, \citenamefont {Zhang},\ and\ \citenamefont {Lu}}]{Hou2025}%
  \BibitemOpen
  \bibfield  {author} {\bibinfo {author} {\bibfnamefont {Y.}~\bibnamefont
  {Hou}}, \bibinfo {author} {\bibfnamefont {J.}~\bibnamefont {Zhou}}, \bibinfo
  {author} {\bibfnamefont {M.}~\bibnamefont {Xue}}, \bibinfo {author}
  {\bibfnamefont {M.}~\bibnamefont {Yu}}, \bibinfo {author} {\bibfnamefont
  {Y.}~\bibnamefont {Han}}, \bibinfo {author} {\bibfnamefont {Z.}~\bibnamefont
  {Zhang}},\ and\ \bibinfo {author} {\bibfnamefont {Y.}~\bibnamefont {Lu}},\
  }\bibfield  {title} {\bibinfo {title} {Strain engineering of twisted bilayer
  graphene: The rise of strain-twistronics},\ }\href
  {https://doi.org/10.1002/smll.202311185} {\bibfield  {journal} {\bibinfo
  {journal} {Small}\ }\textbf {\bibinfo {volume} {21}},\ \bibinfo {pages}
  {2311185} (\bibinfo {year} {2025})}\BibitemShut {NoStop}%
\bibitem [{\citenamefont {Bi}\ \emph {et~al.}(2019)\citenamefont {Bi},
  \citenamefont {Yuan},\ and\ \citenamefont {Fu}}]{Bi2019}%
  \BibitemOpen
  \bibfield  {author} {\bibinfo {author} {\bibfnamefont {Z.}~\bibnamefont
  {Bi}}, \bibinfo {author} {\bibfnamefont {N.~F.~Q.}\ \bibnamefont {Yuan}},\
  and\ \bibinfo {author} {\bibfnamefont {L.}~\bibnamefont {Fu}},\ }\bibfield
  {title} {\bibinfo {title} {Designing flat bands by strain},\ }\href
  {https://doi.org/10.1103/PhysRevB.100.035448} {\bibfield  {journal} {\bibinfo
   {journal} {Phys. Rev. B}\ }\textbf {\bibinfo {volume} {100}},\ \bibinfo
  {pages} {035448} (\bibinfo {year} {2019})}\BibitemShut {NoStop}%
\bibitem [{\citenamefont {Pereira}\ \emph {et~al.}(2009)\citenamefont
  {Pereira}, \citenamefont {Castro~Neto},\ and\ \citenamefont
  {Peres}}]{Pereira2009}%
  \BibitemOpen
  \bibfield  {author} {\bibinfo {author} {\bibfnamefont {V.~M.}\ \bibnamefont
  {Pereira}}, \bibinfo {author} {\bibfnamefont {A.~H.}\ \bibnamefont
  {Castro~Neto}},\ and\ \bibinfo {author} {\bibfnamefont {N.~M.~R.}\
  \bibnamefont {Peres}},\ }\bibfield  {title} {\bibinfo {title} {Tight-binding
  approach to uniaxial strain in graphene},\ }\href
  {https://doi.org/10.1103/PhysRevB.80.045401} {\bibfield  {journal} {\bibinfo
  {journal} {Phys. Rev. B}\ }\textbf {\bibinfo {volume} {80}},\ \bibinfo
  {pages} {045401} (\bibinfo {year} {2009})}\BibitemShut {NoStop}%
\bibitem [{\citenamefont {Sun}\ \emph {et~al.}(2022)\citenamefont {Sun},
  \citenamefont {Liu}, \citenamefont {Du},\ and\ \citenamefont
  {Guo}}]{Sun2022}%
  \BibitemOpen
  \bibfield  {author} {\bibinfo {author} {\bibfnamefont {J.}~\bibnamefont
  {Sun}}, \bibinfo {author} {\bibfnamefont {T.}~\bibnamefont {Liu}}, \bibinfo
  {author} {\bibfnamefont {Y.}~\bibnamefont {Du}},\ and\ \bibinfo {author}
  {\bibfnamefont {H.}~\bibnamefont {Guo}},\ }\bibfield  {title} {\bibinfo
  {title} {Strain-induced pseudo magnetic field in the
  $\ensuremath{\alpha}\ensuremath{-}{T}_{3}$ lattice},\ }\href
  {https://doi.org/10.1103/PhysRevB.106.155417} {\bibfield  {journal} {\bibinfo
   {journal} {Phys. Rev. B}\ }\textbf {\bibinfo {volume} {106}},\ \bibinfo
  {pages} {155417} (\bibinfo {year} {2022})}\BibitemShut {NoStop}%
\bibitem [{\citenamefont {Ma}\ \emph {et~al.}(2019)\citenamefont {Ma},
  \citenamefont {Xu}, \citenamefont {Shen}, \citenamefont {MacNeill},
  \citenamefont {Fatemi}, \citenamefont {Chang}, \citenamefont {Mier~Valdivia},
  \citenamefont {Wu}, \citenamefont {Du}, \citenamefont {Hsu}, \citenamefont
  {Fang}, \citenamefont {Gibson}, \citenamefont {Watanabe}, \citenamefont
  {Taniguchi}, \citenamefont {Cava}, \citenamefont {Kaxiras}, \citenamefont
  {Lu}, \citenamefont {Lin}, \citenamefont {Fu}, \citenamefont {Gedik},\ and\
  \citenamefont {Jarillo-Herrero}}]{Ma2019}%
  \BibitemOpen
  \bibfield  {author} {\bibinfo {author} {\bibfnamefont {Q.}~\bibnamefont
  {Ma}}, \bibinfo {author} {\bibfnamefont {S.~Y.}\ \bibnamefont {Xu}}, \bibinfo
  {author} {\bibfnamefont {H.}~\bibnamefont {Shen}}, \bibinfo {author}
  {\bibfnamefont {D.}~\bibnamefont {MacNeill}}, \bibinfo {author}
  {\bibfnamefont {V.}~\bibnamefont {Fatemi}}, \bibinfo {author} {\bibfnamefont
  {T.-R.}\ \bibnamefont {Chang}}, \bibinfo {author} {\bibfnamefont
  {A.}~\bibnamefont {Mier~Valdivia}}, \bibinfo {author} {\bibfnamefont
  {S.}~\bibnamefont {Wu}}, \bibinfo {author} {\bibfnamefont {Z.}~\bibnamefont
  {Du}}, \bibinfo {author} {\bibfnamefont {C.~H.}\ \bibnamefont {Hsu}},
  \bibinfo {author} {\bibfnamefont {S.}~\bibnamefont {Fang}}, \bibinfo {author}
  {\bibfnamefont {Q.~D.}\ \bibnamefont {Gibson}}, \bibinfo {author}
  {\bibfnamefont {K.}~\bibnamefont {Watanabe}}, \bibinfo {author}
  {\bibfnamefont {T.}~\bibnamefont {Taniguchi}}, \bibinfo {author}
  {\bibfnamefont {R.~J.}\ \bibnamefont {Cava}}, \bibinfo {author}
  {\bibfnamefont {E.}~\bibnamefont {Kaxiras}}, \bibinfo {author} {\bibfnamefont
  {H.~Z.}\ \bibnamefont {Lu}}, \bibinfo {author} {\bibfnamefont
  {H.}~\bibnamefont {Lin}}, \bibinfo {author} {\bibfnamefont {L.}~\bibnamefont
  {Fu}}, \bibinfo {author} {\bibfnamefont {N.}~\bibnamefont {Gedik}},\ and\
  \bibinfo {author} {\bibfnamefont {P.}~\bibnamefont {Jarillo-Herrero}},\
  }\bibfield  {title} {\bibinfo {title} {Observation of the nonlinear hall
  effect under time-reversal-symmetric conditions},\ }\href
  {https://doi.org/10.1038/s41586-018-0807-6} {\bibfield  {journal} {\bibinfo
  {journal} {Nature}\ }\textbf {\bibinfo {volume} {565}},\ \bibinfo {pages}
  {337} (\bibinfo {year} {2019})}\BibitemShut {NoStop}%
\bibitem [{\citenamefont {Kang}\ \emph {et~al.}(2019)\citenamefont {Kang},
  \citenamefont {Li}, \citenamefont {Sohn}, \citenamefont {Shan},\ and\
  \citenamefont {Mak}}]{Kang2019}%
  \BibitemOpen
  \bibfield  {author} {\bibinfo {author} {\bibfnamefont {K.}~\bibnamefont
  {Kang}}, \bibinfo {author} {\bibfnamefont {T.}~\bibnamefont {Li}}, \bibinfo
  {author} {\bibfnamefont {E.}~\bibnamefont {Sohn}}, \bibinfo {author}
  {\bibfnamefont {J.}~\bibnamefont {Shan}},\ and\ \bibinfo {author}
  {\bibfnamefont {K.~F.}\ \bibnamefont {Mak}},\ }\bibfield  {title} {\bibinfo
  {title} {Nonlinear anomalous hall effect in few-layer \text{WTe}$_2$},\
  }\href {https://doi.org/10.1038/s41563-019-0294-7} {\bibfield  {journal}
  {\bibinfo  {journal} {Nature Materials}\ }\textbf {\bibinfo {volume} {18}},\
  \bibinfo {pages} {324} (\bibinfo {year} {2019})}\BibitemShut {NoStop}%
\bibitem [{\citenamefont {Okamoto}\ and\ \citenamefont
  {Xiao}(2018)}]{Okamoto2018}%
  \BibitemOpen
  \bibfield  {author} {\bibinfo {author} {\bibfnamefont {S.}~\bibnamefont
  {Okamoto}}\ and\ \bibinfo {author} {\bibfnamefont {D.}~\bibnamefont {Xiao}},\
  }\bibfield  {title} {\bibinfo {title} {Transition-metal oxide (111)
  bilayers},\ }\href {https://doi.org/10.7566/JPSJ.87.041006} {\bibfield
  {journal} {\bibinfo  {journal} {Journal of the Physical Society of Japan}\
  }\textbf {\bibinfo {volume} {87}},\ \bibinfo {pages} {041006} (\bibinfo
  {year} {2018})}\BibitemShut {NoStop}%
\bibitem [{\citenamefont {Tassi}\ and\ \citenamefont
  {Bercioux}(2024)}]{Tassi2024}%
  \BibitemOpen
  \bibfield  {author} {\bibinfo {author} {\bibfnamefont {C.}~\bibnamefont
  {Tassi}}\ and\ \bibinfo {author} {\bibfnamefont {D.}~\bibnamefont
  {Bercioux}},\ }\bibfield  {title} {\bibinfo {title} {Implementation and
  characterization of the dice lattice in the electron quantum simulator},\
  }\href {https://doi.org/https://doi.org/10.1002/apxr.202400038} {\bibfield
  {journal} {\bibinfo  {journal} {Advanced Physics Research}\ }\textbf
  {\bibinfo {volume} {3}},\ \bibinfo {pages} {2400038} (\bibinfo {year}
  {2024})}\BibitemShut {NoStop}%
\bibitem [{\citenamefont {Khajetoorians}\ \emph {et~al.}(2019)\citenamefont
  {Khajetoorians}, \citenamefont {Wegner}, \citenamefont {Otte}, \citenamefont
  {Lutz},\ and\ \citenamefont {Heinrich}}]{Khajetoorians2019}%
  \BibitemOpen
  \bibfield  {author} {\bibinfo {author} {\bibfnamefont {A.~A.}\ \bibnamefont
  {Khajetoorians}}, \bibinfo {author} {\bibfnamefont {D.}~\bibnamefont
  {Wegner}}, \bibinfo {author} {\bibfnamefont {A.~F.}\ \bibnamefont {Otte}},
  \bibinfo {author} {\bibfnamefont {C.~J.}\ \bibnamefont {Lutz}},\ and\
  \bibinfo {author} {\bibfnamefont {A.~J.}\ \bibnamefont {Heinrich}},\
  }\bibfield  {title} {\bibinfo {title} {Creating designer quantum states of
  matter atom-by-atom},\ }\href {https://doi.org/10.1038/s42254-019-0108-5}
  {\bibfield  {journal} {\bibinfo  {journal} {Nature Reviews Physics}\ }\textbf
  {\bibinfo {volume} {1}},\ \bibinfo {pages} {703} (\bibinfo {year}
  {2019})}\BibitemShut {NoStop}%
\bibitem [{\citenamefont {Doennig}\ \emph {et~al.}(2013)\citenamefont
  {Doennig}, \citenamefont {Pickett},\ and\ \citenamefont
  {Pentcheva}}]{Doenning2013}%
  \BibitemOpen
  \bibfield  {author} {\bibinfo {author} {\bibfnamefont {D.}~\bibnamefont
  {Doennig}}, \bibinfo {author} {\bibfnamefont {W.~E.}\ \bibnamefont
  {Pickett}},\ and\ \bibinfo {author} {\bibfnamefont {R.}~\bibnamefont
  {Pentcheva}},\ }\bibfield  {title} {\bibinfo {title} {Massive symmetry
  breaking in ${\mathrm{laalo}}_{3}/{\mathrm{srtio}}_{3}(111)$ quantum wells: A
  three-orbital strongly correlated generalization of graphene},\ }\href
  {https://doi.org/10.1103/PhysRevLett.111.126804} {\bibfield  {journal}
  {\bibinfo  {journal} {Phys. Rev. Lett.}\ }\textbf {\bibinfo {volume} {111}},\
  \bibinfo {pages} {126804} (\bibinfo {year} {2013})}\BibitemShut {NoStop}%
\bibitem [{\citenamefont {Soni}\ \emph {et~al.}(2020)\citenamefont {Soni},
  \citenamefont {Kaushal}, \citenamefont {Okamoto},\ and\ \citenamefont
  {Dagotto}}]{Soni2020}%
  \BibitemOpen
  \bibfield  {author} {\bibinfo {author} {\bibfnamefont {R.}~\bibnamefont
  {Soni}}, \bibinfo {author} {\bibfnamefont {N.}~\bibnamefont {Kaushal}},
  \bibinfo {author} {\bibfnamefont {S.}~\bibnamefont {Okamoto}},\ and\ \bibinfo
  {author} {\bibfnamefont {E.}~\bibnamefont {Dagotto}},\ }\bibfield  {title}
  {\bibinfo {title} {Flat bands and ferrimagnetic order in electronically
  correlated dice-lattice ribbons},\ }\href
  {https://doi.org/10.1103/PhysRevB.102.045105} {\bibfield  {journal} {\bibinfo
   {journal} {Phys. Rev. B}\ }\textbf {\bibinfo {volume} {102}},\ \bibinfo
  {pages} {045105} (\bibinfo {year} {2020})}\BibitemShut {NoStop}%
\bibitem [{\citenamefont {Geng}\ \emph {et~al.}(2026)\citenamefont {Geng},
  \citenamefont {Wang}, \citenamefont {Guo},\ and\ \citenamefont
  {et~al.}}]{Geng2026}%
  \BibitemOpen
  \bibfield  {author} {\bibinfo {author} {\bibfnamefont {S.}~\bibnamefont
  {Geng}}, \bibinfo {author} {\bibfnamefont {X.}~\bibnamefont {Wang}}, \bibinfo
  {author} {\bibfnamefont {R.}~\bibnamefont {Guo}},\ and\ \bibinfo {author}
  {\bibnamefont {et~al.}},\ }\bibfield  {title} {\bibinfo {title} {Experimental
  realization of dice-lattice flat band at the fermi level in layered electride
  ycl},\ }\href {https://doi.org/10.1038/s41467-026-69049-0} {\bibfield
  {journal} {\bibinfo  {journal} {Nature Communications}\ }\textbf {\bibinfo
  {volume} {17}},\ \bibinfo {pages} {2213} (\bibinfo {year}
  {2026})}\BibitemShut {NoStop}%
\bibitem [{\citenamefont {Qiao}\ \emph {et~al.}(2025)\citenamefont {Qiao},
  \citenamefont {Han}, \citenamefont {Jiao}, \citenamefont {Zheng},
  \citenamefont {Lu},\ and\ \citenamefont {Zhang}}]{Qiao2025}%
  \BibitemOpen
  \bibfield  {author} {\bibinfo {author} {\bibfnamefont {S.~X.}\ \bibnamefont
  {Qiao}}, \bibinfo {author} {\bibfnamefont {Y.~L.}\ \bibnamefont {Han}},
  \bibinfo {author} {\bibfnamefont {N.}~\bibnamefont {Jiao}}, \bibinfo {author}
  {\bibfnamefont {M.-M.}\ \bibnamefont {Zheng}}, \bibinfo {author}
  {\bibfnamefont {H.-Y.}\ \bibnamefont {Lu}},\ and\ \bibinfo {author}
  {\bibfnamefont {P.}~\bibnamefont {Zhang}},\ }\bibfield  {title} {\bibinfo
  {title} {Msene: A large family of two-dimensional transition metal sulfides
  with mxene structure},\ }\href {https://doi.org/10.1103/PhysRevB.111.L041404}
  {\bibfield  {journal} {\bibinfo  {journal} {Phys. Rev. B}\ }\textbf {\bibinfo
  {volume} {111}},\ \bibinfo {pages} {L041404} (\bibinfo {year}
  {2025})}\BibitemShut {NoStop}%
\bibitem [{\citenamefont {Papadopoulou}\ \emph {et~al.}(2022)\citenamefont
  {Papadopoulou}, \citenamefont {Chroneos},\ and\ \citenamefont
  {Christopoulos}}]{Papadopoulou2022}%
  \BibitemOpen
  \bibfield  {author} {\bibinfo {author} {\bibfnamefont {K.~A.}\ \bibnamefont
  {Papadopoulou}}, \bibinfo {author} {\bibfnamefont {A.}~\bibnamefont
  {Chroneos}},\ and\ \bibinfo {author} {\bibfnamefont {S.-R.~G.}\ \bibnamefont
  {Christopoulos}},\ }\bibfield  {title} {\bibinfo {title} {Ion incorporation
  on the zr$_2$cs$_2$ mxene monolayer towards better-performing rechargeable
  ion batteries},\ }\href {https://doi.org/10.1016/j.jallcom.2022.166240}
  {\bibfield  {journal} {\bibinfo  {journal} {Journal of Alloys and Compounds}\
  }\textbf {\bibinfo {volume} {922}},\ \bibinfo {pages} {166240} (\bibinfo
  {year} {2022})}\BibitemShut {NoStop}%
\bibitem [{\citenamefont {Cheng}\ \emph {et~al.}(2020)\citenamefont {Cheng},
  \citenamefont {Yin}, \citenamefont {Lu}, \citenamefont {He}, \citenamefont
  {Wang},\ and\ \citenamefont {Xianlong}}]{Cheng2020}%
  \BibitemOpen
  \bibfield  {author} {\bibinfo {author} {\bibfnamefont {S.}~\bibnamefont
  {Cheng}}, \bibinfo {author} {\bibfnamefont {H.}~\bibnamefont {Yin}}, \bibinfo
  {author} {\bibfnamefont {Z.}~\bibnamefont {Lu}}, \bibinfo {author}
  {\bibfnamefont {C.}~\bibnamefont {He}}, \bibinfo {author} {\bibfnamefont
  {P.}~\bibnamefont {Wang}},\ and\ \bibinfo {author} {\bibfnamefont
  {G.}~\bibnamefont {Xianlong}},\ }\bibfield  {title} {\bibinfo {title}
  {Predicting large-chern-number phases in a shaken optical dice lattice},\
  }\href {https://doi.org/10.1103/PhysRevA.101.043620} {\bibfield  {journal}
  {\bibinfo  {journal} {Phys. Rev. A}\ }\textbf {\bibinfo {volume} {101}},\
  \bibinfo {pages} {043620} (\bibinfo {year} {2020})}\BibitemShut {NoStop}%
\bibitem [{\citenamefont {Andrijauskas}\ \emph {et~al.}(2015)\citenamefont
  {Andrijauskas}, \citenamefont {Anisimovas}, \citenamefont {Ra\ifmmode
  \check{c}\else \v{c}\fi{}i\ifmmode~\bar{u}\else \={u}\fi{}nas}, \citenamefont
  {Mekys}, \citenamefont {Kudria\ifmmode~\check{s}\else \v{s}\fi{}ov},
  \citenamefont {Spielman},\ and\ \citenamefont {Juzeli\ifmmode~\bar{u}\else
  \={u}\fi{}nas}}]{Andrijauskas2015}%
  \BibitemOpen
  \bibfield  {author} {\bibinfo {author} {\bibfnamefont {T.}~\bibnamefont
  {Andrijauskas}}, \bibinfo {author} {\bibfnamefont {E.}~\bibnamefont
  {Anisimovas}}, \bibinfo {author} {\bibfnamefont {M.}~\bibnamefont {Ra\ifmmode
  \check{c}\else \v{c}\fi{}i\ifmmode~\bar{u}\else \={u}\fi{}nas}}, \bibinfo
  {author} {\bibfnamefont {A.}~\bibnamefont {Mekys}}, \bibinfo {author}
  {\bibfnamefont {V.}~\bibnamefont {Kudria\ifmmode~\check{s}\else
  \v{s}\fi{}ov}}, \bibinfo {author} {\bibfnamefont {I.~B.}\ \bibnamefont
  {Spielman}},\ and\ \bibinfo {author} {\bibfnamefont {G.}~\bibnamefont
  {Juzeli\ifmmode~\bar{u}\else \={u}\fi{}nas}},\ }\bibfield  {title} {\bibinfo
  {title} {Three-level haldane-like model on a dice optical lattice},\ }\href
  {https://doi.org/10.1103/PhysRevA.92.033617} {\bibfield  {journal} {\bibinfo
  {journal} {Phys. Rev. A}\ }\textbf {\bibinfo {volume} {92}},\ \bibinfo
  {pages} {033617} (\bibinfo {year} {2015})}\BibitemShut {NoStop}%
\bibitem [{\citenamefont {M\"oller}\ and\ \citenamefont
  {Cooper}(2012)}]{Moller2012}%
  \BibitemOpen
  \bibfield  {author} {\bibinfo {author} {\bibfnamefont {G.}~\bibnamefont
  {M\"oller}}\ and\ \bibinfo {author} {\bibfnamefont {N.~R.}\ \bibnamefont
  {Cooper}},\ }\bibfield  {title} {\bibinfo {title} {Correlated phases of
  bosons in the flat lowest band of the dice lattice},\ }\href
  {https://doi.org/10.1103/PhysRevLett.108.045306} {\bibfield  {journal}
  {\bibinfo  {journal} {Phys. Rev. Lett.}\ }\textbf {\bibinfo {volume} {108}},\
  \bibinfo {pages} {045306} (\bibinfo {year} {2012})}\BibitemShut {NoStop}%
\bibitem [{\citenamefont {Rizzi}\ \emph {et~al.}(2006)\citenamefont {Rizzi},
  \citenamefont {Cataudella},\ and\ \citenamefont {Fazio}}]{Rizzi2006}%
  \BibitemOpen
  \bibfield  {author} {\bibinfo {author} {\bibfnamefont {M.}~\bibnamefont
  {Rizzi}}, \bibinfo {author} {\bibfnamefont {V.}~\bibnamefont {Cataudella}},\
  and\ \bibinfo {author} {\bibfnamefont {R.}~\bibnamefont {Fazio}},\ }\bibfield
   {title} {\bibinfo {title} {Phase diagram of the bose-hubbard model with
  \text{T}$_{3}$ symmetry},\ }\href
  {https://doi.org/10.1103/PhysRevB.73.144511} {\bibfield  {journal} {\bibinfo
  {journal} {Phys. Rev. B}\ }\textbf {\bibinfo {volume} {73}},\ \bibinfo
  {pages} {144511} (\bibinfo {year} {2006})}\BibitemShut {NoStop}%
\bibitem [{\citenamefont {Bercioux}\ \emph {et~al.}(2009)\citenamefont
  {Bercioux}, \citenamefont {Urban}, \citenamefont {Grabert},\ and\
  \citenamefont {H\"ausler}}]{Bercioux2009}%
  \BibitemOpen
  \bibfield  {author} {\bibinfo {author} {\bibfnamefont {D.}~\bibnamefont
  {Bercioux}}, \bibinfo {author} {\bibfnamefont {D.~F.}\ \bibnamefont {Urban}},
  \bibinfo {author} {\bibfnamefont {H.}~\bibnamefont {Grabert}},\ and\ \bibinfo
  {author} {\bibfnamefont {W.}~\bibnamefont {H\"ausler}},\ }\bibfield  {title}
  {\bibinfo {title} {Massless dirac-weyl fermions in a \text{T}$_{3}$ optical
  lattice},\ }\href {https://doi.org/10.1103/PhysRevA.80.063603} {\bibfield
  {journal} {\bibinfo  {journal} {Phys. Rev. A}\ }\textbf {\bibinfo {volume}
  {80}},\ \bibinfo {pages} {063603} (\bibinfo {year} {2009})}\BibitemShut
  {NoStop}%
\bibitem [{\citenamefont {Battilomo}\ \emph {et~al.}(2019)\citenamefont
  {Battilomo}, \citenamefont {Scopigno},\ and\ \citenamefont
  {Ortix}}]{Battilomo2019}%
  \BibitemOpen
  \bibfield  {author} {\bibinfo {author} {\bibfnamefont {R.}~\bibnamefont
  {Battilomo}}, \bibinfo {author} {\bibfnamefont {N.}~\bibnamefont
  {Scopigno}},\ and\ \bibinfo {author} {\bibfnamefont {C.}~\bibnamefont
  {Ortix}},\ }\bibfield  {title} {\bibinfo {title} {Berry curvature dipole in
  strained graphene: A fermi surface warping effect},\ }\href
  {https://doi.org/10.1103/PhysRevLett.123.196403} {\bibfield  {journal}
  {\bibinfo  {journal} {Phys. Rev. Lett.}\ }\textbf {\bibinfo {volume} {123}},\
  \bibinfo {pages} {196403} (\bibinfo {year} {2019})}\BibitemShut {NoStop}%
\bibitem [{\citenamefont {Son}\ \emph {et~al.}(2019)\citenamefont {Son},
  \citenamefont {Kim}, \citenamefont {Ahn}, \citenamefont {Lee},\ and\
  \citenamefont {Lee}}]{Son2019}%
  \BibitemOpen
  \bibfield  {author} {\bibinfo {author} {\bibfnamefont {J.}~\bibnamefont
  {Son}}, \bibinfo {author} {\bibfnamefont {K.-H.}\ \bibnamefont {Kim}},
  \bibinfo {author} {\bibfnamefont {Y.~H.}\ \bibnamefont {Ahn}}, \bibinfo
  {author} {\bibfnamefont {H.-W.}\ \bibnamefont {Lee}},\ and\ \bibinfo {author}
  {\bibfnamefont {J.}~\bibnamefont {Lee}},\ }\bibfield  {title} {\bibinfo
  {title} {Strain engineering of the berry curvature dipole and valley
  magnetization in monolayer ${\mathrm{mos}}_{2}$},\ }\href
  {https://doi.org/10.1103/PhysRevLett.123.036806} {\bibfield  {journal}
  {\bibinfo  {journal} {Phys. Rev. Lett.}\ }\textbf {\bibinfo {volume} {123}},\
  \bibinfo {pages} {036806} (\bibinfo {year} {2019})}\BibitemShut {NoStop}%
\bibitem [{\citenamefont {Wang}\ and\ \citenamefont {Qian}(2019)}]{Wang2019}%
  \BibitemOpen
  \bibfield  {author} {\bibinfo {author} {\bibfnamefont {H.}~\bibnamefont
  {Wang}}\ and\ \bibinfo {author} {\bibfnamefont {X.}~\bibnamefont {Qian}},\
  }\bibfield  {title} {\bibinfo {title} {Ferroelectric nonlinear anomalous hall
  effect in few-layer wte$_2$},\ }\href
  {https://doi.org/10.1038/s41524-019-0257-1} {\bibfield  {journal} {\bibinfo
  {journal} {npj Computational Materials}\ }\textbf {\bibinfo {volume} {5}},\
  \bibinfo {pages} {119} (\bibinfo {year} {2019})}\BibitemShut {NoStop}%
\bibitem [{\citenamefont {Sodemann}\ and\ \citenamefont
  {Fu}(2015)}]{Sodemann2015}%
  \BibitemOpen
  \bibfield  {author} {\bibinfo {author} {\bibfnamefont {I.}~\bibnamefont
  {Sodemann}}\ and\ \bibinfo {author} {\bibfnamefont {L.}~\bibnamefont {Fu}},\
  }\bibfield  {title} {\bibinfo {title} {Quantum nonlinear hall effect induced
  by berry curvature dipole in time-reversal invariant materials},\ }\href
  {https://doi.org/10.1103/PhysRevLett.115.216806} {\bibfield  {journal}
  {\bibinfo  {journal} {Phys. Rev. Lett.}\ }\textbf {\bibinfo {volume} {115}},\
  \bibinfo {pages} {216806} (\bibinfo {year} {2015})}\BibitemShut {NoStop}%
\bibitem [{\citenamefont {Thouless}(1997)}]{Thouless1997}%
  \BibitemOpen
  \bibfield  {author} {\bibinfo {author} {\bibfnamefont {D.~J.}\ \bibnamefont
  {Thouless}},\ }\bibfield  {title} {\bibinfo {title} {Topological quantum
  numbers in nonrelativistic physics},\ }\href
  {https://doi.org/10.1142/S0217979297001623} {\bibfield  {journal} {\bibinfo
  {journal} {International Journal of Modern Physics B}\ }\textbf {\bibinfo
  {volume} {11}},\ \bibinfo {pages} {3319} (\bibinfo {year}
  {1997})}\BibitemShut {NoStop}%
\bibitem [{\citenamefont {Debnath}\ and\ \citenamefont
  {Basu}(2025)}]{Debnath2025}%
  \BibitemOpen
  \bibfield  {author} {\bibinfo {author} {\bibfnamefont {S.}~\bibnamefont
  {Debnath}}\ and\ \bibinfo {author} {\bibfnamefont {S.}~\bibnamefont {Basu}},\
  }\bibfield  {title} {\bibinfo {title} {Magnons on a dice lattice: Topological
  features and transport properties},\ }\href
  {https://doi.org/10.1103/PhysRevB.111.155418} {\bibfield  {journal} {\bibinfo
   {journal} {Phys. Rev. B}\ }\textbf {\bibinfo {volume} {111}},\ \bibinfo
  {pages} {155418} (\bibinfo {year} {2025})}\BibitemShut {NoStop}%
\bibitem [{\citenamefont {Lahiri}\ and\ \citenamefont
  {Basu}(2024)}]{Lahiri2024}%
  \BibitemOpen
  \bibfield  {author} {\bibinfo {author} {\bibfnamefont {S.}~\bibnamefont
  {Lahiri}}\ and\ \bibinfo {author} {\bibfnamefont {S.}~\bibnamefont {Basu}},\
  }\bibfield  {title} {\bibinfo {title} {Second order topology in a band
  engineered chern insulator},\ }\href
  {https://doi.org/10.1038/s41598-024-52321-y} {\bibfield  {journal} {\bibinfo
  {journal} {Scientific Reports}\ }\textbf {\bibinfo {volume} {14}},\ \bibinfo
  {pages} {1880} (\bibinfo {year} {2024})}\BibitemShut {NoStop}%
\bibitem [{\citenamefont {Illes}(2017)}]{Illes2017}%
  \BibitemOpen
  \bibfield  {author} {\bibinfo {author} {\bibfnamefont {E.}~\bibnamefont
  {Illes}},\ }\emph {\bibinfo {title} {Properties of the \text{$\alpha$-$T_3$}
  Model}},\ \href {http://hdl.handle.net/10214/11512} {\bibinfo {type} {Phd
  thesis}},\ \bibinfo  {school} {University of Guelph} (\bibinfo {year}
  {2017})\BibitemShut {NoStop}%
\bibitem [{\citenamefont {Raoux}\ \emph {et~al.}(2014)\citenamefont {Raoux},
  \citenamefont {Morigi}, \citenamefont {Fuchs}, \citenamefont {Pi\'echon},\
  and\ \citenamefont {Montambaux}}]{Raoux2014}%
  \BibitemOpen
  \bibfield  {author} {\bibinfo {author} {\bibfnamefont {A.}~\bibnamefont
  {Raoux}}, \bibinfo {author} {\bibfnamefont {M.}~\bibnamefont {Morigi}},
  \bibinfo {author} {\bibfnamefont {J.-N.}\ \bibnamefont {Fuchs}}, \bibinfo
  {author} {\bibfnamefont {F.}~\bibnamefont {Pi\'echon}},\ and\ \bibinfo
  {author} {\bibfnamefont {G.}~\bibnamefont {Montambaux}},\ }\bibfield  {title}
  {\bibinfo {title} {From dia- to paramagnetic orbital susceptibility of
  massless fermions},\ }\href {https://doi.org/10.1103/PhysRevLett.112.026402}
  {\bibfield  {journal} {\bibinfo  {journal} {Phys. Rev. Lett.}\ }\textbf
  {\bibinfo {volume} {112}},\ \bibinfo {pages} {026402} (\bibinfo {year}
  {2014})}\BibitemShut {NoStop}%
\bibitem [{\citenamefont {Illes}\ \emph {et~al.}(2015)\citenamefont {Illes},
  \citenamefont {Carbotte},\ and\ \citenamefont {Nicol}}]{Illes2015}%
  \BibitemOpen
  \bibfield  {author} {\bibinfo {author} {\bibfnamefont {E.}~\bibnamefont
  {Illes}}, \bibinfo {author} {\bibfnamefont {J.~P.}\ \bibnamefont
  {Carbotte}},\ and\ \bibinfo {author} {\bibfnamefont {E.~J.}\ \bibnamefont
  {Nicol}},\ }\bibfield  {title} {\bibinfo {title} {Hall quantization and
  optical conductivity evolution with variable berry phase in the
  $\ensuremath{\alpha}\text{\ensuremath{-}}{T}_{3}$ model},\ }\href
  {https://doi.org/10.1103/PhysRevB.92.245410} {\bibfield  {journal} {\bibinfo
  {journal} {Phys. Rev. B}\ }\textbf {\bibinfo {volume} {92}},\ \bibinfo
  {pages} {245410} (\bibinfo {year} {2015})}\BibitemShut {NoStop}%
\bibitem [{\citenamefont {Islam}\ \emph {et~al.}(2024)\citenamefont {Islam},
  \citenamefont {Bhattacharyya},\ and\ \citenamefont {Basu}}]{Islam2024}%
  \BibitemOpen
  \bibfield  {author} {\bibinfo {author} {\bibfnamefont {M.}~\bibnamefont
  {Islam}}, \bibinfo {author} {\bibfnamefont {K.}~\bibnamefont
  {Bhattacharyya}},\ and\ \bibinfo {author} {\bibfnamefont {S.}~\bibnamefont
  {Basu}},\ }\bibfield  {title} {\bibinfo {title} {Electron-phonon coupling
  induced topological phase transition in an
  $\ensuremath{\alpha}\text{\ensuremath{-}}{T}_{3}$ haldane-holstein model},\
  }\href {https://doi.org/10.1103/PhysRevB.110.045426} {\bibfield  {journal}
  {\bibinfo  {journal} {Phys. Rev. B}\ }\textbf {\bibinfo {volume} {110}},\
  \bibinfo {pages} {045426} (\bibinfo {year} {2024})}\BibitemShut {NoStop}%
\bibitem [{\citenamefont {Islam}\ and\ \citenamefont {Basu}(2025)}]{Islam2025}%
  \BibitemOpen
  \bibfield  {author} {\bibinfo {author} {\bibfnamefont {M.}~\bibnamefont
  {Islam}}\ and\ \bibinfo {author} {\bibfnamefont {S.}~\bibnamefont {Basu}},\
  }\bibfield  {title} {\bibinfo {title} {Conductance properties of
  $\ensuremath{\alpha}\text{\ensuremath{-}}{T}_{3}$ corbino disks},\ }\href
  {https://doi.org/10.1088/1361-648X/adcdb3} {\bibfield  {journal} {\bibinfo
  {journal} {J. Phys.: Condens. Matter}\ }\textbf {\bibinfo {volume} {37}},\
  \bibinfo {pages} {205302} (\bibinfo {year} {2025})}\BibitemShut {NoStop}%
\bibitem [{\citenamefont {Koshino}(2015)}]{Koshino2015}%
  \BibitemOpen
  \bibfield  {author} {\bibinfo {author} {\bibfnamefont {M.}~\bibnamefont
  {Koshino}},\ }\bibfield  {title} {\bibinfo {title} {Interlayer interaction in
  general incommensurate atomic layers},\ }\href
  {https://doi.org/10.1088/1367-2630/17/1/015014} {\bibfield  {journal}
  {\bibinfo  {journal} {New Journal of Physics}\ }\textbf {\bibinfo {volume}
  {17}},\ \bibinfo {pages} {015014} (\bibinfo {year} {2015})}\BibitemShut
  {NoStop}%
\bibitem [{\citenamefont {Georgi}(1999)}]{Georgi1999}%
  \BibitemOpen
  \bibfield  {author} {\bibinfo {author} {\bibfnamefont {H.}~\bibnamefont
  {Georgi}},\ }\href@noop {} {\emph {\bibinfo {title} {Lie Algebras in Particle
  Physics: From Isospin to Unified Theories}}},\ \bibinfo {edition} {2nd}\
  ed.,\ \bibinfo {series} {Frontiers in Physics}, Vol.~\bibinfo {volume} {54}\
  (\bibinfo  {publisher} {Perseus Books},\ \bibinfo {address} {Reading, MA},\
  \bibinfo {year} {1999})\BibitemShut {NoStop}%
\end{thebibliography}%

\clearpage
\onecolumngrid
\widetext


\setcounter{equation}{0}
\setcounter{figure}{0}
\setcounter{table}{0}

\renewcommand{\theequation}{S\arabic{equation}}
\renewcommand{\thefigure}{S\arabic{figure}}
\renewcommand{\thetable}{S\arabic{table}}
\renewcommand{\bibnumfmt}[1]{[S#1]}
\renewcommand{\citenumfont}[1]{S#1}

\newcommand{\bk}{\boldsymbol\kappa}
\newcommand{\SI}{Supplementary Material}

\newcommand{\beginsupplement}{%
  \setcounter{equation}{0}
  \renewcommand{\theequation}{S\arabic{equation}}%
  \setcounter{table}{0}
  \renewcommand{\thetable}{S\arabic{table}}%
  \setcounter{figure}{0}
  \renewcommand{\thefigure}{S\arabic{figure}}%
  \setcounter{section}{0}
  \renewcommand{\thesection}{S\Roman{section}}%
  \setcounter{subsection}{0}
  \renewcommand{\thesubsection}{S\Roman{section}.\Alph{subsection}}%
}

\beginsupplement

\begin{center}
    {\large \textbf{Supplemental Material: Probing topological phase transitions via nonlinear Hall response in strained moiré dice lattice}}\\[6pt]
    Gourab Paul$^{1}$, Srijata Lahiri$^{1}$, Bilal Tanatar$^{2}$, and Saurabh Basu$^{1}$\\[6pt]
    {
    $^{1}$Department of Physics, Indian Institute of Technology Guwahati, Guwahati-781039, Assam, India\\
    $^{2}$Department of Physics, Bilkent University, 06800 Bilkent, Ankara, Türkiye
    }
\end{center}

\vspace{0.5cm}
\tableofcontents

\section{The tight-binding model of monolayer dice lattice}
\begin{figure}[h]
    \centering
    \includegraphics[width=0.25\linewidth]{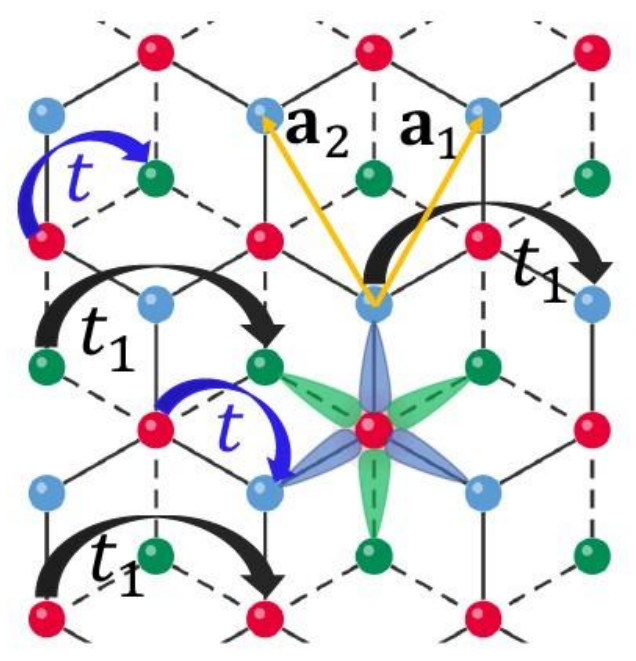}
    \caption{Schematic diagram of the dice lattice. Blue, red, and green dots represent the $A$, $B$, and $C$ sublattice atoms, respectively. The primitive lattice vectors are denoted by $\mathbf{a}_1$ and $\mathbf{a}_2$. The NN hopping amplitudes between $A\rightarrow B$ and $B\rightarrow C$ sites are labeled as $t$, while the NNN hopping amplitude among the $A\rightarrow A$, $B\rightarrow B$, and $C\rightarrow C$ sites is represented by $t_1$.}
    \label{fig:Figure-7}
\end{figure}
The Bravais lattice vectors of the dice lattice are given by $\mathbf{a}_1 = a \left( \frac{1}{2}, \frac{\sqrt{3}}{2} \right)$ and $\mathbf{a}_2 = a \left( -\frac{1}{2}, \frac{\sqrt{3}}{2} \right)$, where $a = \sqrt{3}d$ is the lattice constant and $d$ is the nearest-neighbor (NN) bond length. As in graphene, we set $d = 1.42$~\AA. The NN hopping amplitude between the $A\rightarrow B$ and $B\rightarrow C$ sublattices is denoted by $t$, while the hopping amplitude for next-nearest-neighbor (NNN) interactions between the $A\rightarrow A$, $B\rightarrow B$, and $C\rightarrow C$ sublattices is denoted by $t_1$. The schematic representation of the dice lattice is shown in Fig.~(\ref{fig:Figure-7}). We take the $B$ site as the origin. The relative position vectors of the three NN $A$ sites with respect to the $B$ site are given by:
$\boldsymbol{\tau}^A_1 = (0, d) = \frac{1}{3}(\mathbf{a}_1 + \mathbf{a}_2)$,
$\boldsymbol{\tau}^A_2 = \left( -\frac{d\sqrt{3}}{2}, -\frac{d}{2} \right) = \boldsymbol{\tau}^A_1 - \mathbf{a}_1$, and
$\boldsymbol{\tau}^A_3 = \left( \frac{d\sqrt{3}}{2}, -\frac{d}{2} \right) = \boldsymbol{\tau}^A_1 -\mathbf{a}_2$.
The three nearest-neighbor vectors from the $B$ site to the $C$ sites are simply the negative of those to the $A$ sites:
$\boldsymbol{\tau}^C_i = -\boldsymbol{\tau}^A_i$ for $i = 1, 2, 3$.
The corresponding reciprocal lattice vectors are given by,
$\mathbf{b}_1 = \frac{4\pi}{\sqrt{3}a} \left( \frac{\sqrt{3}}{2}, \frac{1}{2} \right)$ and
$\mathbf{b}_2 = \frac{4\pi}{\sqrt{3}a} \left( -\frac{\sqrt{3}}{2}, \frac{1}{2} \right)$.
The $K_+$ and $K_-$ valleys are located at $\left( \frac{4\pi}{3a}, 0 \right)$ and $\left( -\frac{4\pi}{3a}, 0 \right)$, respectively.
\par In absence of NNN hopping $t_1$, the momentum-space Hamiltonian in the sublattice basis $\left( A_{\mathbf{k}} \quad B_{\mathbf{k}} \quad C_{\mathbf{k}} \right)^{T}$ is given by~\cite{Illes2017}
\begin{equation}
    H(\mathbf{k}) =
\begin{pmatrix}
0 & f(\mathbf{k}) & 0 \\
f^*(\mathbf{k}) & 0 & f(\mathbf{k}) \\
0 &  f^*(\mathbf{k}) & 0
\end{pmatrix} \label{alpha_t_3_Ham},
\end{equation}
where the off-diagonal term is defined as, $f(\mathbf{k}) = - t(1 + e^{-i \mathbf{k}.\mathbf{a}_1} + e^{-i\mathbf{k}.\mathbf{a}_2})$. Near the $K$ valley, the low-energy expansion of $f(\textbf{k})$ yields~\cite{Raoux2014, Illes2015}
\[
f(q_x, q_y) = \frac{\sqrt{3} a t}{2} (q_x - i q_y) = v_F (q_x - i q_y),
\]
where $v_F = \frac{\sqrt{3} a t}{2}$ is the Fermi velocity and $\mathbf{q}$ is measured relative to the $K_+$ point. Substituting this form into Eq.~\eqref{alpha_t_3_Ham}, we obtain the effective low-energy Hamiltonian~\cite{Islam2024,Islam2025}
\begin{equation}
H^\prime(q_x,q_y) = v_F
\begin{pmatrix}
0 & q_x - i q_y & 0 \\
q_x + i q_y & 0 &  q_x - i q_y \\
0 & q_x + i q_y & 0
\end{pmatrix}.
\end{equation}
Now including the NNN hopping amplitude $t_1$, the effective low-energy Hamiltonian near the $K_+$ valley can be written as
\begin{equation}
H(q_x,q_y) = v_F
\begin{pmatrix}
0 & q_x - i q_y & 0 \\
q_x + i q_y & 0 & q_x - i q_y)\\
0 & q_x + i q_y & 0
\end{pmatrix} + t_2 \begin{pmatrix}
q^2 & 0 & 0 \\
0 & q^2 & 0 \\
0 & 0 & q^2
\end{pmatrix}
\end{equation}
where $t_2 = \frac{3}{4}t_1a^2 = \frac{\sqrt{3} a t_1}{2 t} v_F$ and $q^2 = q_x^2+q_y^2$.
\section{The Interlayer Hamiltonian of unstrained TBDL}
\noindent \textbf{${K}_+$ valley}:\\
In the vicinity of $K_+$, the matrix elements of the interlayer Hamiltonian are given by ~\cite{Koshino2015, Ma2024, Paul2026}
\begin{equation}
    U_{\tilde{X},X}(\mathbf{q}, \tilde{\mathbf{q}}) = \sum_{\mathbf{G}, \tilde{\mathbf{G}}} w_{\tilde{X},X}(\mathbf{q} + \mathbf{K}_+ + \mathbf{G}) \, e^{-i \mathbf{G} \cdot \boldsymbol{\tau}_X + i \tilde{\mathbf{G}} \cdot \boldsymbol{\tau}_{\tilde{X}}}
    \delta_{\mathbf{q} + \mathbf{K}_+ + \mathbf{G}, \tilde{\mathbf{q}} + \mathbf{\tilde{K}}_+ + \tilde{\mathbf{G}}}. \label{Interlayer_Ham}
\end{equation}
Here, $\mathbf{G} = m_1 \mathbf{b}_1 + m_2 \mathbf{b}_2$ and $\tilde{\mathbf{G}} = m_1 \tilde{\mathbf{b}}_1 + m_2 \tilde{\mathbf{b}}_2$, where $\mathbf{b}_1$ ($\tilde{\mathbf{b}}_1$) and $\mathbf{b}_2$ ($\tilde{\mathbf{b}}_2$) are the reciprocal lattice vectors of layer-1 (layer-2), corresponding to its primitive lattice vectors  $\mathbf{a}_1$ ($\tilde{\mathbf{a}}_1$) and $\mathbf{a}_2$ ($\tilde{\mathbf{a}}_2$).
Now, we consider the A–B (Bernal) stacking configuration~\cite{Ma2024}, under which the sublattice position vector of the atoms in both layers are given as, $\mathbf{\tau}_A = \frac{1}{3}\left( \mathbf{a}_1 + \mathbf{a}_2\right)$, $\mathbf{\tau}_B = 0$, $\mathbf{\tau}_C = -\frac{1}{3}\left( \mathbf{a}_1 + \mathbf{a}_2\right)$, $\mathbf{\tau}_{\tilde{A}} = \mathbf{\tau}_0 + s \hat{e}_z - \frac{1}{3} \left( \tilde{\mathbf{a}}_1 + \tilde{\mathbf{a}}_2\right)$, $\mathbf{\tau}_{\tilde{B}} = \mathbf{\tau}_0 + s \hat{e}_z + \frac{1}{3} \left( \tilde{\mathbf{a}}_1 + \tilde{\mathbf{a}}_2\right)$, and $\mathbf{\tau}_{\tilde{C}} = \mathbf{\tau}_0 + s \hat{e}_z$. Here, $s$ denotes the interlayer separation, and $\mathbf{\tau}_0$ is the relative in-plane translation vector of layer-2 with respect to layer-1. For simplicity, we consider $\mathbf{\tau}_0 = 0$ in our analysis. Now replacing $\mathbf{G}$ and $\tilde{\mathbf{G}}$ in Eq.~\eqref{Interlayer_Ham}, the interlayer Hamiltonian takes the form
\begin{equation}
    U_{\tilde{X},X}(\mathbf{q}, \tilde{\mathbf{q}}) = \sum_{m_1, m_2} w_{\tilde{X},X}(\mathbf{q} + \mathbf{K}_+ + m_1 \mathbf{b}_1 + m_2 \mathbf{b}_2) \, e^{-i (m_1 \mathbf{b}_1 + m_2 \mathbf{b}_2). \boldsymbol{\tau}_X + i (m_1 \tilde{\mathbf{b}}_1 + m_2 \tilde{\mathbf{b}}_2) . \boldsymbol{\tau}_{\tilde{X}}}
    \delta_{\mathbf{q} + \mathbf{K}_+ + m_1 \mathbf{b}_1 + m_2 \mathbf{b}_2, \tilde{\mathbf{q}} + \mathbf{\tilde{K}}_+ + m_1 \tilde{\mathbf{b}}_1 + m_2 \tilde{\mathbf{b}}_2}. \label{Interlayer_Ham_after}
\end{equation}
The term $\delta_{\mathbf{q} + \mathbf{K}_+ + m_1 \mathbf{b}_1 + m_2 \mathbf{b}_2, \tilde{\mathbf{q}} + \mathbf{\tilde{K}}_+ + m_1 \tilde{\mathbf{b}}_1 + m_2 \tilde{\mathbf{b}}_2}$ in Eq.~\eqref{Interlayer_Ham_after} implies that interlayer scattering between layer-1 and layer-2 is only allowed for specific combinations of $(m_1, m_2)$, for which the coupling $w_{X,\tilde{X}}(\mathbf{q} + \mathbf{K}_+ + m_1 \mathbf{b}_1 + m_2 \mathbf{b}_2)$ is maximized.
The scattering matrices $U_{\mathbf{q}_b}$, $U_{\mathbf{q}_{tr}}$, and $U_{\mathbf{q}_{tl}}$ correspond to momentum transfers $\mathbf{q}_b$, $\mathbf{q}_{tr}$, and $\mathbf{q}_{tl}$, arising from the combinations of $(m_1, m_2) = (0,0),\ (0,1),\ \text{and}\ (-1,0)$, respectively. Now the interlayer Hamiltonian corresponding to $K_+$ valley is given by
\begin{eqnarray}
U(\mathbf{q},\tilde{\mathbf{q}}) 
&=& 
\begin{pmatrix}
U_{\tilde{A},A} & U_{\tilde{B},A} & U_{\tilde{C},A}\\
U_{\tilde{A},B} & U_{\tilde{B},B} & U_{\tilde{C},B}\\
U_{\tilde{A},C}& U_{\tilde{B},C} & U_{\tilde{C},C} \nonumber
\end{pmatrix}\\ \nonumber \\
&=&
\begin{pmatrix}
w_1 & w_2 & w_3\\
w_2 & w_1 & w_2 \\
w_3 & w_2 & w_1
\end{pmatrix}\delta_{\mathbf{q}-\tilde{\mathbf{q}}-\mathbf{q}_b} + 
\begin{pmatrix}
w_1 e^{i\phi}& w_2 & w_3e^{-i\phi}\\
w_2e^{-i\phi} & w_1 e^{i\phi} & w_2 \\
w_3 & w_2 e^{-i\phi}& w_1e^{i\phi}
\end{pmatrix}  \delta_{\mathbf{q}-\tilde{\mathbf{q}}-\mathbf{q}_{tr}}+
\begin{pmatrix}
w_1 e^{-i\phi}& w_2 & w_3e^{i\phi}\\
w_2e^{i\phi} & w_1 e^{-i\phi} & w_2 \\
w_3 & w_2 e^{i\phi}& w_1e^{-i\phi}
\end{pmatrix} \delta_{\mathbf{q}-\tilde{\mathbf{q}}-\mathbf{q}_{tl}}\nonumber\\ \nonumber\\
&=& U_{\mathbf{q}_b} \delta_{\mathbf{q}-\tilde{\mathbf{q}}-\mathbf{q}_b} + U_{\mathbf{q}_{tr}} \delta_{\mathbf{q}-\tilde{\mathbf{q}}-\mathbf{q}_{tr}} + U_{\mathbf{q}_{tl}} \delta_{\mathbf{q}-\tilde{\mathbf{q}}-\mathbf{q}_{tl}} \label{Interlayer_Coupling}
\end{eqnarray}
The matrices $U_{\mathbf{q}_b}$, $U_{\mathbf{q}_{tr}}$, $U_{\mathbf{q}_{tl}}$, which describe the interlayer coupling at $K_+$ valley associated with each momentum transfer, take the following forms: 
$U_{\mathbf{q}_b} =  \begin{pmatrix}
w_1 & w_2 & w_3\\
w_2 & w_1 & w_2 \\
w_3 & w_2 & w_1
\end{pmatrix},$ 
$U_{\mathbf{q}_{tr}} =  \begin{pmatrix}
w_1 e^{i\phi}& w_2 & w_3e^{-i\phi}\\
w_2e^{-i\phi} & w_1 e^{i\phi} & w_2 \\
w_3 & w_2 e^{-i\phi}& w_1e^{i\phi}
\end{pmatrix},$
$U_{\mathbf{q}_{tl}} =  \begin{pmatrix}
w_1 e^{-i\phi}& w_2 & w_3e^{i\phi}\\
w_2e^{i\phi} & w_1 e^{-i\phi} & w_2 \\
w_3 & w_2 e^{i\phi}& w_1e^{-i\phi}
\end{pmatrix}$ (with $\phi = 2\pi/3$).\\\\

\noindent \textbf{${K}_-$ valley}:\\
In the vicinity of $K_-$, the matrix elements of the interlayer Hamiltonian are given by ~\cite{Koshino2015, Ma2024, Paul2026}
\begin{equation}
    U_{\tilde{X},X}(\mathbf{q}, \tilde{\mathbf{q}}) = \sum_{m_1, m_2} w_{\tilde{X},X}(\mathbf{q} + \mathbf{K}_- + m_1 \mathbf{b}_1 + m_2 \mathbf{b}_2) \, e^{-i (m_1 \mathbf{b}_1 + m_2 \mathbf{b}_2). \boldsymbol{\tau}_X + i (m_1 \tilde{\mathbf{b}}_1 + m_2 \tilde{\mathbf{b}}_2) . \boldsymbol{\tau}_{\tilde{X}}}
    \delta_{\mathbf{q} + \mathbf{K}_- + m_1 \mathbf{b}_1 + m_2 \mathbf{b}_2, \tilde{\mathbf{q}} + \mathbf{\tilde{K}}_- + m_1 \tilde{\mathbf{b}}_1 + m_2 \tilde{\mathbf{b}}_2}. \label{Interlayer_Ham_after}
\end{equation}
The term $\delta_{\mathbf{q} + \mathbf{K}_- + m_1 \mathbf{b}_1 + m_2 \mathbf{b}_2, \tilde{\mathbf{q}} + \mathbf{\tilde{K}}_- + m_1 \tilde{\mathbf{b}}_1 + m_2 \tilde{\mathbf{b}}_2}$ in Eq.~\eqref{Interlayer_Ham_after} implies that interlayer scattering between layer-1 and layer-2 is only allowed for specific combinations of $(m_1, m_2)$, for which the coupling $w_{X,\tilde{X}}(\mathbf{q} + \mathbf{K}_- + m_1 \mathbf{b}_1 + m_2 \mathbf{b}_2)$ is maximized.
The scattering matrices $U_{-\mathbf{q}_b}$, $U_{-\mathbf{q}_{tr}}$, and $U_{-\mathbf{q}_{tl}}$ correspond to momentum transfers $\mathbf{q}_b$, $\mathbf{q}_{tr}$, and $\mathbf{q}_{tl}$, arising from the combinations of $(m_1, m_2) = (0,0),\ (0,-1),\ \text{and}\ (1,0)$, respectively. Now the interlayer Hamiltonian corresponding to $K_+$ valley is given by
\begin{eqnarray}
U(\mathbf{q},\tilde{\mathbf{q}}) 
&=& 
\begin{pmatrix}
U_{\tilde{A},A} & U_{\tilde{B},A} & U_{\tilde{C},A}\\
U_{\tilde{A},B} & U_{\tilde{B},B} & U_{\tilde{C},B}\\
U_{\tilde{A},C}& U_{\tilde{B},C} & U_{\tilde{C},C} \nonumber
\end{pmatrix}\\ \nonumber \\
&=&
\begin{pmatrix}
w_1 & w_2 & w_3\\
w_2 & w_1 & w_2 \\
w_3 & w_2 & w_1
\end{pmatrix}\delta_{\mathbf{q}-\tilde{\mathbf{q}}+\mathbf{q}_b} + 
\begin{pmatrix}
w_1 e^{-i\phi}& w_2 & w_3e^{i\phi}\\
w_2e^{i\phi} & w_1 e^{-i\phi} & w_2 \\
w_3 & w_2 e^{i\phi}& w_1e^{-i\phi}
\end{pmatrix}  \delta_{\mathbf{q}-\tilde{\mathbf{q}}+\mathbf{q}_{tr}}+
\begin{pmatrix}
w_1 e^{i\phi}& w_2 & w_3e^{-i\phi}\\
w_2e^{-i\phi} & w_1 e^{i\phi} & w_2 \\
w_3 & w_2 e^{-i\phi}& w_1e^{i\phi}
\end{pmatrix} \delta_{\mathbf{q}-\tilde{\mathbf{q}}+\mathbf{q}_{tl}}\nonumber\\ \nonumber\\
&=& U_{-\mathbf{q}_b} \delta_{\mathbf{q}-\tilde{\mathbf{q}}+\mathbf{q}_b} + U_{-\mathbf{q}_{tr}} \delta_{\mathbf{q}-\tilde{\mathbf{q}}+\mathbf{q}_{tr}} + U_{-\mathbf{q}_{tl}} \delta_{\mathbf{q}-\tilde{\mathbf{q}}+\mathbf{q}_{tl}} \label{Interlayer_Coupling}
\end{eqnarray}
The matrices $U_{-\mathbf{q}_b}$, $U_{-\mathbf{q}_{tr}}$, $U_{-\mathbf{q}_{tl}}$, which describe the interlayer coupling at $K_-$ valley associated with each momentum transfer, take the following forms: 
$U_{-\mathbf{q}_b} =  \begin{pmatrix}
w_1 & w_2 & w_3\\
w_2 & w_1 & w_2 \\
w_3 & w_2 & w_1
\end{pmatrix},$ 
$U_{-\mathbf{q}_{tr}} =  \begin{pmatrix}
w_1 e^{-i\phi}& w_2 & w_3e^{i\phi}\\
w_2e^{i\phi} & w_1 e^{-i\phi} & w_2 \\
w_3 & w_2 e^{i\phi}& w_1e^{-i\phi}
\end{pmatrix},$
$U_{-\mathbf{q}_{tl}} =  \begin{pmatrix}
w_1 e^{i\phi}& w_2 & w_3e^{-i\phi}\\
w_2e^{-i\phi} & w_1 e^{i\phi} & w_2 \\
w_3 & w_2 e^{-i\phi}& w_1e^{i\phi}
\end{pmatrix}$ (with $\phi = 2\pi/3$).\\

Now, taking both valleys into account, the general expression for the interlayer Hamiltonian associated with the $K_\zeta$ valley can be written as
\begin{eqnarray}
U_{\zeta}(\mathbf{q},\tilde{\mathbf{q}}) 
&=& 
\begin{pmatrix}
U_{\tilde{A},A} & U_{\tilde{B},A} & U_{\tilde{C},A}\\
U_{\tilde{A},B} & U_{\tilde{B},B} & U_{\tilde{C},B}\\
U_{\tilde{A},C}& U_{\tilde{B},C} & U_{\tilde{C},C} \nonumber
\end{pmatrix}\\ \nonumber \\
&=&
\begin{pmatrix}
w_1 & w_2 & w_3\\
w_2 & w_1 & w_2 \\
w_3 & w_2 & w_1
\end{pmatrix}\delta_{\mathbf{q}-\tilde{\mathbf{q}}-\zeta \mathbf{q}_b} + 
\begin{pmatrix}
w_1 e^{i\zeta\phi}& w_2 & w_3e^{-i\zeta\phi}\\
w_2e^{-i\zeta\phi} & w_1 e^{i\zeta\phi} & w_2 \\
w_3 & w_2 e^{-i\zeta\phi}& w_1e^{i\zeta\phi}
\end{pmatrix}  \delta_{\mathbf{q}-\tilde{\mathbf{q}}-\zeta\mathbf{q}_{tr}}+
\begin{pmatrix}
w_1 e^{-i\zeta\phi}& w_2 & w_3e^{i\zeta\phi}\\
w_2e^{i\zeta\phi} & w_1 e^{-i\zeta\phi} & w_2 \\
w_3 & w_2 e^{i\zeta\phi}& w_1e^{-i\zeta\phi}
\end{pmatrix} \delta_{\mathbf{q}-\tilde{\mathbf{q}}-\zeta\mathbf{q}_{tl}}\nonumber\\ \nonumber\\
&=& U_{\zeta\mathbf{q}_b} \delta_{\mathbf{q}-\tilde{\mathbf{q}}-\zeta\mathbf{q}_b} + U_{\zeta\mathbf{q}_{tr}} \delta_{\mathbf{q}-\tilde{\mathbf{q}}-\zeta\mathbf{q}_{tr}} + U_{\zeta\mathbf{q}_{tl}} \delta_{\mathbf{q}-\tilde{\mathbf{q}}-\zeta\mathbf{q}_{tl}} \label{Interlayer_Coupling}
\end{eqnarray}
\section{The intralayer and interlayer Hamiltonian for strained TBDL}
Including both NN and NNN hoppings along with a staggered onsite mass term $\Delta S_3$, the low energy intralayer Hamiltonian of the constituent monolayers of TBDL can be described in terms of massless Dirac fermions, where the top (bottom) layer of the TBDL is rotated by an angle $\theta/2$ ($-\theta/2$), as
\begin{eqnarray}
 H_{t/b, \zeta} ({\theta/2})= v_F \mathcal{R}_{\mp \theta/2} \boldsymbol{q}.(\zeta S_1,S_2) + t_2 \mathcal{I}_3q^2 + \Delta S_3,   
\end{eqnarray}
where $v_F = 6326.1$~meV$\cdot$\AA, $t_2$ = $\frac{\sqrt{3} a t_1}{2 t} v_F$, $\mathcal{I}_3$ =  $\begin{pmatrix}
1 & 0 & 0\\
0 & 1 & 0 \\
0 & 0 & 1
\end{pmatrix}$, 
and $\mathcal{R}_{\theta}$ denotes the rotational operator. The momentum $\mathbf{q} = \mathbf{k} - \mathbf{K}_{\zeta}$ is defined relative to the original Brillouin zone (BZ) corner $\mathbf{K}_{\zeta}$ 
of the monolayer dice lattice and $S_i$ ($i = 1, 2, 3$) denotes the matrix representation of the spin-1 operators in the $S_3$ eigenbasis~\cite{Georgi1999}.

The effect of strain can be captured by the linear strain tensor $\mathbf{\mathcal{E}}$, which maps an arbitrary coordinate $\mathbf{r}$ to a new coordinate given by
\begin{equation}
\mathbf{r}^{\prime} = (\mathbb{I} + \mathbf{\mathcal{E}}) \, \mathbf{r}.
\end{equation}
Consequently, the corresponding transformation in momentum space is expressed as
\begin{equation}
\mathbf{q}^\prime = (\mathbb{I}+ \mathbf{\mathcal{E}}^T)^{-1} \mathbf{q} \;\approx\; (\mathbb{I} - \mathbf{\mathcal{E}}^T)\, \mathbf{q},
\end{equation}
where the approximation holds to linear order in the strain tensor. From this point onward, a prime on any quantity denotes that it is evaluated in the presence of strain.

For a uniaxial strain of magnitude $\mathcal{E}_p$ applied along a direction making an angle $\Phi$ with respect to the zigzag axis, the strain tensor takes the explicit form~\cite{Bi2019,Pereira2009}
\begin{eqnarray}
\mathbf{\mathcal{E}} &=& R_\Phi
\begin{pmatrix}
\mathcal{E}_p & 0 \\
0 & -\sigma \mathcal{E}_p
\end{pmatrix}
R^{-1}_\Phi \nonumber \\
&=& \mathcal{E}_p
\begin{pmatrix}
\cos^2 \Phi - \sigma \sin^2 \Phi & (1+\sigma)\cos \Phi \sin \Phi \\
(1+\sigma)\cos \Phi \sin \Phi & \sin^2 \Phi - \sigma \cos^2 \Phi
\end{pmatrix},
\end{eqnarray}
where $\sigma = 0.165$ is the Poisson ratio of the dice lattice, which is also taken to be the same as graphene~\cite{Pereira2009}.
The applied strain influences both the bottom layer Hamiltonian and the interlayer tunneling. In particular, for the bottom layer, strain generates an effective gauge field~\cite{Bi2019,Sun2022} given by
\begin{equation}
\mathbf{A} = \frac{\beta}{d}
\left( \mathcal{E}_{xx} - \mathcal{E}_{yy}\,, -2\mathcal{E}_{xy} \right),
\end{equation}
where $\beta = 1.57$ is the Grüneisen parameter~\cite{Bi2019}, which is also taken to be the same as graphene. Upon performing the substitution $\mathbf{q} \to \mathbf{q} + \zeta \mathbf{A}$, the strained bottom-layer Hamiltonian can be written as
\begin{eqnarray}
 H_{b,\zeta}^{\prime} =  v_F \,{R}_{\theta/2} \left[ (\mathbb{I} + \mathbf{\mathcal{E}}^T)\,\mathbf{q}^{\prime} + \zeta \mathbf{A} \right].(\zeta S_1, S_2) + \Delta S_3.
\end{eqnarray}
The positions of the Dirac points in the bottom layer in the presence of this gauge field, are shifted to~\cite{Bi2019}
\begin{equation}
\mathbf{\mathcal{D}}_{\zeta} = \left(\mathbb{I} - \mathbf{\mathcal{E}}^{T}\right)\,\mathbf{K}_{b,\zeta} - \zeta\,\mathbf{A}. 
\end{equation}

The interlayer tunneling is likewise affected by strain, as it modifies the momentum transfer between the layers. 
Because the momentum in the bottom layer is transformed, the corresponding transferred momenta become
\begin{align}
\zeta \mathbf{q}_b^\prime &=
{R}_{-\frac{\theta}{2}} (\mathbb{I} - \mathbf{\mathcal{E}}^{T})\mathbf{K}_\zeta
- {R}_{\frac{\theta}{2}} \mathbf{K}_\zeta, \nonumber \\
\zeta\mathbf{q}_{tr}^\prime &= {R}_{-\frac{\theta}{2}} (\mathbb{I} - \mathbf{\mathcal{E}}^{T})(\mathbf{K}_\zeta + \zeta \mathbf{b}_2)
- {R}_{\frac{\theta}{2}} (\mathbf{K}_\zeta + \zeta\mathbf{b}_2), \nonumber\\
\zeta \mathbf{q}_{tl}^\prime &= {R}_{-\frac{\theta}{2}} (\mathbb{I} - \mathbf{\mathcal{E}}^{T})(\mathbf{K}_\zeta -\zeta\mathbf{b}_1)
- {R}_{\frac{\theta}{2}} (\mathbf{K}_\zeta -\zeta\mathbf{b}_1). 
\end{align}
Under these modifications, the strained interlayer Hamiltonian can be expressed as
\begin{eqnarray}
U^{\prime}_{\zeta}(\mathbf{q},\tilde{\mathbf{q}}) 
&=& 
U^{\prime}_{\zeta\mathbf{q}^{\prime}_b} \delta_{\mathbf{q}-\tilde{\mathbf{q}}-\zeta\mathbf{q}^{\prime}_b} + U^{\prime}_{\zeta\mathbf{q}^{\prime}_{tr}} \delta_{\mathbf{q}-\tilde{\mathbf{q}}-\zeta\mathbf{q}^{\prime}_{tr}} + U^{\prime}_{\zeta\mathbf{q}^{\prime}_{tl}} \delta_{\mathbf{q}-\tilde{\mathbf{q}}-\zeta\mathbf{q}^{\prime}_{tl}} .\label{Interlayer_Coupling1}
\end{eqnarray}
The hopping matrices remain unchanged and preserve their original structure:
\begin{eqnarray}
U^{\prime \tilde{\alpha}\beta}_{\zeta \mathbf{q}'_b} 
&=& w_{\tilde{\alpha}\beta} 
= U^{\tilde{\alpha}\beta}_{\zeta \mathbf{q}_b}, \\
U^{\prime \tilde{\alpha}\beta}_{\zeta \mathbf{q}'_{tr}} 
&=& w_{\tilde{\alpha}\beta} 
e^{i \zeta \left(-\mathbf{b}_2 \cdot \boldsymbol{\tau}_\beta 
+ \tilde{\mathbf{b}}^{\prime}_2 \cdot \boldsymbol{\tau}'_{\tilde{\alpha}} \right)} \nonumber \\
&=& w_{\tilde{\alpha}\beta} 
e^{i \zeta \left(-\mathbf{b}_2 \cdot \boldsymbol{\tau}_\beta 
+ \tilde{\mathbf{b}}_2 \cdot \boldsymbol{\tau}_{\tilde{\alpha}} \right)}
= U^{\tilde{\alpha}\beta}_{\zeta \mathbf{q}_{tr}}, \\
U^{\prime \tilde{\alpha}\beta}_{\zeta \mathbf{q}'_{tl}} 
&=& w_{\tilde{\alpha}\beta} 
e^{i \zeta \left(\mathbf{b}_1 \cdot \boldsymbol{\tau}_\beta 
- \tilde{\mathbf{b}}^{\prime}_1 \cdot \boldsymbol{\tau}'_{\tilde{\alpha}} \right)} \nonumber \\
&=& w_{\tilde{\alpha}\beta} 
e^{i \zeta \left(\mathbf{b}_1 \cdot \boldsymbol{\tau}_\beta 
- \tilde{\mathbf{b}}_1 \cdot \boldsymbol{\tau}_{\tilde{\alpha}} \right)}
= U^{\tilde{\alpha}\beta}_{\zeta \mathbf{q}_{tl}}.
\end{eqnarray}
Here, $\alpha,\beta = A,B,$ and $C$ denote the sublattice indices, with $w_{\tilde{A}A} = w_{\tilde{B}B} = w_{\tilde{C}C}=w_1$, $w_{\tilde{A}B} = w_{\tilde{B}A} =w_{\tilde{B}C} = w_{\tilde{C}A}= w_2$, and $w_{\tilde{A}C} = w_{\tilde{C}A}=w_3$.
In obtaining these results, we used $\tilde{\mathbf{b}}^{\prime}_{1,2} \cdot \boldsymbol{\tau}'_{\tilde{\alpha}} = \tilde{\mathbf{b}}_{1,2} \cdot \boldsymbol{\tau}_{\tilde{\alpha}}$, 
and neglected any strain dependence of the tunneling amplitudes.
\section{Band structure of strained TBDL}
In this section, we show the band structures of TBDL at the magic angle $\theta = 1.08^\circ$ in the presence of different strain values are plotted along the high-symmetry path $\mathrm{K_t}\rightarrow\mathrm{K_b}\rightarrow\Gamma\rightarrow\mathrm{K_t}$, where the lowest band at the middle band edge is highlighted in red color. Panel (a) representing the energy dispersions in the chiral limit of the system, implemented by setting $w_1=w_3=0$ and $w_2=110.7$~meV, while panel (b) corresponds to the the broken chiral limit of the system, obtained by setting $w_1=w_3=60$~meV and $w_2=110.7$~meV. The calculations for both the panels are carried out using the Fermi velocity $v_F = 6326.1$~meV$\cdot$\AA\ and $t_2 = 0.001\,v_F$, with $\Delta = 170$~meV for panels (a) and $\Delta = 60$~meV for panel (b). The energy dispersions are shown for the three distinct increasing strain values, $\mathcal{E}_p = 0.2\%$, $0.4\%$, and $0.6\%$, represented by solid, dashed and, dotted lines, respectively.
\begin{figure}[h]
    \centering
\includegraphics[width=0.65\textwidth]{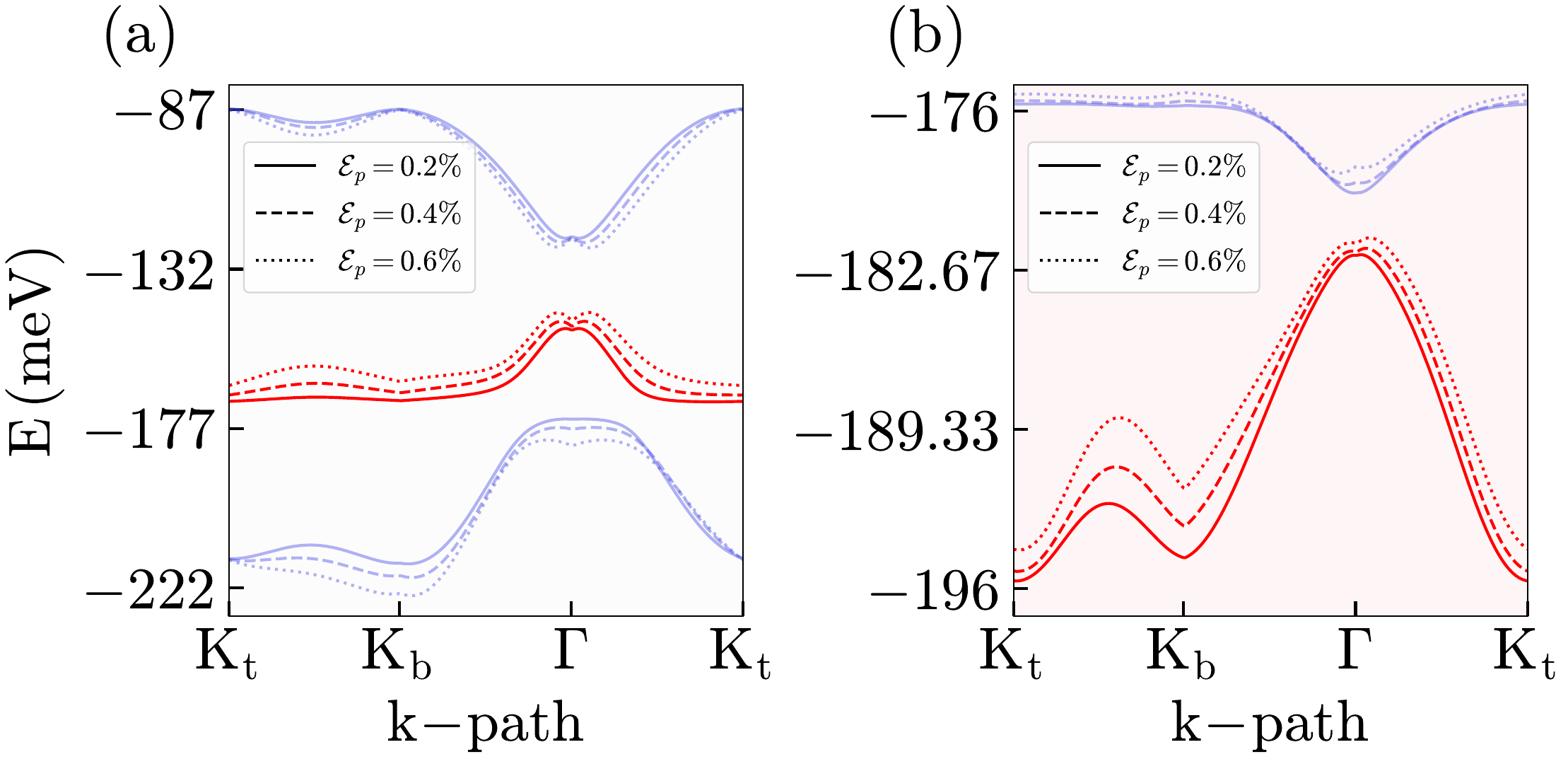}
    \caption{Band structures of TBDL at the magic angle $\theta = 1.08^\circ$ in the presence of different strain values.}\label{fig:4}
\end{figure}
\section{Berry curvature and Berry curvature dipole distribution}
\begin{figure}[h]
    \centering
\includegraphics[width=0.64\textwidth]{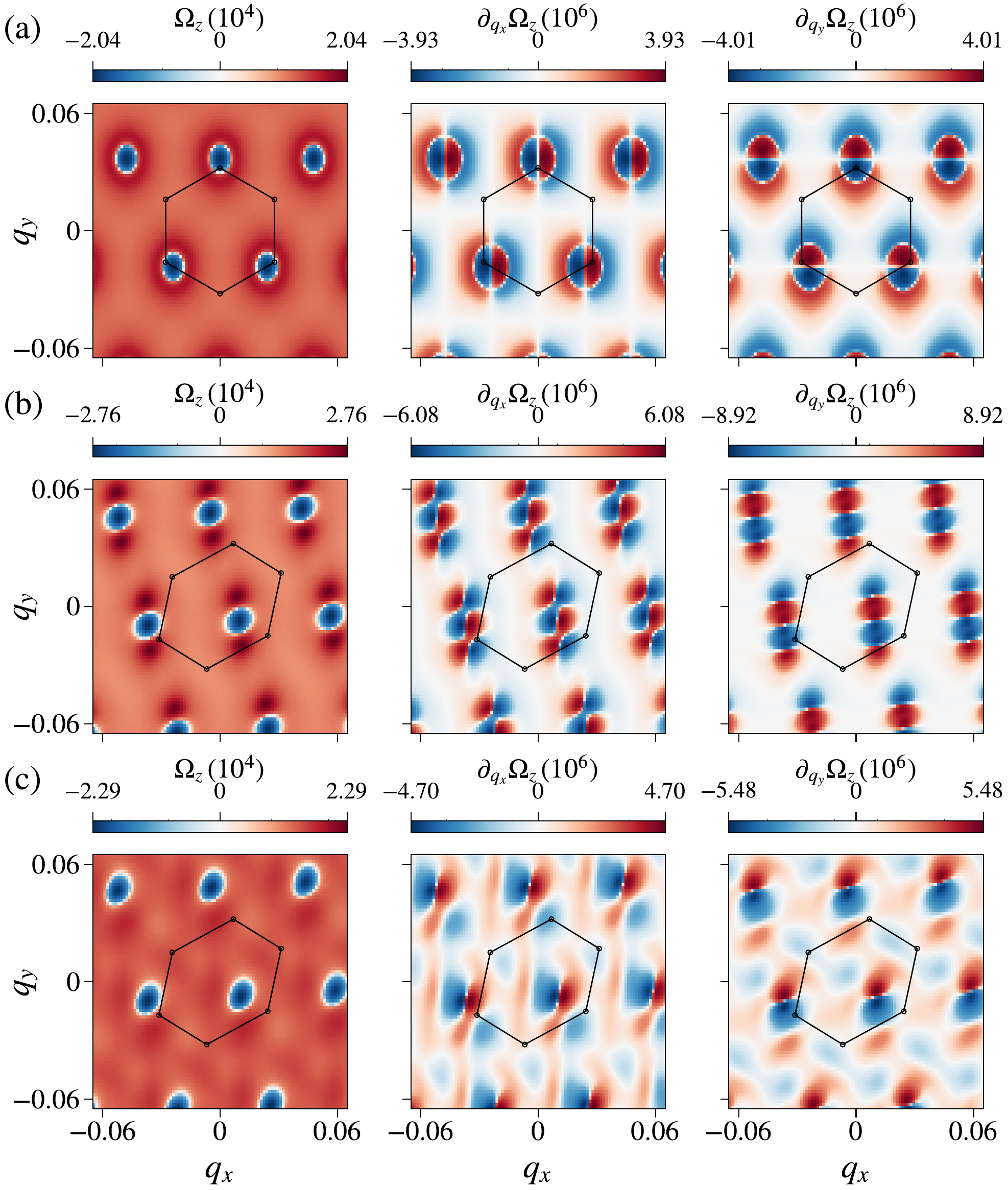}
    \caption{The distribution of the Berry curvature $\Omega_z$ and the BCD density $\partial_{q_a}\Omega_z$ ($a = x, y$) for the lowest band at the middle band edge, plotted in the logarithmic scale, is shown in panels (a)-(c) for three distinct cases. Panel (a) represents the unstrained configuration of the system with preserved $\mathcal{C}_3$ symmetry in the chiral limit ($w_1 = w_3 =0$ and $w_2=110.7$~meV), whereas panels (b) and (c) correspond to the broken $\mathcal{C}_3$ symmetry due to the presence of a strain $\mathcal{E}_p = 0.4\%$, with panel (b) corresponding to the chiral limit and panel (c) representing the broken chiral symmetry regime ($w_1 = w_3 =60$~meV and $w_2=110.7$~meV) of the system. The calculations for all the panels are carried out using the Fermi velocity $v_F = 6326.1$~meV$\cdot$\AA\ and $t_2 = 0.001\,v_F$, with $\Delta = 170$~meV for panels (a) and (b), and $\Delta = 60$~meV for panel (c). Momenta are measured in units of \AA${^{-1}}$, Berry curvature in units of \AA$^{2}$, while the dipole density in units of \AA$^{3}$.}\label{fig:3}
\end{figure}

The distribution of the Berry curvature ($\Omega_z$) and the BCD density ($\partial_{q_b}\Omega_z$ with $b = x, y$) in the $K_+$ valley, both in the presence and absence of $\mathcal{C}_3$ symmetry, are presented in Figs.~\ref{fig:3}[(a)–(c)] for the lowest band at the middle band edge (corresponding to the band index $n=162$). Fig.~\ref{fig:3}(a) and Fig.~\ref{fig:3}(b) correspond to the chiral limit of the system ($w_1 = w_3 =0$ and $w_2=110.7$~meV), with panel (a) representing the case with preserved $\mathcal{C}_3$ symmetry and panel (b) depicting the case with broken $\mathcal{C}_3$ symmetry. While Fig.~\ref{fig:3}(c), on the other hand, corresponds to the broken chiral symmetry regime ($w_1 = w_3 =60$~meV and $w_2=110.7$~meV), where $\mathcal{C}_3$ symmetry of the system remains broken. We emphasize that the presence of TRS in TBDL, the Berry curvature satisfies $\Omega_{z,K_+}(\mathbf{q}) = -\Omega_{z,K_-}(-\mathbf{q})$. In presence of $\mathcal{C}_3$ symmetry (i.e., in the absence of strain), the Berry curvature and the BCD density shows symmetric patterns in the MBZ [see Fig.~\ref{fig:3}(a)]. Interestingly, we find that, in the presence of $\mathcal{C}_3$ symmetry, although each valley in TBDL exhibits large Berry curvature, but the total BCD (BCD density integrated over the whole BZ) vanishes identically. However, in any realistic systems strain can be applied to break the $\mathcal{C}_3$ symmetry, which in turn gives rise to a finite BCD in the system. In the presence of strain the BCD density shows asymmetric distributions [see Fig.~\ref{fig:3}(b) and Fig.~\ref{fig:3}(c)], which generate a net dipole within the moiré unit cell.
\section{Chern phase diagram of the band at the lower edge of the middle subband in unstrained TBDL}
\begin{figure}[h]
    \centering
\includegraphics[width=0.45\textwidth]{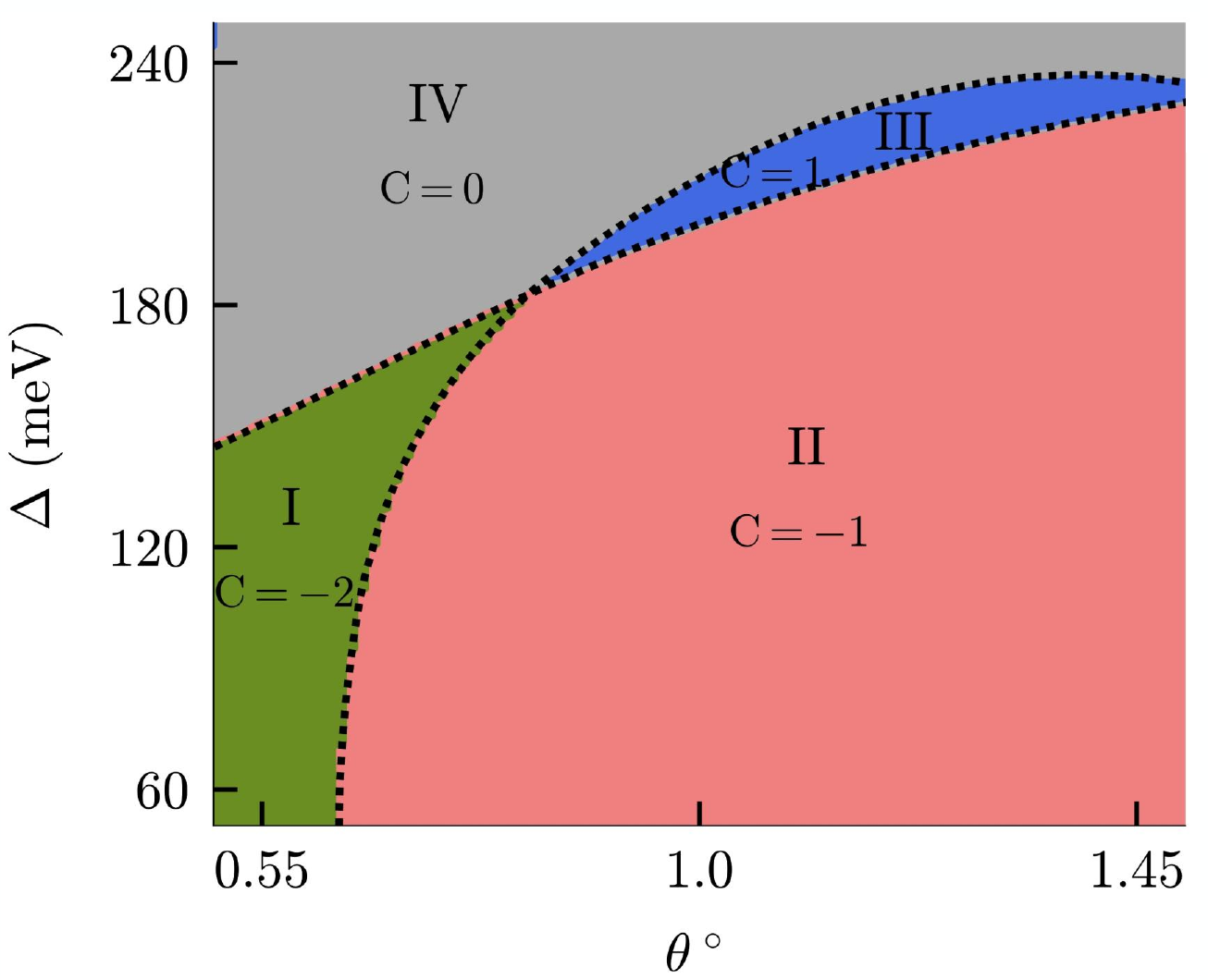}
    \caption{Chern number phase diagram for the band index $n=162$. The figure shows emergent topological phases for the lowest eigenstate of the middle sub-band of unstrained TBDL, as a function of the onsite mass $\Delta$ and the twist angle $\theta$.}\label{Fig:Phase3}
\end{figure}
In this section, we analyze the topological phases that emerge corresponding to the $K_+$ valley in unstrained TBDL as a function of the onsite potential $\Delta$ and the twist angle $\theta$. 
The phase diagram depicting the Chern number of the eigenstate with eigenvalue at the lower edge of the middle sub-band (band index $n=162$), in the $\Delta-\theta$ plane and within the chiral symmetry regime, implemented by setting $w_1 = w_3 = 0$ and $w_2 = 110.7$~meV, is shown in Fig. \ref{Fig:Phase3}. The emergence of multiple topological phases corresponding to Chern number $C=1, -1$ and $-2$ is observed, separated by distinct transition lines.
We observe that for high values of $\Delta$, the band transitions into a completely trivial phase while for its lower values, the $C=-1$ phase is predominant.

\section{Probing distinct topological phase transitions marked by Chern number changes $C=-2$ $\rightarrow$ $C=-1$ and $C=-2$ $ \rightarrow$ $C=0$}
\begin{figure}[h]
    \centering
\includegraphics[width=0.9                             \textwidth]{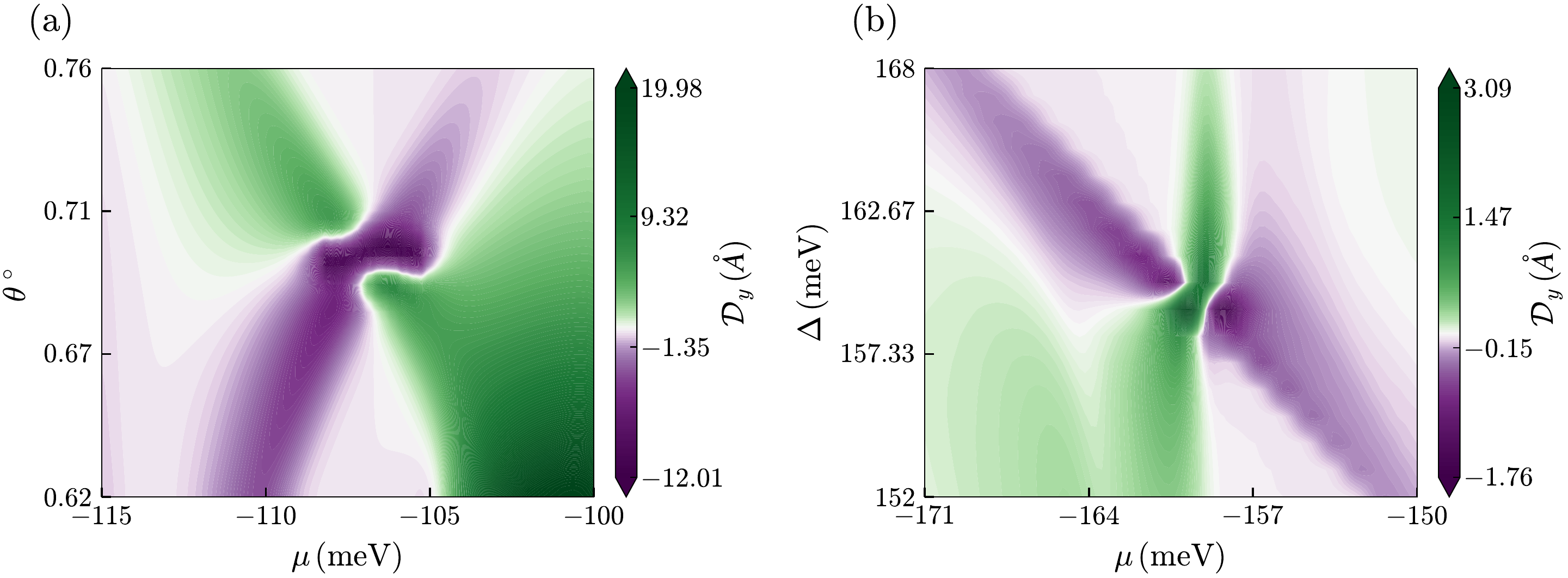}
    \caption{Topological phase transitions identified through the variation of BCD in (a) the $\theta$--$\mu$ plane, with $\Delta$ fixed at 120~meV, corresponding to the transition from Phase (I) to Phase (II), and (b) the $\Delta$--$\mu$ plane, with $\theta$ fixed at $0.6^\circ$, corresponding to the transition from Phase (I) to Phase (IV).}\label{fig:Phase4}
\end{figure}
In this section, we present the phase diagram of the BCD component $\mathcal{D}_y$ in the $\theta$--$\mu$ plane for a fixed $\Delta=120$~meV, as illustrated in Fig.~\ref{fig:Phase4}(a), and in the $\Delta$--$\mu$ plane, shown in Fig.~\ref{fig:Phase4}(b), while keeping the twist angle fixed at $\theta=0.6^\circ$. Fig.~\ref{fig:Phase4}(a) corresponds to the topological phase transition from Phase (I) to Phase (II), whereas Fig.~\ref{fig:Phase4}(b) represents the transition from Phase (I) to Phase (IV). The butterfly like structures in the colormap of $\mathcal{D}_y$ indicate points of phase transition which can be correlated directly with the Chern phase plot in Fig.~3 of the main manuscript. The lower two lobes of each butterfly pattern along a horizontal line (with $\theta$ fixed in Fig.~\ref{fig:Phase4}(a) and $\Delta$ fixed in Fig.~\ref{fig:Phase4}(b)) represent the values of $\mathcal{D}_y$ for the corresponding band at a given chemical potential $\mu$. On the other hand, the sign reversal of $\mathcal{D}_y$ along a vertical line (with fixed $\mu$) reflects the change in the sign of the BCD for the same band as a function of either $\theta$ or $\Delta$, respectively. As evident from Fig.~\ref{fig:Phase4}(a) and Fig.~\ref{fig:Phase4}(b), the phase transitions occur near $\theta=0.698^\circ$ and $\Delta=159.3\mathrm{~meV}$, respectively. These results highlight the experimental significance of the BCD as an effective probe for identifying topological phase transitions.
\section{Influence of temperature on the BCD in the chiral limit}
\begin{figure}[h]
    \centering
\includegraphics[width=0.4                                \textwidth]{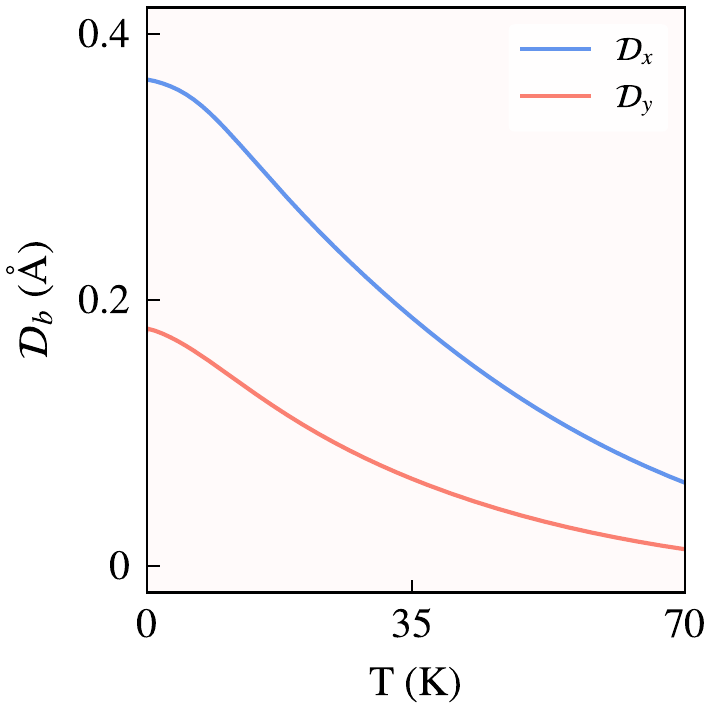}
    \caption{Temperature dependence of the BCD components at the magic angle $\theta = 1.08^\circ$ in the chiral limit of the system, implemented by setting $w_1 = w_3 = 0$ and $w_2 = 110.7$~meV, with $\Delta = 170$~meV at a strain $\mathcal{E}_p = 0.4\%$. The chemical potential is fixed at $\mu = -165.9$~meV.}\label{fig:6}
\end{figure}
So far, all the results we discussed have been obtained at a fixed temperature of $T = 4$~K. Now in-order to see the temperature dependence of BCD in TBDL, in this section we plot the BCD components, $\mathcal{D}_x$ and $\mathcal{D}_y$, as a function of temperature in the chiral limit of the system at the magic angle $\theta = 1.08^\circ$, as shown in Fig.~\ref{fig:6}. Here, we fix the chemical potential at $\mu = -165.9$~meV. It is evident from the Fig.~\ref{fig:6} that, as the temperature increases the dipole components gradually decreases. This can be attributed to the thermal broadening of the Fermi Dirac distribution function, thereby the reduction in its slope suppresses the magnitude of the dipole components. This behavior indicates that one could detect a strong NLH response at the low temperatures in TBDL. This temperature dependence of the BCD components in TBDL exhibits similar kind of behavior to that reported in the previous experimental studies on TBG~\cite{Duan2022} and twisted bilayer WTe$_2$~\cite{Cao2025}.
\end{document}